\documentclass[a4paper,11pt]{article}
\pdfoutput=1 

\usepackage{jcappub} 

\usepackage[T1]{fontenc} 

\title{\boldmath Combining Full-Shape and BAO Analyses of Galaxy Power Spectra:\\
\Large A 1.6\% CMB-independent constraint on $H_0$}

\def\beq{\begin{eqnarray}}
\def\eeq{\end{eqnarray}}
\newcommand{\Mpch}{h^{-1}\mathrm{Mpc}}
\newcommand{\hMpc}{h\,\mathrm{Mpc}^{-1}}

\usepackage{bm}
 
\let\vec\bm

\newcommand{\resub}[1]{#1}
\numberwithin{equation}{section}

\author[a,1]{Oliver H.\,E. Philcox\note{Corresponding author.}}
\author[b,c]{Mikhail M. Ivanov}
\author[d]{Marko Simonovi\'{c}}
\author[e]{Matias Zaldarriaga}

\affiliation[a]{Department of Astrophysical Sciences, Princeton University,\\ Princeton, NJ 08540, USA}
\affiliation[b]{Center for Cosmology and Particle Physics, Department of Physics, New York University,\\ New York, NY 10003, USA}
\affiliation[c]{Institute for Nuclear Research of the Russian Academy of Sciences,\\ 60th October Anniversary Prospect, 7a, 117312 Moscow, Russia}
\affiliation[d]{Theoretical Physics Department, CERN,\\ 1 Esplanade des Particules, Geneva 23, CH-1211, Switzerland}
\affiliation[e]{School of Natural Sciences, Institute for Advanced Study,\\ 1 Einstein Drive, Princeton, NJ 08540, USA}

\emailAdd{ohep2@alumni.cam.ac.uk}
\emailAdd{mi1271@nyu.edu}
\emailAdd{marko.simonovic@cern.ch}
\emailAdd{matiasz@ias.edu}

\abstract{We present cosmological constraints from a joint analysis of the pre- and post-reconstruction galaxy power spectrum multipoles from the final data release of the Baryon Oscillation Spectroscopic Survey (BOSS). Geometric constraints are obtained from the positions of BAO peaks in reconstructed spectra, which are analyzed in combination with the unreconstructed spectra in a full-shape (FS) likelihood using a joint covariance matrix, giving stronger parameter constraints than FS-only or BAO-only analyses. We introduce a new method for obtaining constraints from reconstructed spectra based on a correlated theoretical error, which is shown to be simple, robust, and applicable to any flavor of density-field reconstruction. Assuming $\Lambda$CDM with massive neutrinos, we analyze clustering data from two redshift bins  $z_\mathrm{eff}=0.38,0.61$ and obtain $1.6\%$ constraints on the Hubble constant $H_0$, using only a single prior on the current baryon density $\omega_b$ from Big Bang Nucleosynthesis (BBN) and no knowledge of the power spectrum slope $n_s$. This gives $H_0 = 68.6\pm1.1\,\mathrm{km\,s}^{-1}\mathrm{Mpc}^{-1}$, with the inclusion of BAO data sharpening the measurement by $40\%$, representing one of the strongest current constraints on $H_0$ independent of cosmic microwave background data, comparable with recent constraints using BAO data in combination with other data-sets. Restricting to the best-fit slope $n_s$ from Planck (but without additional priors on the spectral shape), we obtain a $1\%$ $H_0$ measurement of $67.8\pm 0.7\,\mathrm{km\,s}^{-1}\mathrm{Mpc}^{-1}$. Finally, we find strong constraints on the cosmological parameters from a joint analysis of the FS, BAO, and Planck data. This sets new bounds on the sum of neutrino masses $\sum m_\nu < 0.14\,\mathrm{eV}$ (at $95\%$ confidence) and the effective number of relativistic degrees of freedom $N_\mathrm{eff} = 2.90^{+0.15}_{-0.16}$, though contours are not appreciably narrowed by the inclusion of BAO data.}

\begin{document}


\maketitle
\flushbottom

\section{Introduction}

Since the dawn of civilization, cultures have striven to understand essential properties of the Universe; its composition, evolution and structure. In the current cosmological paradigm, $\Lambda$CDM, many of these questions have been reduced to determining a small set of numbers controlling the relative proportions of cosmological components and the expansion history. Constraining such parameters, however, is non-trivial and a subject of much debate. Different data-sets have sometimes yielded inconsistent results, especially for the Hubble parameter $H_0$, which encodes the Universe's expansion rate. In particular, analyses of two of the great pillars of precision cosmology, the Cosmic Microwave Background (CMB; e.g.\,\citep{2018arXiv180706209P}) and the distances to Type Ia Supernovae (SNe) calibrated from local distance ladders (e.g.\,\citep{2019ApJ...876...85R,2019NatRP...2...10R}), are not in agreement, leading some to claim the existence of a `tension' between the early- and late-Universe, prompting a swathe of new physics to be invented (see Ref.\,\citep{2019arXiv190803663K} for a review). 

To resolve such controversy, independent probes are required, and a particularly promising one lies within the analysis of Large Scale Structure (LSS) information, through redshift-space galaxy surveys. Current surveys provide measurements of the angular positions and redshifts of galaxies across large cosmological volumes, and with upcoming surveys such as the Large Synoptic Sky Telescope (LSST) \citep{2009arXiv0912.0201L}, SPHEREx \citep{2014arXiv1412.4872D}, Euclid \citep{2011arXiv1110.3193L} and the Dark Energy Spectroscopic Instrument (DESI) \citep{2011AJ....142...72E} the volume of data will only continue to grow. Constraints on cosmology are most commonly obtained by considering the two-point clustering statistics of the galaxies, either in configuration \citep{2017MNRAS.466.2242B,2017MNRAS.464.3409B} or Fourier \citep{2017MNRAS.469.1369S,2017MNRAS.464.1640S,2017MNRAS.464.1168R} space, though recent papers have begun to explore inclusion of the three-point bispectrum \citep{2017MNRAS.465.1757G,2018MNRAS.478.4500P,2019JCAP...11..034C}. Through these statistics, a number of features can be probed, including the Alcock-Pacyznski effect \citep{1979Natur.281..358A}, allowing measurement of angular and radial distance scales, and redshift-space distortions (RSD) \citep{1987MNRAS.227....1K}, from the non-linear conversion between configuration- and redshift-space, allowing constraints to be placed on
the amplitude of velocity fluctuations $f\sigma_8$ \citep{2017MNRAS.469.1369S} and cosmological-scale tests of General Relatively \citep{2008Natur.451..541G}, for example.

Historical analyses of galaxy power spectra have focused around the signature of Baryon Acoustic Oscillations (BAO); an imprint of sound waves in the pre-recombination Universe \citep{1970ApJ...162..815P,1970Ap&SS...7....3S}, which create a `standard-ruler', allowing determination of the distance-redshift relation at the effective sample redshift (e.g.\,\citep{1996ApJ...471...30H,2003ApJ...598..720S,2005ApJ...633..560E}). This reduces the analysis to simply measuring the positions of BAO harmonics in the observed spectrum, though this is complicated by various non-linear effects (e.g.\,\citep{1999MNRAS.304..851M,2005ApJ...633..575S,2010ApJ...720.1650S,2011ApJ...734...94M}). Such constraints are significantly sharpened by the process of density-field reconstruction \citep{2007ApJ...664..675E,2012MNRAS.427.2132P,2014JCAP...02..042S,2017PhRvD..96b3505S,2018MNRAS.478.1866H}, which reduces the information loss afforded by non-linear effects by displacing galaxies by an estimate of their large scale bulk flow. This has been applied in a number of cosmological analyses (e.g. \citep{2012MNRAS.427.3435A,2012MNRAS.427.2132P,2014MNRAS.441...24A,2017MNRAS.464.3409B}).

Going beyond the BAO peak, the broadband power spectrum contains information about a number of physical effects, yet its analysis is complicated by difficulties in its modelling. Creating a consistent model has been a subject of much work, but recent advances have led to the development of the `Effective Field Theory of Large Scale Structure' (hereafter EFT) \citep{2012JHEP...09..082C,2012JCAP...07..051B}, with extensions incorporating galaxy bias \cite{2015JCAP...11..007S,2009JCAP...08..020M,2014JCAP...08..056A,2015JCAP...05..019L}, redshift-space distortions \citep{2014arXiv1409.1225S,2016arXiv161009321P} and higher-order corrections \citep{2014JCAP...07..057C,2015PhRvD..92l3007B}. Of particular interest is the description of long-wavelength modes (bulk flows), which cannot be treated perturbatively. Introduction of infra-red resummation schemes into the EFT alleviates this problem, allowing for an accurate theoretical model into the non-linear regime \citep{2015JCAP...02..013S,2018JCAP...05..019S,2015PhRvD..92d3514B,2016JCAP...07..028B,2018JCAP...07..053I,2016JCAP...03..057V}. Recently this model has been applied to current observational data \citep{2019arXiv190905277I,2019arXiv191208208I,2019arXiv190905271D,Colas:2019ret}, allowing constraints to be placed on cosmology using all the information contained in the power spectrum on quasi-linear scales.

In this work, we aim to produce stronger constraints on cosmology by combining the two analysis techniques discussed above, utilizing both information from the full-shape (FS) of the galaxy power spectrum and the sharp location of the BAO peak after density field reconstruction. The most obvious approach
to this would be to model the galaxy power spectrum \textit{after} reconstruction; this turns out to be a difficult task since reconstruction significantly modifies the broadband shape of the power spectrum, leading to complex perturbation theory models, with strong dependence on the particular reconstruction algorithm and its assumptions \citep{2019JCAP...02..027S}. Whilst this is a subject of continuing research, thus far, consistent perturbative models have only been computed (assuming the simplest reconstruction framework) for the matter field in real \citep{2017PhRvD..96d3513H} and redshift \citep{2020PhRvD.101d3510H} space, as well as the galaxy spectrum in a Lagrangian framework \citep{2019JCAP...09..017C}. A consistent, and simply computable, EFT for reconstructed spectra is yet to be derived. 
Furthermore, it has been recently demonstrated in Ref.~\cite{2019arXiv191208208I}
that the reconstructed power spectrum of 
the BOSS data can \textit{only} provide geometric information through the positions of the BAO peaks.
 It was shown in
 Ref.\,\citep{2019JCAP...11..034C} that even in the hypothetical case where the
  post-reconstruction non-linear damping scale
  $\Sigma_\mathrm{NL}$ is known precisely, the
  reconstructed power spectrum shape is only able to improve measurements of the physical
  baryon density $\omega_b$, which will still be significantly weakly constrained
 compared to the BBN or the Planck limits even in the era of future LSS surveys. 

Motivated by this, we opt to extract only BAO information from the reconstructed spectra and use this to inform an FS analysis of the unreconstructed power spectra with a joint covariance, using the techniques developed in Ivanov \textit{et al.} \citep{2019arXiv190905277I}, and applying the method to the final data release of the Baryon Oscillation Spectroscopic Survey (BOSS) \citep{2017MNRAS.470.2617A}. This allows for a full Markov Chain Monte Carlo (MCMC) exploration of the cosmological parameter space, and in particular, we are able to place strong constraints on the expansion rate $H_0$ \resub{from low redshift data} $(z\sim 0.5)$ \resub{within $\Lambda$CDM}, without applying any prior information from Planck, obtaining a value inconsistent with with SN analyses \citep{2019ApJ...876...85R}. \resub{Our $H_0$ constraints are of similar precision to those obtained using BAO data in combination with weak-lensing \citep{2018MNRAS.480.3879A} and Lyman-$\alpha$ information \citep{2019JCAP...10..044C}.}

This paper has the following structure. We begin by discussing our motivations for the analysis (Sec.\,\ref{subsec: motivations}), before outlining the theoretical model used to analyze reconstructed and unreconstructed power spectra in Secs.\,\ref{subsec: recon-analysis}\,\&\,\ref{subsec: unrecon-analysis} respectively. In Sec.\,\ref{sec: data-and-priors} we review the data-sets (both observational and mock) used, and discuss our choice of priors for cosmological and nuisance parameters. The results of the BAO analysis are presented in Sec.\,\ref{sec: results-bao}, before we discuss the joint FS+BAO analysis in Sec.\,\ref{sec: results-joint}, including \resub{CMB-independent} cosmological constraints. In Sec.\,\ref{sec: results-Planck}, we discuss the combination of CMB with galaxy surveys and give the results of parameter inferences from the combination of all three data-sets (BAO, FS and Planck), before concluding with a summary in Sec.\,\ref{sec: conclusion}. Appendices \ref{appen: patchy-tests}\,\&\,\ref{appen: individual-chunks} contain supplementary material regarding tests of the pipeline on mock data and analyses of each data chunk separately. 

For the casual reader who has less interest in technical details, we recommend a perusal of Sec.\,\ref{subsec: motivations} to understand our motivations and rough pipeline, before skipping to the main exposition of results in Secs.\,\ref{sec: results-joint} and \ref{sec: results-Planck}. The key cosmological parameters obtained from the FS+BAO analyses are presented in Tab.\,\ref{table0}, and Figs.\,\ref{fig:wns}\,\&\,\ref{fig:fix-ns}, with joint constraints with Planck given in Tab.\,\ref{tab:Planck} and Figs.\,\ref{fig:mnuPl}\,\&\,\ref{fig:neffPl}.

\section{Methodology and Implementation}\label{sec: analysis}
Here, we discuss both the theoretical underpinnings and practical application of our approach in detail. We begin with a few words of background concerning power spectrum reconstruction.

\subsection{Motivation and Theoretical Background}\label{subsec: motivations}
When analyzing observational galaxy power spectra, the primary goal has recently been
to measure the position of the prominent BAO peak (e.g.\,\citep{1996ApJ...471...30H,2003ApJ...598..720S,2005ApJ...633..560E}). This is usually done via constraining the Alcock-Paczynski (hereafter AP) parameters $\vec\alpha\equiv\{\alpha_\parallel, \alpha_\perp\}$ \citep{1979Natur.281..358A}, which measure the radial and angular distances by means of
geometric distortions induced by an incorrectly assumed fiducial cosmology used in co-ordinate conversion.\footnote{We stress that one does not need to generate these distortions on purpose. 
Even if the fiducial cosmology exactly matches the true one, 
the distortions will be contained in the trial 
theoretical templates 
that are used to fit the data during MCMC scans 
and hence the AP effect will 
still be effective.
} The two parameters separately measure distortions parallel and perpendicular to the line-of-sight (hereafter LoS), and encode the Hubble parameter $H(z)$, the sound horizon at the redshift of decoupling, $r_s(z_d)$, and angular diameter distance $D_A(z)$ via
\beq\label{eq: AP-params}
    \alpha_\parallel &=& \frac{H^\mathrm{fid}(z)r_s^\mathrm{fid}(z_d)}{H(z)r_s(z_d)},\quad
    \alpha_\perp = \frac{D_A(z)r_s^\mathrm{fid}(z_d)}{D_A^\mathrm{fid}(z)r_s(z_d)},
\eeq
where the superscript `$\mathrm{fid}$' indicates the values in some (unimportant) fiducial cosmology and $z$ is the effective redshift of the sample. If our aim is a precise measurement of $\vec\alpha$, the precision is greatly improved by reconstructing the galaxy field, first proposed by Eisenstein \textit{et al.} \citep{2007ApJ...664..675E}. Many variants exist (e.g.\,\citep{2012MNRAS.427.2132P,2017PhRvD..96b3505S,2018MNRAS.478.1866H,2019MNRAS.484.3818S}), all based on the notion that, by shifting galaxies closer to their initial (Lagrangian) positions, we are able to reduce the effects of non-linear structure formation and redshift-space distortions and reduce the information leakage from the power spectrum to higher-order statistics \citep{2015PhRvD..92l3522S}. In its most basic form, reconstruction consists of smoothing the late time density field by some (Gaussian) kernel $W(k) = e^{-k^2\Sigma_\mathrm{smooth}^2/2}$ on scale $\Sigma_\mathrm{smooth}$, then shifting the galaxies and a set of uniformly distributed particles by their negative Zel'dovich displacements. This has been used in a number of studies (e.g.\,\citep{,2012MNRAS.427.3435A,2012MNRAS.427.2132P,2014MNRAS.441...24A,2017MNRAS.464.3409B}), allowing stronger constraints to be placed on $\vec\alpha$ than with the unreconstructed spectra.

It has recently been shown \citep{2019arXiv190905277I,2019arXiv190905271D} (and earlier \citep{2017MNRAS.464.1640S,2017MNRAS.467.2085G,2017MNRAS.469.1369S}) that the full shape (FS) of the galaxy power spectrum (in addition to the BAO peak) can be used to place strong constraints on cosmological parameters, by comparison with accurate theoretical models, based on the Effective Field Theory of Large Scale Structure (hereafter EFT) \citep{2012JCAP...07..051B,2012JHEP...09..082C}. In previous analyses, this has been applied only to unreconstructed power spectra; one may na\"{i}vely expect stronger cosmological constraints when using the \textit{reconstructed} power spectra, which contain information from both the two-point and higher-point correlators. In practice, this is difficult to achieve, since the process of reconstruction distorts the broadband spectral shape. Whilst a number of works have attempted to model this in perturbation theory \citep{2017PhRvD..96d3513H,2019JCAP...09..017C,2020PhRvD.101d3510H}, it is laborious even in the simplest of reconstruction schemes. In addition, to fully analyze the reconstructed power spectrum at one-loop order, we require both resummation of long-wavelength modes, and an effective treatment of small-scale physics, neither of which have yet been considered. In addition, implicit assumptions in the reconstruction procedure, such as the fiducial cosmology and bias, can have non-negligible impacts on the broadband shape \citep{2019JCAP...02..027S}, further complicating the analysis, though we note that these are \textit{not} expected to noticeably affect the BAO peak position. 

To extract maximal information from galaxy power spectra, we propose a joint method, motivated by the following observations:
\begin{itemize}
    \item The reconstructed power spectrum can be used to place strong constraints on the AP parameters, regardless of the precise details of the reconstruction method;
    \item The constraints on $\vec\alpha$ from the reconstructed spectra are largely independent of non-linear damping scale $\Sigma_\mathrm{NL}$ (defined in Eq.\,\ref{eq: IR-resum-LO}). Thus a more precise theoretical model for the reconstructed field will \textit{not} sharpen our constraints on $\vec\alpha$;
    \item The full shape of the unreconstructed power spectrum can be used to place strong constraints on cosmological parameters, using the mildly non-linear models from EFT;
    \item The full shape of the reconstructed power spectrum is difficult to model and modified significantly by the flavor of reconstruction and modeling assumptions.
\end{itemize}
The main approach of this paper is therefore:
\begin{enumerate}
    \item Use the reconstructed power spectra to place constraints on the AP parameters $\vec\alpha$ (hereafter the `BAO analysis'), in particular obtaining a best-fit $\vec\alpha$ for the given data-set.
    \item Generate a joint covariance matrix between the unreconstructed spectra and the best-fit AP parameters. This can be done from mocks or by basic theoretical calculations.
    \item Use the full shape (FS) of the unreconstructed spectra, together with the best-fit $\vec\alpha$ and the joint covariance, to produce tight constraints on cosmological parameters (hereafter the `FS analysis'). Here, $\vec\alpha$ is treated as an \textit{additional observable}, which constrains the model alongside the unreconstructed spectra.
\end{enumerate}
By combining both measurements, we can make use of both the sharp BAO peak following reconstruction, and the undistorted broadband shape, without requiring complex (and computationally intensive) new modeling. \resub{In a sense, our methods are analogous to those used for CMB analysis; the broadband shape of the galaxy power spectrum constrains physical parameters that define the cosmological sound horizon $r_s$. Geometric information encoded in the Alcock-Paczynski parameters then set the measurement of $H_0$, in particular from the BAO peak position.}

\subsection{BAO Analysis: Extracting Alcock-Paczynski Parameters from Reconstructed Spectra}\label{subsec: recon-analysis}
Within the reconstructed power spectra, the key information encoding the AP parameters lies within the wiggly parts of the spectra, which, whilst prominent at small wavenumber, are hidden by the broadband spectrum at higher $k$. In order to extract $\vec\alpha$ it is thus critical to (a) model the full reconstructed spectrum into the quasi-linear regime, (b) separate the wiggle and no-wiggle parts of the spectrum or (c) marginalize over the unknown broadband component. Whilst the optimal approach would be (a), modeling this spectrum beyond linear theory is difficult, as previously discussed. Fortunately, as shown below, such modeling does \textit{not} strongly affect the efficacy of our AP constraints, since they are sourced only by the wiggle part, and the perturbative computations mainly constrain the broadband spectral shape. In previous works including Refs.\,\citep{2016MNRAS.460.2453S,2017MNRAS.464.3409B,2020MNRAS.tmp..347H}, approach (c) is adopted, with a number of free polynomial parameters added to marginalize over the spectral shape. In this work, we use a somewhat different approach, based on a \textit{theoretical error} template, as proposed by Baldauf \textit{et al.} \citep{2016arXiv160200674B}, which effectively marginalizes over the broadband shape by introducing a large additional covariance with correlation length larger than the BAO scale. This will be discussed further in Sec.\,\ref{subsec: theoretical-error}.

In this section, we will require some fiducial cosmology against which the AP parameters are calibrated. \resub{Following \citep{2017MNRAS.464.3409B}}, we assume a flat $\Lambda$CDM universe with $\{\Omega_m = 0.31,\Omega_bh^2 = 0.022, h = 0.676, \sigma_8 = 0.824, n_s = 0.96, \sum m_\nu = 0.06\,\mathrm{eV}, Y_\mathrm{He} = 0.2454, z_\mathrm{reio}=11.357, r^\mathrm{fid}_s(z_d) = 147.78\,\mathrm{Mpc}\}$, where $r_s(z_d)$ is the sound horizon at decoupling. It is worth mentioning that in standard cosmological models, only the \textit{isotropic} AP parameter $\alpha_\mathrm{iso} = \alpha_\parallel^{1/3}\alpha_\perp^{2/3}$ is important, which provides a strong distance measurement (useful for $H_0$), but, at low redshift, only a weak constraint on $\Omega_m$ \citep{2019JCAP...10..044C,2019arXiv190905277I}. For non-minimal cosmological models however, anisotropic warping can become important, thus we here include both $\alpha_\parallel$ and $\alpha_\perp$ for full generality.

\subsubsection{Theoretical Model for $P_\ell^\mathrm{rec}(k)$}\label{subsec: recon-theoretical-model}
Before considering any shape marginalization, we require a theoretical model for the reconstructed power spectra that is accurate in the linear regime ($k\lesssim0.1\hMpc$) and correctly treats the BAO wiggles at higher $k$. As a starting point, recall the familiar Kaiser power spectrum \citep{1987MNRAS.227....1K}, which is accurate for (unreconstructed) galaxy redshift-space power spectra on linear scales;
\beq\label{eq: P-kaiser}
    P_\mathrm{Kaiser}(k,\mu;z) = \left[b(z)+f(z)\mu^2\right]^2P_\mathrm{lin}(k;z),
\eeq
where $\mu = \hat{\vec k}\cdot\hat{\vec n}$ is the angle between the momentum vector $\vec k$ and the (local) line of sight $\hat{\vec n}$, $b$ is the local linear bias (which is a free parameter in the analysis) and $f$ is the logarithmic growth factor (defined as $f = d\log D(a)/d\log a$ for linear growth factor $D$ and $a = (1+z)^{-1}$). $P_\mathrm{lin}$ is the linear power spectrum computable via \texttt{CAMB} \citep{2011ascl.soft02026L} or \texttt{CLASS} \citep{2011JCAP...07..034B}. From this point forwards, the dependencies on redshift will be left implicit, and quantities assumed to be evaluated at the effective redshift of the galaxy sample.

Whilst Eq.\,\ref{eq: P-kaiser} holds for a standard galaxy power spectrum, it is not appropriate to use for \textit{reconstructed} fields, due to the density-field smoothing and anisotropic reconstruction applied therein. A simple calculation along the lines of Refs.\,\citep{2016MNRAS.460.2453S,2019JCAP...09..017C,2020PhRvD.101d3510H}, then shows that the relevant tree-level power spectrum is in fact
\beq
    P^\mathrm{rec}_\mathrm{tree}(k,\mu) = \begin{cases}\left[b+f\mu^2\right]^2P_\mathrm{lin}(k) & \text{"Rec-Sym"}\\ \left[b+f\mu^2\left(1-W(k)\right)\right]^2P_\mathrm{lin}(k) & \text{"Rec-Iso",}\end{cases}
\eeq
where $W(k)$ is the Gaussian smoothing kernel and "Rec-Sym" and "Rec-Iso", defined in Ref.\, \citep{2016MNRAS.460.2453S}, correspond to (a) shifting the galaxies and random particles by the same amount or (b) shifting the galaxies by an additional factor of $(1+f)$ \resub{along the line of sight}.\footnote{Note that our `Rec-Iso' result agrees with Ref.\,\citep{2016MNRAS.460.2453S} (Appen.\,A) and \resub{corrects a minor typographical error} in Ref.\,\citep{2019JCAP...09..017C} (Eq.\,4.11).} The latter scheme is used to remove RSD on \resub{large} scales, and will be assumed henceforth. \resub{As noted in \citep{2015MNRAS.450.3822W}, this may be sub-optimal for BAO analyses, but is chosen here for better comparison with BOSS. Furthermore, our model is largely insensitive to the broadband shape of the reconstructed spectra.}

Though this model contains the correct linear physics, it is not yet a correct treatment of the BAO wiggles, from which the AP parameters will be extracted. To do this, we must carefully consider the long wavelength (infrared; IR) modes that cannot be treated perturbatively, with some IR resummation procedure \cite{2015JCAP...02..013S,2015PhRvD..92d3514B,2018JCAP...05..019S,2018JCAP...07..053I}. At leading order, this modifies the linear spectrum to
\beq\label{eq: IR-resum-LO}
    P_\mathrm{lin}(k)\rightarrow P_\mathrm{IR\,res,\,LO}(k,\mu) \equiv P_\mathrm{nw}(k) + e^{-k^2\Sigma^2(\mu)}P_\mathrm{w}(k),
\eeq
where $P_\mathrm{nw}$ and $P_\mathrm{w}$ are the no-wiggle (broadband) and wiggly parts of the linear spectrum respectively.\footnote{Note that we have an additional factor of $1/2$ in the exponent compared with Ref.\,\citep{2019arXiv190905277I}, matching Ref.\,\cite{2017MNRAS.464.3409B}.} The smoothing kernel $\Sigma^2$ may be written
\beq
    \Sigma^2(\mu) = \Sigma^2_\mathrm{NL}\left[1+f\mu^2(2+f)\right]
\eeq
\citep{2015PhRvD..92d3514B,2016JCAP...07..028B,2018JCAP...07..053I}, for logarithmic growth rate $f$,\footnote{We ignore the sub-leading $\delta\Sigma^2$ term of Ref.\,\cite{2018JCAP...07..053I} arising from RSD. For reconstructed fields, the full form may differ somewhat even at linear order, though this will not affect our determination of $\vec\alpha$.} where the amplitude $\Sigma^2_\mathrm{NL}$ may be predicted for \textit{unreconstructed} spectra as
\beq\label{eq: sigma-nl-unrec}
    \Sigma^2_\mathrm{NL,unrec} = \frac{1}{6\pi^2}\int_0^{k_S} dq\,P_\mathrm{nw}(q)\left[1-j_0(qr_s(z_d))+2j_2(qr_s(z_d))\right],
\eeq
for sound horizon scale at decoupling $r_s(z_d)$, spherical Bessel functions $j_\ell$ and cut-off momentum $k_S$. For reconstructed spectra, this is more complex, requiring a self-consistent IR-resummed theory model. For this work, we allow $\Sigma_\mathrm{NL}$ to be a free parameter for the reconstructed field analysis, noting that, for the purpose of extracting $\vec\alpha$, the main function of a better theoretical model is to precisely constrain $\Sigma_\mathrm{NL}$.

Collecting results, we arrive at the model
\beq\label{eq: rec-spectral-model}
    P^\mathrm{rec}_\mathrm{fid}(k,\mu) = \left[b+f\mu^2\left(1-W(k)\right)\right]^2P_\mathrm{nw}(k)\left[1+\left(\mathcal{O}_\mathrm{lin}(k)-1\right)e^{-k^2\Sigma^2(\mu)}\right],
\eeq
where
\beq
    \mathcal{O}_\mathrm{lin}(k) \equiv \frac{P_\mathrm{lin}(k)}{P_\mathrm{nw}(k)}.
\eeq
In practice, the smooth function $P_\mathrm{nw}(k)$ is computed by fitting the fiducial linear power spectrum $P_\mathrm{lin}(k)$ to the combination of an Eisenstein \& Hu spectrum \citep{1998ApJ...496..605E} and five \resub{(fixed)} polynomial terms, as in Ref.\,\citep{2017MNRAS.464.3409B}. This model is similar to that of Refs.\,\citep{2017MNRAS.464.3409B} \& \citep{2020PhRvD.101d3510H} (ignoring the shape marginalization at this point), though we note that we do not include terms to account for Finger-of-God (FoG) effects \resub{or shot-noise}. This is justified since these do not affect the observed power spectrum at small $k$, and, at large $k$, contribute a broadband term which is degenerate with the theoretical error considered below.

Whilst the above model is appropriate for the reconstructed power spectrum at small $k$, it assumes that the observed cosmology matches the fiducial one, and is therefore useless for cosmological analyses. To ameliorate this, we introduce the AP scaling parameters defined in Eq.\,\ref{eq: AP-params}, which relate the observed wavenumbers parallel and perpendicular to the LoS ($k_\parallel$, $k_\perp$) to the true wavenumbers ($k'_\parallel$, $k'_\perp$) via $k'_\parallel = k_\parallel/\alpha_\parallel$, $k'_\perp = k_\perp/\alpha_\perp$, or alternatively
\beq\label{eq: AP-momentum-effects}
    k' &=& \frac{k}{\alpha_\perp}\left[1+\mu^2\left(\frac{1}{F^2}-1\right)\right]^{1/2}\\\nonumber
    \mu' &=& \frac{\mu}{F}\left[1+\mu^2\left(\frac{1}{F^2}-1\right)\right]^{-1/2},
\eeq
where $F = \alpha_\parallel/\alpha_\perp$. This gives the reconstructed power spectrum
\beq\label{eq: AP-model-effects}
    P^\mathrm{rec}_\mathrm{model}(k,\mu) =\left(\frac{r_s^\mathrm{fid}}{r_s}\right)^3\frac{1}{\alpha_\perp^2\alpha_\parallel} P^\mathrm{rec}_\mathrm{fid}\left(k'(k),\mu'(\mu)\right),
\eeq
where the prefactor accounts for the different volumes of the two cosmologies.\footnote{\resub{This is fully degenerate with the bias parameters and theoretical error, thus does not carry cosmological information.}} From this model, we can define the multipole moments via
\beq
    P^\mathrm{rec}_\ell(k) &=& \frac{2\ell+1}{2}\int_{-1}^1d\mu\,P^\mathrm{rec}_\mathrm{model}(k,\mu)L_\ell(\mu)\\\nonumber
    &=&\left(\frac{r_s^\mathrm{fid}}{r_s}\right)^3\frac{2\ell+1}{2\alpha_\perp^2\alpha_\parallel} \int_{-1}^1d\mu\,P^\mathrm{rec}_\mathrm{fid}\left(k'(k),\mu'(\mu)\right)L_\ell(\mu),
\eeq
where $L_\ell(\mu)$ is the Legendre polynomial of order $\ell$. In practice this is computed by summation over 30 points in $\mu$-space via Gaussian quadrature.

One final ingredient is required to compare the model to observational data; the survey window function. For the BOSS survey this is highly anisotropic, and we follow the treatment of Refs.\,\citep{2017MNRAS.464.3121W}\,\&\,\citep{2017MNRAS.466.2242B}, which transforms the model power spectrum multipoles into correlation function multipoles (via \texttt{FFTLog} \citep{2000MNRAS.312..257H}), multiplies by the window function multipoles then transforms back to harmonic space, practically performing a convolution integral. This gives
\beq\label{eq: window-conv-power}
    \hat\xi_0 &=& \xi_0W_0^2 +\frac{1}{5}\xi_2W_2^2+...\\\nonumber
    \hat\xi_2 &=& \xi_0W_2^2 + \xi_2\left[W_0^2+\frac{2}{7}W_2^2\right]+...\,,
\eeq
where $\xi$ and $\hat{\xi}$ are the model and convolved-model correlation functions respectively, and $W_\ell$ are the (publicly available) window function multipoles. Terms beyond the quadrupole were found to be negligible in Ref.\,\citep{2017MNRAS.464.3121W}. We do not include the `integral constraint' in our formalism, as it affects modes only below $k<0.005\hMpc$ which are not used in our analysis \citep{2017MNRAS.464.3409B}. Note that, since the theory model depends on the AP parameters, the window function convolution must be re-evaluated every time the AP parameters are sampled in the MCMC chain.

\subsubsection{Covariance with Theoretical Error}\label{subsec: theoretical-error}
Though the theoretical model of Sec.\,\ref{subsec: recon-theoretical-model} provides an accurate treatment of BAO wiggles into the quasi-linear regime, we cannot simply fit it to the observed data since our treatment of the broadband spectrum is not correct. In previous analyses \citep{2016MNRAS.460.2453S,2017MNRAS.464.3409B,2020PhRvD.101d3510H}, $\sim 5$ free polynomial parameters were added to the spectral model in order to marginalize over the unknown broadband shape; whilst this has been shown to be effective, it is at the expense of sampling speed and a large increase in the number of parameters. Here, we adopt the theoretical error method proposed by Baldauf \textit{et al.} \citep{2016arXiv160200674B}, as applied in Ref.\,\citep{2019JCAP...11..034C}. The basic premise of this is discussed below.

At the center of any cosmological analysis is a theoretical model for $P(\vec k)$, which, in order to be useful, one needs to trust over some range of wavenumbers. 
In perturbation theory one always computes the theoretical model at a given order of the relevant Taylor expansion.
However, as one moves to higher $k$'s, the neglected higher-order  corrections become more important, and at some point they become bigger than the statistical errors. 
Whilst most analyses simply impose a cut-off $k$ up to which the theory is deemed valid, in reality there is a smooth continuum of error, which, for a model computed to $\ell$-loop order, grows as the size of the $(\ell+1)$-loop contribution. To perform analyses with some degree of rigor, it is thus desirable to include a \textit{theoretical} error, in addition to the usual observational covariance. Practically, this error (which represents the difference between our model and the true (as yet unknown) theory), is not completely undetermined; it must be correlated between neighboring bins, and can be treated as a smooth envelope with some correlation scale $\Delta k$. Since the error primarily relates to the broadband spectrum (given that we have allowed the BAO damping scale to be a free parameter in our model), $\Delta k$ must exceed the BAO scale of $\sim 0.05\hMpc$; we here adopt $\Delta k = 0.1\hMpc$ as in Ref.\,\citep{2019JCAP...11..034C}, noting that our analysis is insensitive to the precise value of $\Delta k$, providing it exceeds both the BAO and binning scales. That the correlation length exceeds the BAO scale is a crucial point in our analysis, since by including the theoretical error we can accurately marginalize over the quasi-linear spectrum yet still constrain the AP parameters in regimes where the BAO signal is weak compared to the broadband.

In Ref.\,\citep{2016arXiv160200674B}, it was shown that including the theoretical error with a Gaussian prior was equivalent to modifying the data covariance matrix \resub{$\mathsf{C}^d$};
\beq
    \mathsf{C}^d\rightarrow\mathsf{C}^d+\mathsf{C}^e,
\eeq
where $\mathsf{C}^e$ is the error covariance, given by
\beq
    \mathsf{C}^e_{ij} = E_iE_j\operatorname{exp}\left[-\frac{\left(k_i-k_j\right)^2}{2\Delta k^2}\right]
\eeq
in bins $i,j$, where $E_i$ is the error envelope and we assume a Gaussian correlation matrix. In our context, the theoretical model is accurate only at linear-order, thus we include an error kernel which scales as the approximate one-loop power spectrum;
\beq\label{eq: theoretical-err-kernel}
    E_\ell(k;z) = 2\times \sqrt{2\ell+1}\left(\frac{D(z)}{D(0)}\right)^2\left(\frac{k}{0.31\hMpc}\right)^{1.8}P_0^\mathrm{rec}(k),
\eeq
where the reconstructed power model is evaluated with the fiducial cosmology. This is based on Ref.\,\citep{2016arXiv160200674B}, who advocate $\left(D(z)/D(0)\right)^2\left(k/0.31\hMpc\right)^{1.8}P_\mathrm{lin}(k)$ as the error kernel for the linear matter power spectrum, which is simply a fit to the one-loop spectrum. Here, we use the reconstructed spectral model rather than linear theory (so as to include bias parameters and damping), and assert that the theoretical error for the $\ell$-th multipole should scale as $\sqrt{2\ell+1}P_0(k)$, which is the scaling of the data covariance at leading order, making the conservative assumption that the theoretical errors of different multipoles are uncorrelated.\footnote{This is more stable than using $E_\ell\sim P_\ell(k)$, since higher multipoles can cross zero at relatively small $k$.} We additionally include a safety factor of 2, to ensure the error is not underestimated. Note that the exact form of the error kernel is not crucial to the analysis; we obtain similar constraints when the kernel is inflated by a factor of 10. In this paper, we restrict to a $k$-space regime where the one-loop error kernel is appropriate, though we note that the analysis could simply be extended to higher $k$ by simply adding in two-loop errors. This would be expected to somewhat sharpen our constraints on the AP parameters. 

\subsubsection{Constraining the AP Parameters}
Given the above theoretical model and covariance, it is a simple matter to place constraints on the AP parameters $\vec\alpha$. This is done by minimizing the likelihood
\beq\label{eq: rec-likelihood}
    -2\log \mathcal{L}_\mathrm{rec} = \left(\vec X_d - \vec X_m\right)^T\left(\mathsf{C}_d+\mathsf{C}_e\right)^{-1}\left(\vec X_d - \vec X_m\right),
\eeq
where $\vec X_d$ and $\vec X_m$ are vectors containing the data and (window-convolved) model $P_\ell(k)$ as in Sec.\,\ref{subsec: recon-theoretical-model} and $\mathsf{C}_d$ and $\mathsf{C}_e$ are the data and error covariances, discussed in Sec.\,\ref{subsec: theoretical-error}. In practice this is done using Markov Chain Monte Carlo (hereafter MCMC) using \texttt{montepython} v3.0 \citep{2013JCAP...02..001A,2018arXiv180407261B} to optimize for $\vec\alpha = \{\alpha_\parallel,\alpha_\perp\}$, with the additional nuisance parameters $\{b, \Sigma_\mathrm{NL}\}$ of Sec.\,\ref{subsec: recon-theoretical-model}.\footnote{\resub{The commonly used $f\sigma_8$ parameter is not required to be free here; any cosmological dependence is marginalized over by the theoretical error (for the broadband part) or captured by the free $\Sigma_\mathrm{NL}$ parameter (for the wiggly part). All remaining cosmological dependence is thus encoded in the AP parameters. Ideally, one may wish to first perform the FS analysis, extract best-fit cosmology, and use it to produce a template for the BAO measurement. This approach is not adopted here.}}

Note that to sample the likelihood, we must place priors on the model parameters; these are discussed in Sec.\,\ref{subsec: priors}. Convergence of the MCMC chains are assessed via the standard Brooks-Gelman and Gelman-Rubin criteria \citep{doi:10.1080/10618600.1998.10474787,gelman1992}. From this, we obtain a best-fit value of $\vec\alpha$, which will be used as an additional observable in the FS analysis of unreconstructed spectrum below. In effect, we condense all BAO information in the reconstructed spectrum into a single observable, $\vec\alpha$, which informs the later analysis. Note that since we only require computation of the best-fit values of two parameters, we do not strictly require an MCMC analysis, and we expect that gradient descent would yield similar results in reduced computation time. Furthermore, due to the greatly reduced parameter space compared to the BOSS analyses, our analysis with theoretical error is significantly faster. We present the results in Sec.\,\ref{sec: results-bao}. 

\subsection{FS Analysis: Extracting Cosmological Parameters from Unreconstructed Spectra}\label{subsec: unrecon-analysis}
Following the determination of the AP parameters in Sec.\,\ref{subsec: recon-analysis}, we may proceed to the analysis of the unreconstructed power spectra, allowing us to place strong constraints on cosmological parameters. Our methodology in this section is based on that of Ivanov \textit{et al.} \citep{2019arXiv190905277I} (also Ref.\,\citep{2019arXiv191208208I} and similar to Ref.\,\citep{2019arXiv190905271D}), which we briefly recapitulate below.

\subsubsection{Theoretical Model for $P_\ell^\mathrm{unrec}(k)$}\label{subsec: P-unrec-model}
Since we are now interested in constraining cosmological parameters beyond $\vec\alpha$, we require a power spectrum model that is able to accurately model the entire full-shape (hereafter FS) of the redshift-space galaxy power spectrum into the quasi-linear regime, including both the broadband and the BAO wiggles. For this, we employ a model based on one-loop perturbation theory (presented in detail in Ref.\,\citep{2019arXiv190905277I}, Appen.\,A.), which has the following schematic form
\beq
    P_{\ell}^\mathrm{unrec}(k) = P_{\ell}^\mathrm{tree}(k)+P_{\ell}^\mathrm{1-loop}(k)+P_\ell^\mathrm{ctr}(k)+P_{\ell}^\mathrm{noise}(k).
\eeq
Here $P^\mathrm{tree}$ represents the tree-level (linear) galaxy power spectrum (equal to the familiar Kaiser spectrum \citep{1987MNRAS.227....1K}), with the corresponding one-loop corrections appearing in $P^\mathrm{1-loop}$. These arise from gravitation- and bias-induced non-linearities, in addition to the conversion from real- to redshift-space, and are discussed in detail in Refs.\,\cite{2016arXiv161009321P,2018arXiv180512394F}. Note that we apply the following basis of bias operators relating the galaxy ($\delta_g$) and matter ($\delta$) overdensity fields
\beq
    \delta_g(\vec x) = b_1\delta(\vec x) + \frac{b_2}{2}\delta^2(\vec x)+b_{\mathcal{G}_2}\mathcal{G}_2(\vec x),
\eeq
where $\mathcal{G}_2$ is the tidal field operator. Whilst this strictly neglects the additional bias parameter $b_{\Gamma_3}$, this was found to be degenerate with other free parameters in Ref.\,\citep{2019arXiv190905277I}. The one-loop terms are non-trivial to compute, depending on a number of convolution-type integrals of the linear power spectrum with pre-determined kernels, though this is possible via the \texttt{FFTLog} procedure \citep{2000MNRAS.312..257H,2018JCAP...04..030S}, implemented in a custom \texttt{CLASS} module. 

An important ingredient of our model is the ultraviolet counterterms, $P^\mathrm{ctr}$, which encapsulate complex short-scale (UV) physics (which cannot be modeled perturbatively) in addition to contributions from the FoG effect, and other degenerate effects, such as higher-derivative biases. These counterterms comprise of a fixed scale dependence (predicted by EFT, either by consideration of the UV divergences of the SPT expressions or by coarse-graining the equations of motion), in addition to a free amplitude, whose magnitude or sign cannot be predicted by analytic theory. We refer the reader to canonical EFT references (\citep{2012JCAP...07..051B,2012JHEP...09..082C,2014arXiv1409.1225S,2015JCAP...11..007S,2016arXiv161009321P}) for further discussion of these effects. In practice, (and considering only the $\ell=0$ monopole and $\ell=2$ quadrupole), this leads to three counterterms with free amplitudes $\{c_0,c_2,\tilde{c}\}$, with the third parametrizing next-to-leading order effects from FoG (which can affect larger scales than other non-linearities). 

Furthermore, the term $P^\mathrm{noise}$ includes any stochastic contributions to the galaxy power spectrum which, to one-loop order, includes only a constant, and direction-independent, Poissonian shot-noise, whose amplitude is a free parameter, $P_\mathrm{shot}$. In addition to the above components, infra-red resummation must be included (as in Sec.\,\ref{subsec: recon-theoretical-model}), which is performed with the approach of Refs.\,\citep{2016JCAP...07..028B,2018JCAP...07..053I}, accurate to one-loop order. Furthermore, the AP parameters must be included by relating the observed and true momentum vectors (as in Eq.\,\ref{eq: AP-momentum-effects}\,\&\,\ref{eq: AP-model-effects}) and the survey window is included via window function convolution, as before. In total, we obtain a model with seven nuisance parameters; $\{b_1, b_2, b_{\mathcal{G}_2}, P_\mathrm{shot}, c_0, c_2, \tilde{c}\}$. Since this is based on one-loop perturbation theory, we expect it to be accurate for $k\lesssim0.25\hMpc$ (with two-loop corrections subdominant \citep{2014JCAP...07..057C,2015PhRvD..92l3007B}), thus we will only use spectral data in this wavenumber range.

\subsubsection{Joint Likelihood of $P^\mathrm{unrec}_\ell(k)$ and AP parameters}\label{eq: unrec-plus-alpha-likelihood}
In previous analyses \citep{2019arXiv190905277I,2019arXiv191208208I}, the theory model outlined in Sec.\,\ref{subsec: P-unrec-model} has been directly compared to observational data via MCMC to constrain cosmology, via a simple Gaussian likelihood. In this work, we aim to combine constraints from both the BAO and FS analyses, thus we require a likelihood that will include the constraints on the AP parameters, discussed in Sec.\,\ref{subsec: recon-analysis}. Whilst one may surmise a number of complex ways in which to do this, we here adopt a simple procedure, considering the joint Gaussian likelihood of $P^\mathrm{rec}_\ell(k)$ and $\vec\alpha$, noting that these are \textit{not} independent.\footnote{Note that there are a number of ways to combine data-sets. Possibly the most robust technique would be to consider the joint likelihood of the reconstructed and unreconstructed spectra using a full theoretical model, which accurately treats both the broadband and BAO wiggles. However, this requires significant theoretical work, and, since the reconstructed spectra are mostly useful for BAO information, we do not believe it will lead to significantly stronger constraints on cosmology.} This yields
\beq\label{eq: unrec-likelihood}
    -2\log \mathcal{L}_\mathrm{unrec} = \left(\vec X_d - \vec X_m\right)^T\mathsf{C}_d^{-1}\left(\vec X_d - \vec X_m\right),
\eeq
where the data-vector $\vec X_d$ contains both the observed (pre-reconstruction) redshift-space galaxy power spectrum multipoles and the best-fit AP parameter vector obtained by fitting the reconstructed data. The associated model $\vec X_m$ similarly contains the model for $P_\ell^\mathrm{unrec}(k)$ and the current value of the AP parameters $\vec\alpha$. Note that we allow $\vec\alpha$ to be a free parameter in the pre-reconstruction analysis; our procedure effectively assigns it a relatively tight prior, which is covariant with the spectrum itself. 

In the above likelihood, we require a joint covariance of $P^\mathrm{unrec}_\ell(k)$ and $\vec\alpha$, labelled $\mathsf{C}_d$.\footnote{For a discussion of the covariance of $\vec\alpha$ obtained from different BOSS analyses, see Ref.\,\citep{2017MNRAS.464.1493S}.} 
Whilst the form of the auto-covariance of $P^\mathrm{unrec}_\ell$ can be well predicted theoretically, the cross-covariance of this with $\vec\alpha$ may seem difficult to estimate, though we show in Sec.\,\ref{subsec: pk-alpha-cov} that it can be well estimated from a simple theoretical model. In practice, this is complicated by the survey window function, which has non-trivial, though small, effect on the BAO wiggles and hence the \resub{cross-covariance}. For an exact treatment, the power spectrum covariance is usually estimated from a suite of simulations,\footnote{Note that semi-analytic geometry-specific treatments are possible, for example Refs.\,\citep{2016MNRAS.457..993P,2020MNRAS.492.1214P} or Ref.\,\citep{2019MNRAS.490.5931P} for the associated correlation function multipoles.} thus we elect to do the same here for the $\vec\alpha$ covariance and cross-terms. This is made possible by fast sampling speed of the BAO analysis, allowing us to estimate $\vec\alpha$ from each of a large number of mocks, as discussed in Sec.\,\ref{subsec: mocks}. 

With the likelihood of Eq.\,\ref{eq: unrec-likelihood} in hand (and the priors of Sec.\,\ref{subsec: priors}), we can constrain desired cosmological parameters (and the seven nuisance parameters of Sec.\,\ref{subsec: P-unrec-model}) using MCMC, practically achieved via the \texttt{montepython} code, as before. For speed, we vary the nuisance parameters more often than the cosmological parameters, noting that the full model must be recomputed each time any cosmological parameter is varied. The results of this analysis will be discussed in Sec.\,\ref{sec: results-joint}.

\section{Data and Priors}\label{sec: data-and-priors}
In this section, we discuss the aspects of our analysis that are specific to the data-set and cosmological model tested; the BOSS DR12 data (with its associated covariance), and the choice of priors.

\subsection{The BOSS DR12 data-set}\label{subsec: data}
The data-set used in this work is that of the twelfth data release (DR12) \citep{2017MNRAS.470.2617A} of the Baryon Oscillation Spectroscopic Survey (BOSS), part of SDSS-III \citep{2011AJ....142...72E,2013AJ....145...10D}. This contains the positions and redshifts of a sample of 1,198,006 galaxies, spanning $0.2<z<0.75$, which, for this analysis, are divided into two disjunct redshift bins with effective redshifts $z_\mathrm{eff}=0.38$ and $0.61$. As in \citep{2019arXiv190905277I}, these will be referred to as `low-z' and `high-z' respectively. In addition, the sample is divided into two spatial regions (totalling 10,252 $\mathrm{deg}^2$) around the Northern and Southern Galactic Cap (hereafter NGC and SGC), giving a total of four independent data-sets. These have effective volumes (in $\left(h^{-1}\mathrm{Gpc}\right)^3$ assuming $h = 0.676$) of $0.84$, $0.31$, $0.93$ and $0.34$ for low-z NGC, low-z SGC, high-z NGC and high-z SGC respectively. Each region has a unique survey geometry, which is specified by a set of random particles whose distribution matches the selection function and unclustered galaxy number density. The publicly available\footnote{\resub{\href{https://fbeutler.github.io/hub/boss_papers.html}{https://fbeutler.github.io/hub/boss\_papers.html}}} multipoles of these are used to construct the window-convolved theory estimates for the galaxy power spectra, via Eq.\,\ref{eq: window-conv-power}.

For each chunk, we use the publicly available redshift-space galaxy power spectra (both pre- and post-reconstruction). These are generated using the (Fast Fourier Transform-based) estimator of Refs.\,\citep{2015PhRvD..92h3532S,2015MNRAS.453L..11B}, computing the multipole power from the data-set combined with a set of random points. The computation of these, and the associated weights, are described in detail in Refs.\,\citep{2017MNRAS.464.3409B} and \citep{2017MNRAS.464.1168R} respectively. Density field reconstruction has been applied via the method of Padmanabhan \textit{et al.} \citep{2012MNRAS.427.2132P} (based on Ref.\,\citep{2007ApJ...664..675E}), utilizing a random catalog (with the same selection function and unclustered distribution) and smoothing the observed density field with a kernel of width $\Sigma_\mathrm{smooth} = 15\Mpch$ (chosen to give optimal signal-to-noise). The displacement field is estimated via linear theory and the method of finite differences, and the galaxies (but not the randoms) displaced by an additional factor $(1+f)$ along the line-of-sight to remove RSD at lowest order.\footnote{Note that this is the `Rec-Iso' reconstruction scheme, in the language of Ref.\,\citep{2016MNRAS.460.2453S} and Sec.\,\ref{subsec: recon-theoretical-model}.} This process, along with its potential systematic errors, is discussed in detail in Refs.\,\citep{2012MNRAS.427.2132P,2017MNRAS.464.3409B,2011ApJ...734...94M,2015arXiv150906384V}.

Here, we use data from both the monopole ($\ell=0$) and quadrupole ($\ell=2$), in $\Delta k = 0.005\hMpc$ bins, using $k\in [0.0025,0.25]$ ($k\in[0.0025,0.3]$) for the unreconstructed (reconstructed) spectra giving a total of 50 (60) bins for each multipole. Note that we use a larger $k_\mathrm{max}$ for the reconstructed spectra; since our BAO analysis marginalizes over the (non-linear) broadband spectral shape, moving to larger wavenumber only leads to more BAO wiggles being observed, whose positions are not affected by the higher-order terms. This is not true for the unreconstructed data, whose analysis relies on the theoretical model being accurate, and two-loop terms remaining subdominant.\footnote{In practice, the maximum wavenumber for the reconstruction analysis is set by the Nyquist frequency $k_\mathrm{Nyq} = 0.6\hMpc$ and the limits of the publicly available data.} \resub{We refer the reader to Appendices A and B of \cite{2019arXiv190905277I} for further discussion of these effects and validation of the FS machinery on Patchy mocks.}

\begin{figure}
    \centering
    \includegraphics[width=\textwidth]{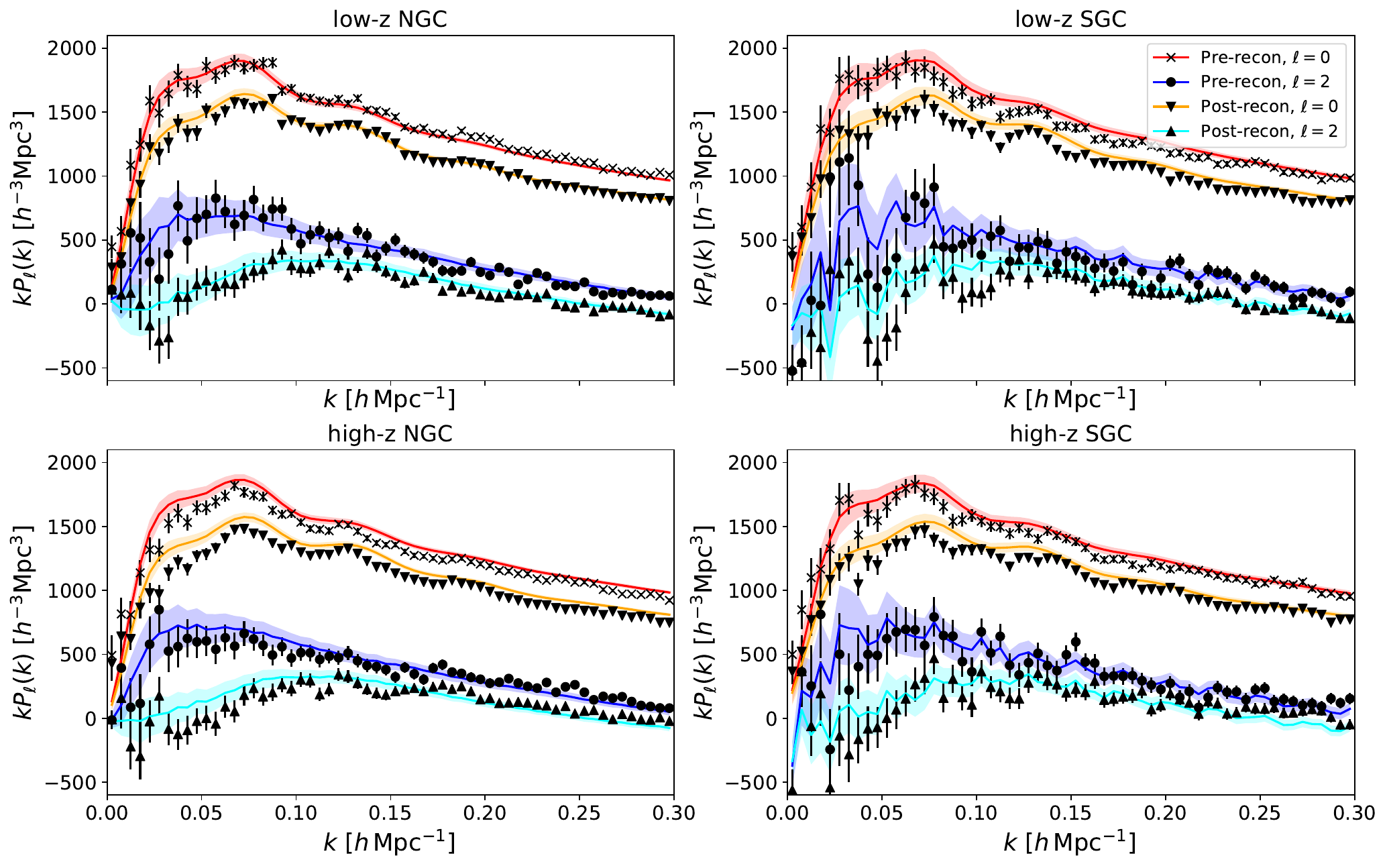}
    \caption{Power spectra used in this analysis, from both BOSS DR12 data, and MultiDark-Patchy mocks. For each separate chunk (at a different sky location and redshift bin), we show both the monopole ($\ell=0$, upper two spectra) and quadrupole ($\ell=2$, lower two spectra), before (darker colors) and after (lighter colors) density field reconstruction. Colored lines and shaded regions indicate the mean and $1\sigma$ variations between 999 mock catalogs in each chunk, with the data shown as black points. Errorbars indicate the square root of the covariance diagonal, estimated from the same set of mocks. We note that reconstruction sharpens the Fourier-space BAO wiggles, whilst slightly reducing the overall amplitude and removing most of the large-scale quadrupole power. Further note that the NGC data appears much smoother, due to the larger effective volumes of these regions.}
    \label{fig: all_pk}
\end{figure}

In Fig.\,\ref{fig: all_pk} we show the power spectrum multipoles for all fields both pre- and post-reconstruction, which are compared to the results from mocks (Sec.\,\ref{subsec: mocks}). The effect of using a larger observational region is clear; the NGC data, with over double the effective volume, is far less noisy than SGC data. We note prominent BAO wiggles in the data-sets (mainly the monopole), whose power is somewhat enhanced after density field reconstruction, which will strengthen the constraints on $\vec\alpha$. In all unreconstructed chunks, there is a non-negligible quadrupole arising from RSD, but the reconstruction is able to significantly reduce this at low $k$, with $P_2$ consistent with zero for $k\lesssim0.1\hMpc$. This is a consequence of shifting the galaxies by an additional factor $(1+f)$ during the reconstruction.

\subsection{MultiDark-Patchy Mocks}\label{subsec: mocks}
To estimate the covariance matrix for the BOSS DR12 spectra, we require accurate and numerous mock catalogs. Here, we use the MultiDark-Patchy (hereafter Patchy) \citep{2014MNRAS.439L..21K,2016MNRAS.456.4156K}, which are generated with approximate Lagrangian perturbation theory techniques, with a stochastic halo bias prescription, calibrated from a high-resolution N-body simulation; BigMultiDark \citep{2016MNRAS.457.4340K}. Halo abundance matching \citep{2016MNRAS.460.1173R} is used to ensure the simulations have the correct redshift evolution and correlation functions, and separate mocks are produced for each of the four BOSS DR12 chunks (Sec.\,\ref{subsec: data}) with the relevant selection functions and windows. These use a fiducial cosmology of $\{\Omega_m = 0.307115, \Omega_b = 0.048206, \sigma_8 = 0.8288, n_s = 0.9611, h = 0.6777\}$, which is slightly different to the BOSS fiducial cosmology, though given that the associated AP parameters are very close to unity ($\vec\alpha = \{0.9991, 0.9983\}$ and $\{0.9998, 0.9987\}$ for low-z and high-z samples respectively) we do not expect this to affect our analysis.

In this work, we use the set of $N_\mathrm{mocks}=999$ Patchy mocks for which density-field reconstruction has been performed, and the corresponding spectra made public. The mean and $1\sigma$ limits of these are shown in Fig.\ref{fig: all_pk}, and we note fair agreement of mocks and model across all wavenumbers. From visual inspection, it may seem that the low-z NGC mock spectra (both pre- and post-reconstruction) exhibit a smaller BAO signature than the data; this has been already discussed in Ref.\,\citep{2017MNRAS.464.3409B}. 
By virtue of this anomaly, the $H_0$ constraints 
extracted from the FS analysis of the actual low-z NGC chunk were $\sim 40\%$ better than the 
results based on the analysis of the Patchy mocks \cite{2019arXiv190905277I}.
When we combine the BAO and 
and FS data from this chunk, the constraints 
will improve only marginally. This suggests an explanation of this anomaly as a `lucky'
particular realization of the data in this chunk,
whose dark matter displacement field happened to have reduced power.
The detailed investigation of such a phenomenon goes beyond the scope of this paper.

Given the set of all mocks, with the $n$-th mock consisting of some data vector $\vec X^{(n)} = \{X_a^{(n)}\}$ (for bin label $a$), we may estimate the covariance of $\vec X$ as
\beq
    \mathrm{cov}\left(X_a,X_b\right) \equiv \mathsf{C}_{ab} &=& \frac{1}{N_\mathrm{mocks}-1}\sum_{n=1}^{N_\mathrm{mocks}}\left(X^{(n)}_a-\overline{X}_a\right)\left(X^{(n)}_b-\overline{X}_b\right),
\eeq
where 
\beq
    \overline{X}_a  = \frac{1}{N_\mathrm{mocks}}\sum_{n=1}^{N_\mathrm{mocks}}X^{(n)}_a
\eeq
is the mean over all mocks. For the BAO analysis (Sec.\,\ref{subsec: recon-analysis}, $\vec X$ is simply the vector of monopole and quadrupole power in each bin, i.e. $\vec X = \{\vec P^\mathrm{rec}_0,\vec P^\mathrm{rec}_2\}$. As described in Sec.\,\ref{subsec: unrecon-analysis}, to place strong constraints on key cosmological parameters using the pre-reconstructed fields, we require the joint covariance between the power spectra and the AP parameter estimates, $\widehat{\vec\alpha}$, which come running an MCMC chain on each of the 999 reconstructed mocks. This gives the data vector $\vec X = \{\vec P^\mathrm{unrec}_0,\vec P^\mathrm{unrec}_2,\widehat{\vec\alpha}\}$.\footnote{Note that $\widehat{\vec\alpha}$ is here taken as the best-fit of the MCMC posterior discussed in Sec.\,\ref{subsec: recon-analysis}.} For use in Gaussian likelihoods, we require the \textit{precision matrix}, $\mathsf{\Psi}_d$, which, in the limit of zero noise, is the inverse of the covariance matrix. With a finite $N_\mathrm{mocks}$, this is a biased estimator, thus we apply the rescaling factor of Hartlap \textit{et al.} \citep{2007A&A...464..399H}, giving
\beq
    \mathsf{\Psi}_d = \frac{N_\mathrm{mocks}-N_\mathrm{bins}-2}{N_\mathrm{mocks}-1}\mathsf{C}_d^{-1},
\eeq
where $N_\mathrm{bins}$ is the number of bins; 102 in the pre-reconstructed case. Note that it is \textit{not} correct to apply this rescaling factor to the reconstructed data, since, in this case, we must invert the sum of the (stochastic) data and (smooth) theory covariance, rather than just the Wishart-distributed sample covariance. The appropriate rescaling factor for the summed matrix is non-trivial and thus ignored. For simplicity, we do not include the parameter error bar inflation of Ref.\,\citep{2014MNRAS.439.2531P}, noting that this contributes only a small $\sim2\%$ error on the derived parameter errors. For a fully robust treatment, we should allow the covariance matrix to vary with cosmology, though this is difficult and computationally infeasible for highly anisotropic surveys such as BOSS, though progress is possible with techniques such as Refs.\,\citep{2019JCAP...01..016L,2019arXiv191002914W}.

\subsection{Priors}\label{subsec: priors}
\subsubsection{BAO Analysis}
When analyzing the reconstructed power spectrum multipoles we vary two cosmological parameters and two nuisance parameters
\beq
    \left\{\alpha_\parallel,\alpha_\perp\right\}\times\left\{b,\Sigma_\mathrm{NL}\right\},
\eeq
as discussed in Sec.\,\ref{subsec: recon-analysis}. Each of the four data-chunks must be analyzed separately (in order to provide joint covariance of $P^\mathrm{unrec}_\ell(k)$ with $\vec\alpha$), thus $b$ and $\Sigma_\mathrm{NL}$ are allowed to vary between chunks. For the biases, this is appropriate since each region has a different selection function, though for the non-linear damping, one may presume that this should be the same for chunks at the same effective redshift. In practice, we allow it to vary since the non-linear damping encodes a variety of effects including higher-order bias terms and the reconstruction efficacy, that can vary between chunks.

Given that the AP parameters are expected to equal unity for a universe with fiducial cosmology, we adopt a uniform prior in $(0.8,1.2)$ for both $\alpha_\parallel$ and $\alpha_\perp$, noting that this spans a wide range of cosmologies. (Recall that the Patchy mocks have a somewhat different cosmology to fiducial, yet $|\alpha_i-1|<0.002$ for $\alpha_i\in\{\alpha_\parallel,\alpha_\perp\}$.) For the bias parameter, which encompasses both the galaxy bias and other (small) constants from reconstruction, we center the prior around the approximate bias obtained from BOSS of $b\approx2$, but otherwise keep its amplitude unconstrained. 

Finally, the prior on the non-linear damping scale $\Sigma_\mathrm{NL}$ is worthy of attention, since unlike the analysis of Ref.\,\citep{2017MNRAS.464.3409B} we allow this to vary freely. In this work we adopt a uniform prior of $\Sigma_\mathrm{NL}\in (1,6)\Mpch$, which encompasses the post-reconstruction values found in Ref.\,\citep{2017MNRAS.464.3409B} and centers on the values of $\Sigma_\mathrm{NL}\sim 3.5\Mpch$ found in initial testing. Furthermore, this is restricted to be less than the pre-reconstruction value of Eq.\,\ref{eq: sigma-nl-unrec}, since any reconstruction should sharpen the BAO peaks and thus reduce $\Sigma_\mathrm{NL}$. The prior width on $\Sigma_\mathrm{NL}$ encodes our knowledge of the post-linear regime; tightening the prior width around some best-fit value thus allows us to judge how the error bars on $\vec\alpha$ are modified by better modeling of the reconstructed spectra, which will be investigated further below.

\subsubsection{FS Analysis}
Our methodology for the FS analysis closely follows that of Ref.\,\citep{2019arXiv190905277I}, to which we refer the reader for further details. In the baseline FS analysis, we fit six cosmological parameters of the minimal $\Lambda$CDM model, including massive neutrinos (hereafter denoted $\nu\Lambda$CDM),
\begin{equation}
    \left\{\omega_b,\omega_{cdm},h,n_s,A^{1/2},\sum m_\nu\right\}\,,
\end{equation}
where $\sum m_\nu$ is the sum of the neutrino masses, $A$ is the dimensionless ratio of primordial spectral amplitude to the best-fit value obtained by Planck \cite{2018arXiv180706209P}, 
\begin{equation}
    A\equiv \frac{A_s}{A_{\rm s,\,Planck}},\,\quad 
    A_{\rm s,\,Planck}=2.099\times 10^{-9}\,,
\end{equation}
$h = H_0/\left(100\,\mathrm{km\,s}^{-1}\mathrm{Mpc}^{-1}\right)$ and $\omega_b$, $\omega_{cdm}$ are the physical baryon and cold dark matter densities, related to $\Omega_b$, $\Omega_{cdm}$ via $\omega_i \equiv \Omega_ih^2$. 

In this work, we impose the Gaussian BBN prior on $\omega_b$ \citep{2018arXiv180706209P,2015JCAP...07..011A,2018ApJ...855..102C,2019JCAP...10..029S} (see a discussion in Ref.~\cite{2019arXiv190905277I} for more detail),
\begin{equation}
    \omega_b = 0.02268 \pm 0.00036\,,
\end{equation}
though we note that, in principle, the FS data can constrain $\omega_b$ without any external input. However, this constraint is expected to be much weaker than the measurements from BBN or Planck, thus, keeping in mind the eventual combination of the FS and Planck data, it is reasonable to impose this prior. We additionally note that this is $1\sigma$ consistent with the value obtained by Planck, yet a factor of $\sim 3$ broader. Use of this prior will allow us to better assess the information 
content of the FS data in combination with Planck, whilst ensuring that our constraints are still independent of the CMB. Note that, since $\omega_{cdm}$ is still free, there is no strong shape prior imposed on our spectra.

Considering the neutrino mass, we will impose the following flat prior
\beq\label{eq: mnuprior}
   0.06~\text{eV}
   < \sum m_\nu <0.18~\text{eV}\,.
\eeq
The motivation behind this is twofold. On one hand, we could fix the neutrino mass to some fiducial value, analogous to the Planck baseline analysis \citep{2018arXiv180706209P}, which fixes it to the lowest mass allowed by oscillation experiments ($\sum m_\nu = 0.06\,\mathrm{eV}$). Whilst this choice is arbitrary, it allows the parameter inference to be significantly expedited. On the other hand, given that the sum of the neutrino masses is currently unknown, we could adopt a loose prior which aims to measure it directly from the BOSS data. Previous analyses \citep{2019arXiv191208208I,Colas:2019ret} have shown however, that only weak constraints are possible from BOSS, ($\sum m_\nu \lesssim 1\,\mathrm{eV}$), which is significantly worse than the current limits obtained from the various combinations of the cosmological data (see Ref.~\cite{2019arXiv191208208I} and references therein). In this case, our MCMC chains will spend a lot of time exploring a range of neutrino masses ruled out by other data-sets, so they are particularly inefficient. To test our prior, one data chunk (low-z NGC) has been analyzed without placing an upper bound on the neutrino mass, and we conclude that, for the BOSS data-set, $\sum m_\nu$ is not degenerate with $H_0$, $\Omega_m$ and $\sigma_8$ (which represent the main results of our analysis). Thus, we decide to scan over the neutrino mass only in the realistic range (Eq.\,\ref{eq: mnuprior}) allowed by other measurements in our baseline analysis, keeping in mind our eventual intention to combine the BOSS, BAO and Planck likelihoods. Note that in contrast to Ref.\,\citep{2019arXiv190905277I}, we use three degenerate neutrinos rather than a single massive state, matching the latest Planck analyses \citep{2018arXiv180706209P}. 

For the remaining cosmological parameters $\{\omega_{cdm}, h, n_s, A\}$, we assume no informative priors, ensuring that our analysis is fully independent of Planck and thus a useful cross-check.

The main difference between the BAO-only and FS analysis is the necessity to model more complex non-linearities, which affect the broadband power spectrum. To this end, we use a model for non-linear bias and redshift-space distortions (discussed in Sec.\,\ref{subsec: unrecon-analysis}) that has seven free parameters;
\beq
    \{b_1,b_2,b_{\mathcal{G}_2},P_\mathrm{shot},c_0^2,c_2^2,\tilde{c}\}.
\eeq
Note that the nuisance parameters are allowed to vary separately for each data-chunk, due to the different sample selection functions and effective redshifts. Performing a joint analysis of all four chunks thus requires $7\times 4 = 28$ nuisance parameters, in addition to the six cosmological parameters. We use the following flat priors for the bias parameters,
 \begin{equation}
   \begin{split}
   &  b_1A^{1/2}\in (1,4)\,,\quad    b_2A^{1/2}\in (-4,2)\,,\\
 &    b_{\mathcal{G}_2}A^{1/2}\in (-3,3)\,,\quad   P_{\rm shot}\in (0,1)\times 10^4 \,[\Mpch]^3\,,\\
 \end{split}   
 \end{equation}
 and the counterterms,
 \begin{equation}
   \begin{split}
   &  c_0,c_2\in (-100,100)\,\quad [\Mpch]^2 \,,\\
 & \tilde{c}\in (-10^3,10^3)\,\quad [\Mpch]^4\,.\\
 \end{split}   
 \end{equation}
The priors for the bias parameters $b_2$ and $b_{\mathcal{G}_2}$ are informed by measurements of biases for dark matter halos that are typical hosts for the BOSS galaxies \citep{2016JCAP...02..018L}, and those on the counterterms are motivated by the requirement that the counterterm contributions be smaller than the tree-level spectra, i.e. that they do not cause the perturbative description to break down. Technically, priors on both the counterterms and bias parameters have no impact on the inferred cosmological parameters for the data cut $k_{\rm max}=0.25$ $\hMpc$ used in the FS analysis. Equivalently, these priors could be waived without changing our main conclusions; their only purpose is to speed up the convergence of our MCMC chains.

Finally, it should be mentioned that we do not impose a prior on the AP parameters $\vec\alpha$, since these are uniquely determined from Eq.\,\ref{eq: AP-params}, given the current and fiducial cosmologies. An effective prior is set on these however from the covariance obtained from the post-reconstruction BAO analysis.

\section{Results of the BAO analysis}\label{sec: results-bao}
Below, we present and discuss the parameter constraints obtained from the analysis outlined in Sec.\,\ref{sec: analysis}, as applied to the data of Sec.\,\ref{sec: data-and-priors}. We will discuss both the intermediate constraints on the AP parameters from the analysis of reconstructed data, and the final constraints on cosmology from the joint analysis.

\subsection{AP Parameters from the
reconstructed BAO}\label{subsec: results-AP}
Before considering the observed AP parameters from the four data-chunks, it is pertinent to consider the dependence on key model parameters; the scale of the theoretical error covariance and the prior width on the BAO damping scale $\Sigma_\mathrm{NL}$. To this end, we perform an MCMC analysis of the reconstructed spectra (as in Sec.\,\ref{subsec: recon-analysis}), (a) inflating the theoretical error kernel $E_\ell$ by a factor of ten, and (b) changing the prior on $\Sigma_\mathrm{NL}$ from a loose prior of $\Sigma_\mathrm{NL}\in(1,6)\Mpch$ to a tight uniform prior in $\Sigma_\mathrm{NL}\in(2.9,3.1)\Mpch$. For these tests, we use only the high-z NGC data chunk, which has the highest volume and is thus expected to be the most constraining. 

\begin{figure}[htb]
    \centering
    \includegraphics[width=0.5\textwidth]{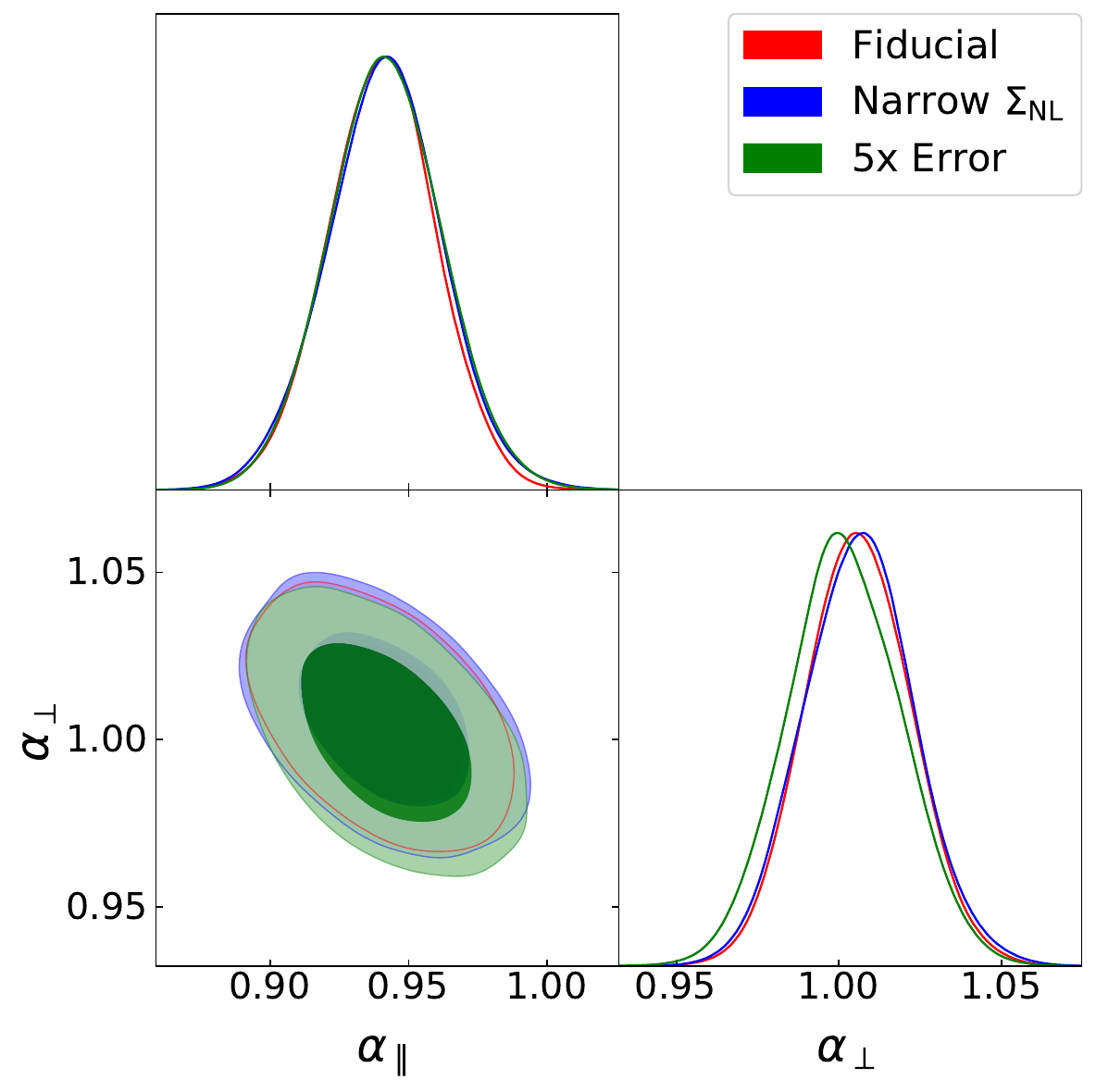}
    \caption{Posterior distribution of the Alcock-Paczynski (AP) parameters $\alpha_\parallel$, $\alpha_\perp$ (defined in Eq.\,\ref{eq: AP-params}) obtained from analysis of the high-z NGC BOSS DR12 power spectrum, after density-field reconstruction. Posterior samples are obtained by minimizing a likelihood consisting of a linear model with an additional theoretical error to account for the poorly-understood post-linear shape of the spectrum, as described in Sec.\,\ref{subsec: recon-analysis}. We show results from three choices of hyperparameters in the analysis; the fiducial choice (red), adopting a much narrower prior on the non-linear damping scale $\Sigma_\mathrm{NL}$ (blue) and inflating the theoretical error kernel (Eq.\,\ref{eq: theoretical-err-kernel}) by a factor of five. These contours were generated from $\sim 10^4$ posterior samples obtained from running 16 MCMC chains in parallel. We note negligible difference in the AP parameters from imposing a tight prior on $\Sigma_\mathrm{NL}$, with a slight bias obtained by inflating the theoretical error.}
    \label{fig: BAO-analysis-param-choices}
\end{figure}

The posterior contours for $\vec\alpha$ obtained from these analyses (as well as with the fiducial set of hyperparameters) are shown in Fig.\,\ref{fig: BAO-analysis-param-choices}, created with the \texttt{getdist} code \citep{2019arXiv191013970L},\footnote{\href{https://getdist.rtfd.io/en/latest}{getdist.rtfd.io/en/latest}} which is part of the \texttt{CosmoMC} package \citep{2002PhRvD..66j3511L,2013PhRvD..87j3529L}. Note that in all cases, we obtain a constraint on $\vec\alpha$ which is significantly sharper than the uniform prior. When the scale of the theoretical error is inflated by a factor of five (and hence the covariance by 25), we observe no obvious bias in $\alpha_\parallel$, though a slight ($-\sim 0.05\sigma$) bias in the $\alpha_\perp$ parameter, though this is significantly below the width of the statistical uncertainty, so may be safely ignored. We attribute this slight shift to the reduction of the BAO information provided by the broadband part of the spectrum, which is reduced with a large theoretical error.

Notably, we observe no change to the AP parameter contours when imposing a tight prior on the BAO damping around its best fit value of $\Sigma_\mathrm{NL}\sim 3\Mpch$, despite the fact that the posterior of $\Sigma_\mathrm{NL}$ is \textit{not} tightly constrained in the fiducial case. This implies that the AP information arises from knowledge of the positions of the BAO harmonic peaks, irrespective of knowledge of their amplitudes. We may thus conclude that better knowledge of the BAO damping will not tighten the constraints on $\vec\alpha$. Furthermore, given that (a) the majority of AP information arises from the position of the BAO peaks and not the broadband spectrum, and (b) better modeling of the reconstructed power spectra will primarily constrain the post-linear broadband shape and $\Sigma_\mathrm{NL}$, it is true that including a full perturbation-theory model of the non-linear reconstructed power spectrum will \textit{not} make a large impact on the AP parameters. Furthermore, reconstruction's primary utility is to sharpen BAO wiggles, thus we do not expect the broadband to carry additional cosmological information. For this reason, when extracting cosmological information from power spectra, reconstruction should be used \textit{only} for the BAO-specific parameters $\{\alpha_\parallel, \alpha_\perp\}$, and it is not beneficial to use a more complex theoretical treatment, since it will not give significantly sharper constraints on $\Sigma_\mathrm{NL}$.

\begin{figure}[htb]
    \centering
    \includegraphics[width=0.8\textwidth]{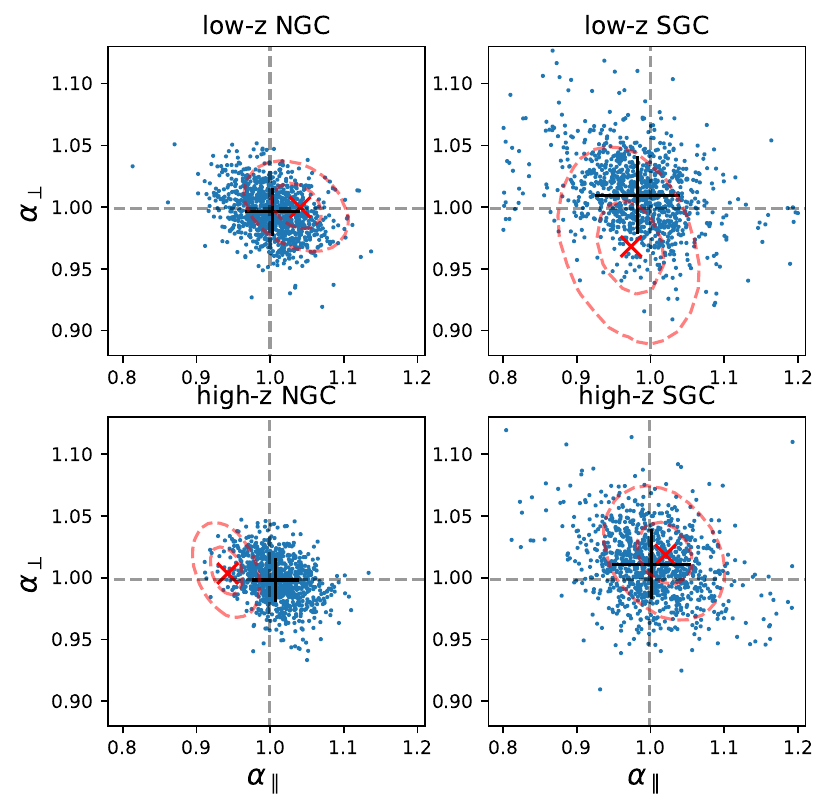}
    \caption{Distribution of the \resub{best-fit} AP parameters obtained from applying the BAO analysis method of Sec.\,\ref{subsec: recon-analysis} to 999 MultiDark-Patchy mock galaxy samples. For each mock, we plot the best-fit value obtained from an MCMC analysis which fits the corresponding power spectrum against a theoretical model and outputs a posterior distribution for $\{\alpha_\parallel, \alpha_\perp\}$. The dotted lines indicate the expected value for the mock cosmology (Sec.\,\ref{subsec: mocks}), with the black cross showing the average and $1\sigma$ deviation across all mocks. 
        Note that the error bar is not normalized by the number of mocks, thus it represents the expected variation from a single mock.
    The red cross and dashed lines show the best-fit obtained (and its 68\% and 95\% confidence levels (CLs)) from analyzing the true BOSS data-set in this chunk (as tabulated in Tab.\,\ref{tab: AP-results}). \resub{A comparison of the posterior contour shapes for mocks and data is shown in Fig.\,\ref{fig: ap-param-shape}.}}
    \label{fig: all-AP-params}
\end{figure}

Having established the validity of our hyperparameters, we can consider the distributions of the AP parameters obtained from analyzing the 999 Patchy mocks. For each mock, and each chunk, the analysis of Sec.\,\ref{subsec: recon-analysis} is performed to obtain a posterior contour on $\vec\alpha$, allowing to set a covariance, as in Sec.\,\ref{subsec: mocks}. We present the distributions of $\vec\alpha$ from the four chunks in Fig.\,\ref{fig: all-AP-params}, plotting the best-fit of the posterior distribution obtained for each chunk and mock. In each case, the distribution is centered close to the expected value (of $\vec\alpha\sim \vec 1$), with the mean consistent with unity at $1\sigma$, and we observe substantial negative correlations between $\alpha_\parallel$ and $\alpha_\perp$, as found in Ref.\,\citep{2017MNRAS.464.3409B}. Notably, the constraints are tighter for the NGC regions; this is expected given their larger effective volume. The distributions of AP parameters shown will be used to create the joint covariance matrix with the observed $P^\mathrm{unrec}_\ell(k)$.

\begin{figure}[htb]
    \centering
    \includegraphics[width=0.8\textwidth]{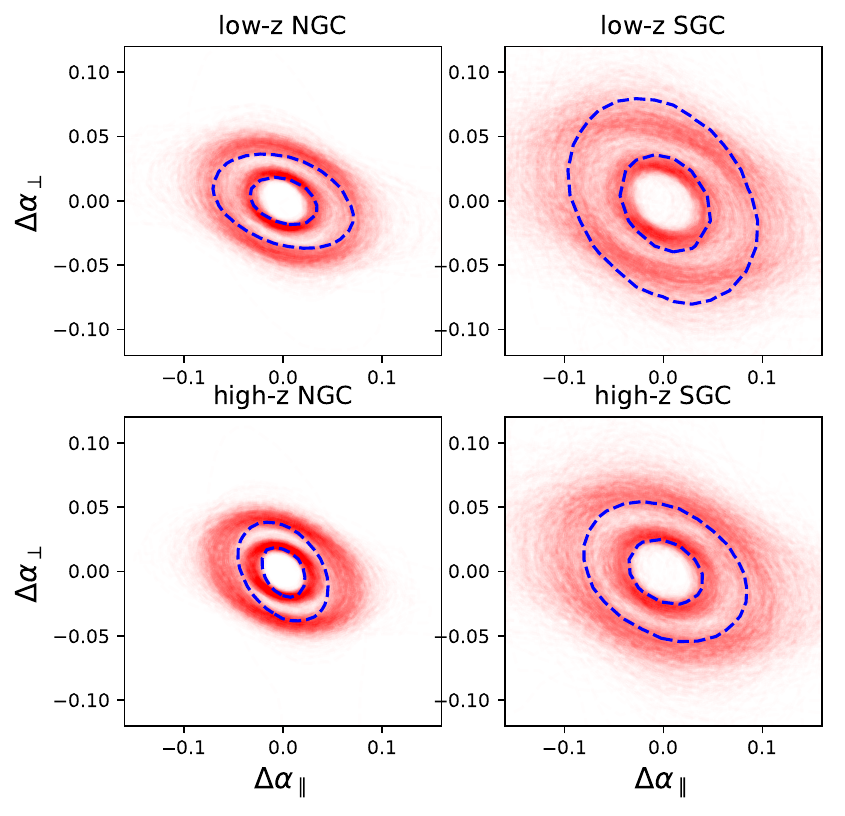}
    \caption{\resub{Posterior distribution shapes for the AP parameter estimates shown in Fig.\,\ref{fig: all-AP-params}. We overplot the 68\% and 95\% CL contours from the BAO analysis of 999 mock galaxy samples in red, shifting the distribution to have zero mean. The corresponding posterior for the data in each patch is shown in blue. Note that the data contours are consistent with a random draw from the set of mock contours.}}
    \label{fig: ap-param-shape}
\end{figure}

Also shown in Fig.\,\ref{fig: all-AP-params} are the best-fit values of $\vec\alpha$ obtained from the reconstruction analysis applied to the true BOSS data, with contours indicating the statistical uncertainty from the MCMC chains; these are broadly consistent with the mock results, and are $2\sigma$ consistent in all cases. The latter results are tabulated in Tab.\,\ref{tab: AP-results}. \resub{In Fig.\,\ref{fig: ap-param-shape}, we additionally plot the posterior contours for $\vec\alpha$ in each chunk from all of the analyzed mocks as well as the data. It is clear that the posterior widths and orientations are consistent between the mocks and the data, as expected.}

Whilst the primary output of this analysis is a set of $\vec\alpha$ parameters for each individual chunk, it is instructive to consider the joint constraints from a combination of the NGC and SGC regions, in order to compare with previous results. To estimate these, we assume the $\vec\alpha$ constraints between NGC and SGC to be Gaussian and independent, allowing the joint mean and covariance to be simply estimated. (Note that the nuisance parameters are assumed to be independent between the samples). Our constraints on the AP parameters are consistent with those from the BOSS DR12 analysis of Ref.\,\citep{2017MNRAS.464.3409B} (which have $\vec\alpha = \{1.028\pm0.030,0.984\pm0.016\}$ for low-z, $\vec\alpha=\{0.964\pm0.022,1.000\pm0.015\}$ for high-z), within the stated errors. \resub{Note that we do not expect exact consistency between our results and those of Ref.\,\citep{2017MNRAS.464.3409B} since (a) we use finer $k$-space bins, (b) our combined NGC+SGC constraints are simply a Gaussian approximation of the joint posterior and (c) we use a different methodology to marginalize over the unknown broadband shape, additionally allowing $\Sigma_\mathrm{NL}$ to remain free. The latter point may explain why our error bars are slightly smaller than the earlier work.}

\begin{table}[htb]
    \centering
    \begin{tabular}{|c||c|cc|cc|}\hline
    Chunk  & $z_\mathrm{eff}$ & \multicolumn{2}{c|}{$\alpha_\parallel$} & \multicolumn{2}{c|}{$\alpha_\perp$}\\\hline
     & & mean $\pm1\sigma$ & best-fit & mean $\pm1\sigma$ & best-fit\\\hline \hline
    \textit{Prior} & -&$\mathit{1.0\pm 0.2}$ & & $\mathit{1.0\pm0.2}$& -\\\hline
    low-z NGC     & $0.38$ & $1.035\pm0.033$ & 1.041& $1.001\pm0.017$ &1.000\\
    low-z SGC     & $0.38$ & $0.971\pm0.043$ & 0.974& $0.969\pm0.036$ &0.968\\
    high-z NGC    & $0.61$ & $0.940\pm0.021$ & 0.942& $1.006\pm0.018$ &1.003\\
    high-z SGC    & $0.61$ & $1.017\pm0.038$ & 1.020& $1.020\pm0.024$ &1.019\\\hline
    low-z NGC+SGC & $0.38$ & $1.008\pm0.026$ & -& $0.997\pm0.015$ &-\\
    high-z NGC+SGC & $0.61$ & $0.957\pm0.018$ & -& $1.012\pm0.014$ &-\\\hline
    \end{tabular}
    \caption{Constraints on the AP parameters obtained from the analysis of BOSS DR12 reconstructed power spectra, obtained from running an MCMC analysis on each of the four data chunks. These are displayed graphically in red in Fig.\,\ref{fig: all-AP-params}, and we indicate the best-fit, mean and $1\sigma$ error bars of each sample. For comparison with BOSS results, we additionally estimate the joint contours on the parameters combining the NGC and SGC data-sets, by assuming the two sets of parameters to follow independent multivariate Gaussian distributions. Since the joint analysis was not performed in full, we do not provide best-fit parameters in these cases.}
    \label{tab: AP-results}
\end{table}

\subsection{Cross-Covariance of AP Parameters and Power Spectra}\label{subsec: pk-alpha-cov}

Before considering the simulated covariance of the pre-reconstruction power spectrum with the AP parameters $\vec\alpha$, it is useful to perform a simple forecast to obtain an idea of the function shape. First, note that the cross-covariance can be written
\beq
    \operatorname{cov}\left(P^\mathrm{unrec}(k,\mu),\hat{\vec\alpha}\right) &\equiv& \left\langle\delta P^\mathrm{unrec}(k,\mu)\delta\hat{\vec\alpha}\right\rangle =  \left\langle\left.\frac{\partial P^\mathrm{unrec}(k,\mu)}{\partial\vec\alpha}\right|_{\vec\alpha=\vec\alpha_0}\delta\hat{\vec\alpha}^T\delta\hat{\vec\alpha}\right\rangle\\\nonumber
    &=& \left.\frac{\partial P^\mathrm{unrec}(k,\mu)}{\partial\vec\alpha}\right|_{\vec\alpha=\vec \alpha_0}\cdot\operatorname{cov}(\hat{\vec\alpha}),
\eeq
using $\left\langle\delta\vec\alpha^T\delta\vec\alpha\right\rangle = \operatorname{cov}(\vec \alpha)$, with $\delta$ indicating the fluctuations around some mean value, e.g. $\delta\vec\alpha = \vec\alpha-\vec\alpha_0$. Next, we note that $\hat{\vec\alpha}$ is only dependent on $P^\mathrm{unrec}$ through $P^\mathrm{rec}$ (from which it is measured) and further, that only the \textit{wiggly} part $P^\mathrm{rec}_w$ is sensitive to $\vec\alpha$, since the broadband part is degenerate with the bias and hidden by theoretical error. Thus;
\beq
    \operatorname{cov}\left(P^\mathrm{unrec}(k,\mu),\hat{\vec\alpha}\right) = \frac{\partial P^\mathrm{unrec}(k,\mu)}{\partial P^\mathrm{rec}(k,\mu)}\cdot\frac{\partial P^\mathrm{rec}_w(k,\mu)}{\partial\vec\alpha}\cdot\operatorname{cov}(\hat{\vec\alpha}).
\eeq
For a simple forecast, we can assume that the derivative of $P^\mathrm{unrec}$ with respect to $P^\mathrm{rec}$ is a unit matrix, and consider only the isotropic AP parameter $\alpha \equiv \alpha_\parallel^{1/3}\alpha_\perp^{2/3}$, related to the power spectrum by $P(k;\alpha) \propto P(k/\alpha)$ (with $\alpha$-dependent prefactors being absorbed into the bias). We adopt the simple wiggly power spectrum model
\beq
    P_w(k;\alpha) \approx 0.05P_{nw}(k) \sin\left(\frac{k\ell_\mathrm{BAO}}{\alpha}\right)e^{-k^2\left(\Sigma_\mathrm{NL}^2+\Sigma_\mathrm{Silk}^2\right)},
\eeq
where $\Sigma_\mathrm{NL}\approx 3\Mpch$ and $\Sigma_\mathrm{Silk}\approx 5\Mpch$ encode non-linear and Silk damping respectively and $\ell_\mathrm{BAO} \approx 105\Mpch$ is the BAO scale (e.g.\,\citep{2015PhRvD..92d3514B}). Note that we have inserted $\alpha$ only in the sinusoidal part (where our algorithm is sensitive). This yields the estimate
\beq\label{eq: cov-alpha-pk-estimate}
    \frac{\operatorname{cov}(P^\mathrm{unrec}(k),\hat{\alpha})}{P^\mathrm{unrec}(k)\operatorname{var}(\hat{\alpha})} = \frac{\partial\log P^\mathrm{unrec}(k)}{\partial\alpha} \approx -0.05\frac{k\ell_\mathrm{BAO}}{\alpha^2}\cos\left(\frac{k\ell_\mathrm{BAO}}{\alpha}\right)e^{-k^2\left(\Sigma^2_\mathrm{NL}+\Sigma^2_\mathrm{Silk}\right)},
\eeq
where we have additionally supposed that $P_{nw}\gg P_w$. We further note that $\operatorname{var}(\hat{\alpha}) \approx 5/9\operatorname{var}(\hat{\alpha}_\perp)$ (for $\operatorname{var}(\hat{\alpha}_\perp)\approx \operatorname{var}(\hat{\alpha}_\parallel)$ and $\vec\alpha\approx \vec 1$), and will hence divide by the factor $5/9$ when comparing to single parameter estimates.

In practice, this is complicated by the hitherto ignored $\mu$-dependence (which is computable via the relations of Eq.\,\ref{eq: AP-momentum-effects}\,\&\,\ref{eq: AP-model-effects}, and may be evaluated at the fiducial value $\vec\alpha_0=\vec{1}$) and the survey window function  which has non-trivial, though small, effect on the BAO wiggles and hence the $\vec\alpha$ derivative. 

\begin{figure}[htb]
    \centering
    \includegraphics[width=0.6\textwidth]{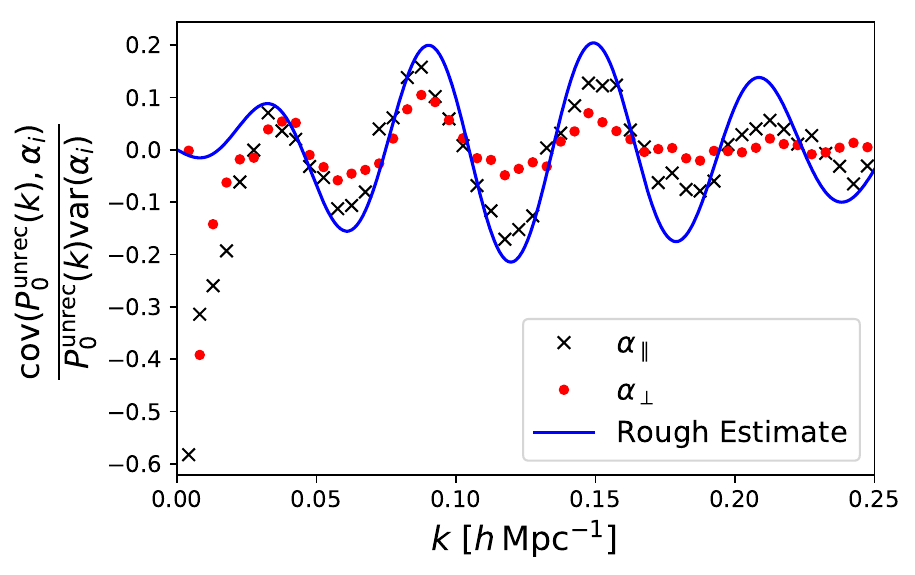}
    \caption{Covariance of the unreconstructed monopole power and the AP parameters for the low-z NGC chunk, using data from 999 Patchy mocks. Black crosses (red circles) show the covariance of the parallel (perpendicular) parameter and we normalize by the unreconstructed power measurements and AP variance in each case. The blue line shows a rough estimate based on a simple model of the post-reconstruction wiggly power spectrum (Eq.\,\ref{eq: cov-alpha-pk-estimate}), and we note that this is capture the functional form well.}
    \label{fig: alpha-pk-covariance}
\end{figure}

In Fig.\,\ref{fig: alpha-pk-covariance}, we show the covariance between the unreconstructed power monopole and $\{\alpha_\parallel,\alpha_\perp\}$ from 999 Patchy mocks, using the values of $\hat{\vec\alpha}$ described in Sec.\,\ref{subsec: results-AP}. For both parallel and perpendicular parameters we note sinusoidal oscillations which decay at large $k$, though with differing behavior at small $k$, where we expect to be strongly affected by the window function. Remarkably, our simple back-of-the-envelope theoretical model is able to capture the functional form well, including both the BAO wiggle and damping, even though it is computed assuming isotropy, ignoring the window function and the full wiggly power spectrum model. Given the success of this model, we thus expect a more thorough (yet still elementary) theoretical treatment to accurately capture this covariance, removing the need to compute the statistic on a large number of mocks. (We expect that the autocovariance of $\hat{\vec\alpha}
$ can be simply estimated via a Fisher forecast, allowing the full covariance to be computed.)

\section{Results of the Joint FS and BAO analysis}\label{sec: results-joint}
We now consider the cosmological constraints obtained from analyzing the pre-reconstruction redshift-space power spectra, armed with the joint covariances of the spectra and the AP parameters, which act as an informative prior on $\vec\alpha$. Before applying the analysis to the real data, it is important to test that our analysis is working correctly; to this end we perform a cosmological analysis of the mean of 999 Patchy mock spectra (which has much reduced statistical error). The results of this are discussed in Appendix \ref{appen: patchy-tests}, and we obtain similar conclusions to the analysis using the true BOSS data. 

\subsection{CMB-Independent Constraints on Cosmology}\label{subsec: cmb-indep-results}
\begin{figure}
    \centering
    \includegraphics[width=\textwidth]{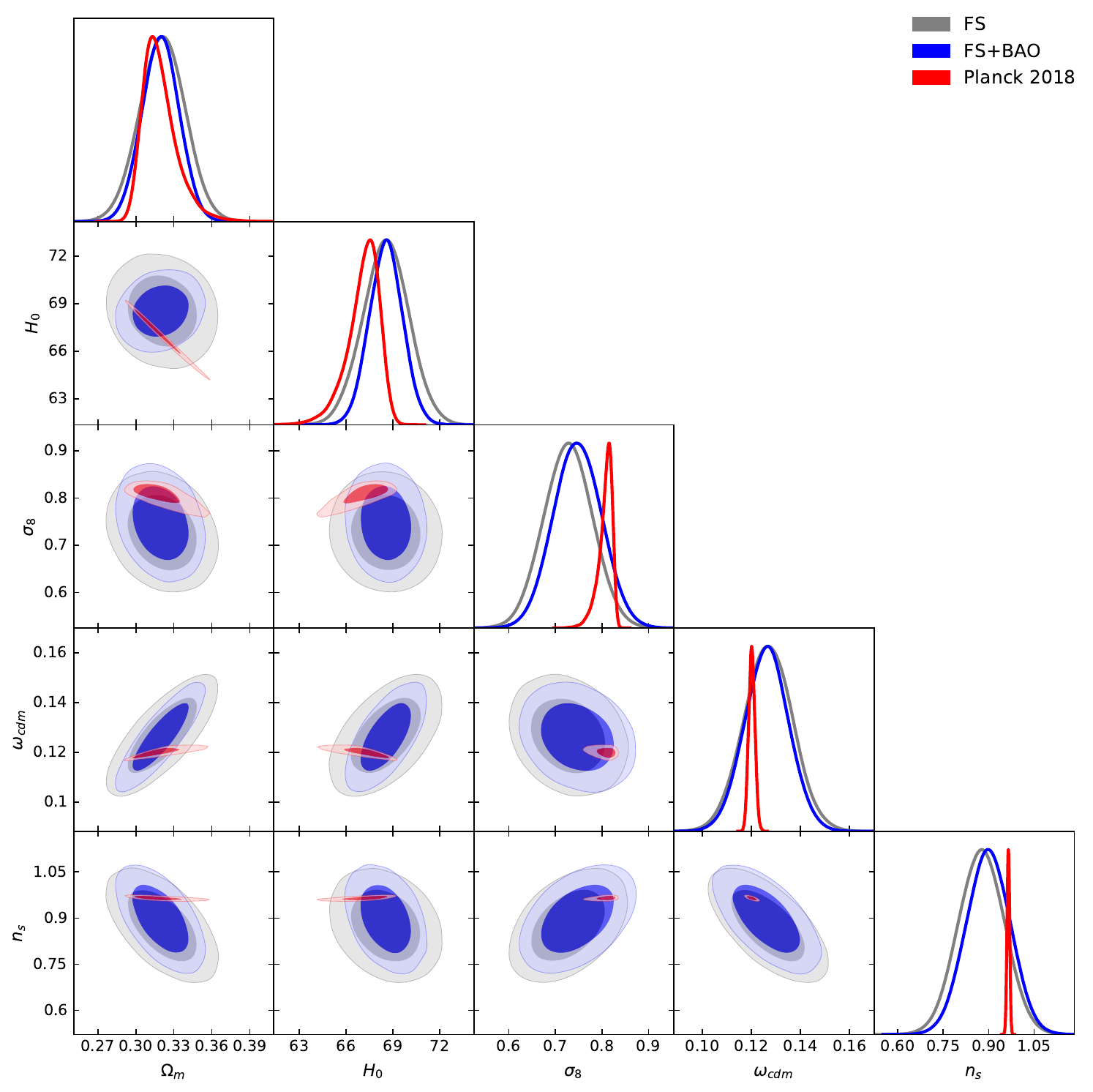}
    \caption{CMB-independent cosmological constraints obtained from this work for the baseline $\nu\Lambda$CDM model, as tabulated in Tab.\,\ref{table0}. The `FS+BAO' data-set refers to the combination of full-shape (FS) modelling of unreconstructed power spectra via a one-loop full-shape model and BAO-modelling of reconstructed power spectra to compute Alcock-Paczynski parameters, incorporating the theoretical error methodology of Ref.\,\citep{2016arXiv160200674B}, with a joint sample covariance used to unite the two approaches. The `FS' data-set (equivalent to the full-shape analysis of Sec.\,\ref{subsec: unrecon-analysis}) was presented in Ref.\,\cite{2019arXiv190905277I} and `Planck 2018' refers to Ref.\,\cite{2018arXiv180706209P}. This plot shows the cosmological constraints obtained from combination of four BOSS DR12 data chunks, which are displayed separately in Fig.\,\ref{fig:wns-separate}.  $H_0$ is quoted in $\mathrm{km}\,\mathrm{s}^{-1}\mathrm{Mpc}^{-1}$ units.}
    \label{fig:wns}
\end{figure}

Emboldened by the above success, we can proceed to analyze the full unreconstructed BOSS power spectra in conjunction with the best-fit AP parameters of Tab.\,\ref{tab: AP-results}. The cosmological constraints obtained are presented in Tab.\,\ref{table0}, with the corresponding contours shown in Fig.\,\ref{fig:wns}, comparing the results of the combined FS and BAO analysis to those obtained from a FS only analysis \citep{2019arXiv190905277I} and those of the final Planck data release \citep{2018arXiv180706209P}, using the TT,TE,EE+lowE+lowl+lensing data. These represent the main results of this paper.

\begin{table*}[t!]
    \centering
      \begin{tabular}{|c||c|c||c|c|} \hline
      & \multicolumn{2}{|c|}{base $\nu\Lambda$CDM} & \multicolumn{2}{|c|}{base $\nu\Lambda$CDM~+ fixed $n_s$}\\  \hline
         \hline
        Parameter  &  FS & FS+BAO    &  FS 
         &   FS+BAO
          \\ [0.2cm]
     \hline 
      $\omega_{cdm}$  & $0.1265_{-0.01}^{+0.01}$
      & $0.1259_{-0.0093}^{+0.009}$ 
      & $0.1113_{-0.0048}^{+0.0047}$  
      & $0.1121_{-0.0041}^{+0.0041}$ 
       \\ \hline
    $n_s$  
    &  $0.8791_{-0.076}^{+0.081}$
    & $0.9003_{-0.071}^{+0.076}$
    & $-$  
    & $-$   \\ 
    \hline
    $H_0$   & $68.55_{-1.5}^{+1.5}$
    & $68.55_{-1.1}^{+1.1}$ 
    & $67.90_{-1.1}^{+1.1}$ 
    & $67.81_{-0.69}^{+0.68}$
    \\ \hline
    $\sigma_8$   & $0.7285_{-0.053}^{+0.055}$
    &$0.7492_{-0.052}^{+0.053}$ 
    & $0.7215_{-0.044}^{+0.044}$
    & $0.7393_{-0.041}^{+0.04}$ \\
    \hline\hline
    $\Omega_m$   & $0.3203_{-0.019}^{+0.018}$ 
    & $0.3189_{-0.015}^{+0.015}$ 
    & $0.2945_{-0.01}^{+0.01}$ 
    & $0.2962_{-0.008}^{+0.0082}$   \\ \hline
    \end{tabular}
    \caption{Mean values and 68\% CL minimum credible intervals for the parameters of the base $\nu\Lambda$CDM cosmology from the joint analysis of unreconstructed and reconstructed power spectra from BOSS DR12, representing the main results of this paper. The left and right columns show the results with the power spectrum tilt ($n_s$) free and constrained by the Planck prior respectively, which correspond to the contours in Figs.\,\ref{fig:wns} and \ref{fig:fix-ns} respectively. Results are displayed in the format ``mean$^{+1\sigma}_{-1\sigma}$.'' (with $H_0$ in $\mathrm{km}\,s^{-1}\mathrm{Mpc}^{-1}$, displaying only cosmological parameters whose measurements are independent of the priors. Note that the left column is fully independent of the CMB.}
    \label{table0}
\end{table*}

The FS+BAO analysis is able to obtain strong constraints on $\Omega_m$ and $H_0$, of comparable strength the latest results from Planck, which yield $H_0 = 67.14^{+1.3}_{-0.72}$, $\Omega_m = 0.3188^{+0.0091}_{-0.016}$ for a similar $\nu\Lambda$CDM model \citep{2018arXiv180706209P}. It is thus clear that we can place strong constraints on these key parameters with \textit{no} information from the CMB. As anticipated in the forecast of Ref.\,\citep{2019arXiv191208208I}, the addition of the BAO data allows one to improve the errorbar on $H_0$ by $\sim 40\%$ compared to the FS-only analysis, due to the breaking of parameter degeneracies. This shows the utility of this method in a CMB-independent analyses. \resub{Notably, several of the parameters are more than $1\sigma$ away from the fiducial cosmology assumed to create the Patchy mocks, and hence the sample covariance matrix. Ref.\,\citep{2016MNRAS.460.4188G} showed this effect to be statistically insignificant for BOSS, thus it is not expected to bias our analysis.}

For other parameters including $n_s$ and $\omega_{cdm}$, we obtain optimal parameters that are consistent with Planck though with much larger errors. In these cases, the improvement from adding BAO information is marginal, since, due to the theoretical error marginalization, the BAO analysis is sensitive to only the wiggly part of the power spectrum, where these parameters have minimal impact. Note that neither the FS or FS+BAO analysis is able to place strong constraints on the power spectrum slope $n_s$, with values far away from the Planck prediction (with far from scale-invariant power spectra) allowed by the BOSS data. This is a result of the paucity of modes in the large-scale regime, which are particularly sensitive to $n_s$.

\begin{figure}
    \centering
    \includegraphics[width=\textwidth]{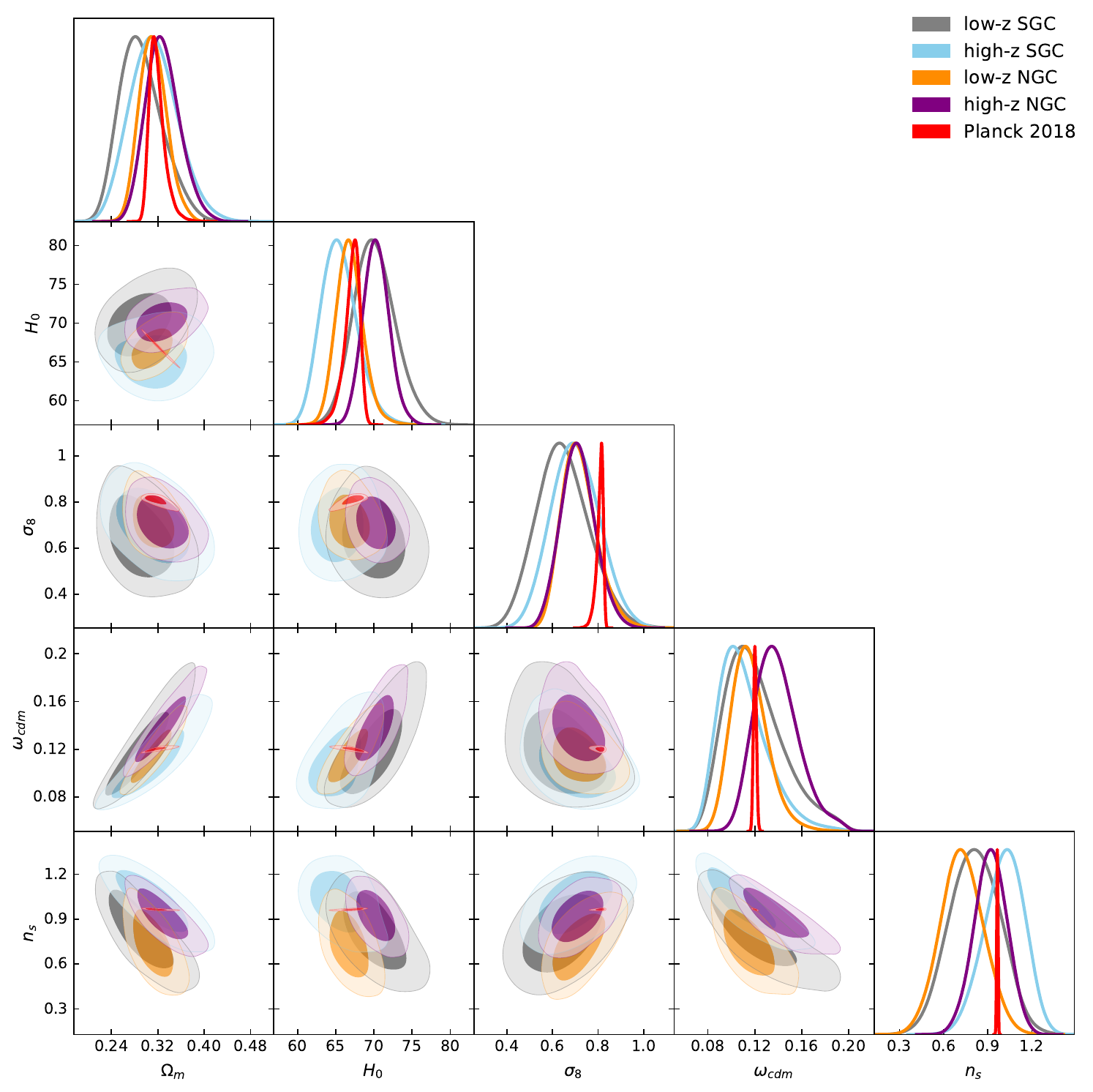}
    \caption{Cosmological constraints obtained from the joint FS and BAO analysis of four disjunct BOSS DR12 data chunks, compared to the constraints from the latest Planck analysis \citep{2018arXiv180706209P}. The joint analyses of all four chunks yields the contours of Fig.\,\ref{fig:wns}.}
    \label{fig:wns-separate}
\end{figure}

In Fig.\,\ref{fig:wns-separate} we show the constraints obtained from analyzing each of the four data chunks separately, with corresponding parameters given in Tab.\,\ref{table:allc} of Appendix \ref{appen: individual-chunks}. Note that, even in the FS+BAO analysis of all four chunks simultaneously, the AP parameters are computed for each chunk independently (via the BAO analysis of \ref{subsec: recon-analysis}), to allow correct computation of joint covariance matrices. Comparing Figs.\,\ref{fig:wns}\,\&\,\ref{fig:wns-separate} allows one to conclude that the cosmological parameters extracted from different BOSS data chunks are strongly compatible with each other. As expected, constraints from the high-z NGC region are the tightest, since this has the highest effective volume.

\subsection{Fixing the Spectral Tilt}\label{subsec: results-fix-tilt}

Given the weak constraints on the spectral tilt $n_s$ obtained from the FS+BAO analysis above, it is instructive to check the impact of imposing the Planck prior on $n_s$. Indeed, this prior is $\sim 20$ times tighter than the BOSS measurement itself, thus if one were to combine the FS+BAO and Planck likelihoods, the $n_s$ measurement would be completely dominated by Planck. Moreover, including this prior can be seen as a result of using a minimal input from Planck, since it does not completely fix the shape information of BOSS. 

\begin{figure}
    \centering
    \includegraphics[width=0.49\textwidth]{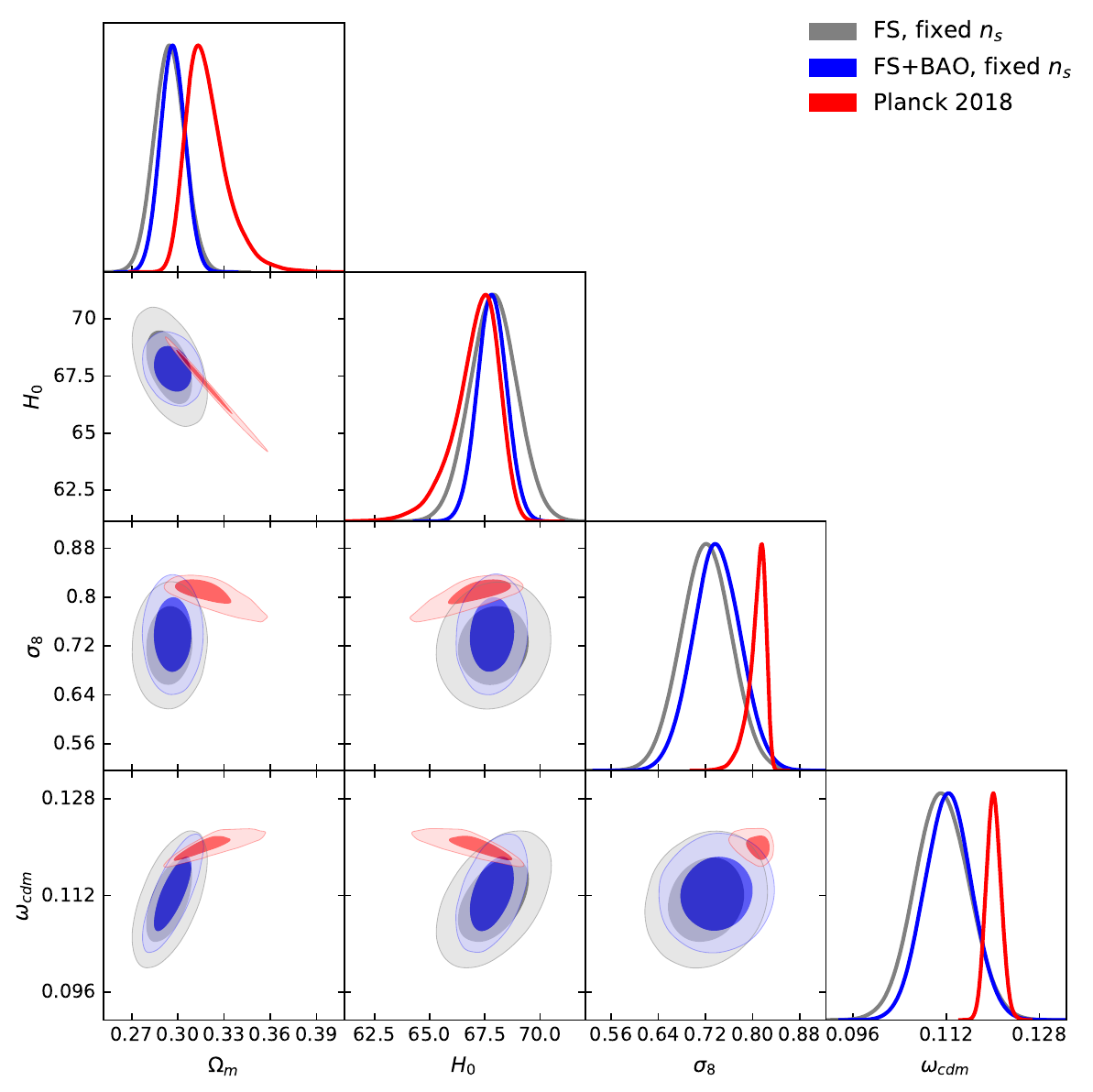}
    \includegraphics[width=0.49\textwidth]{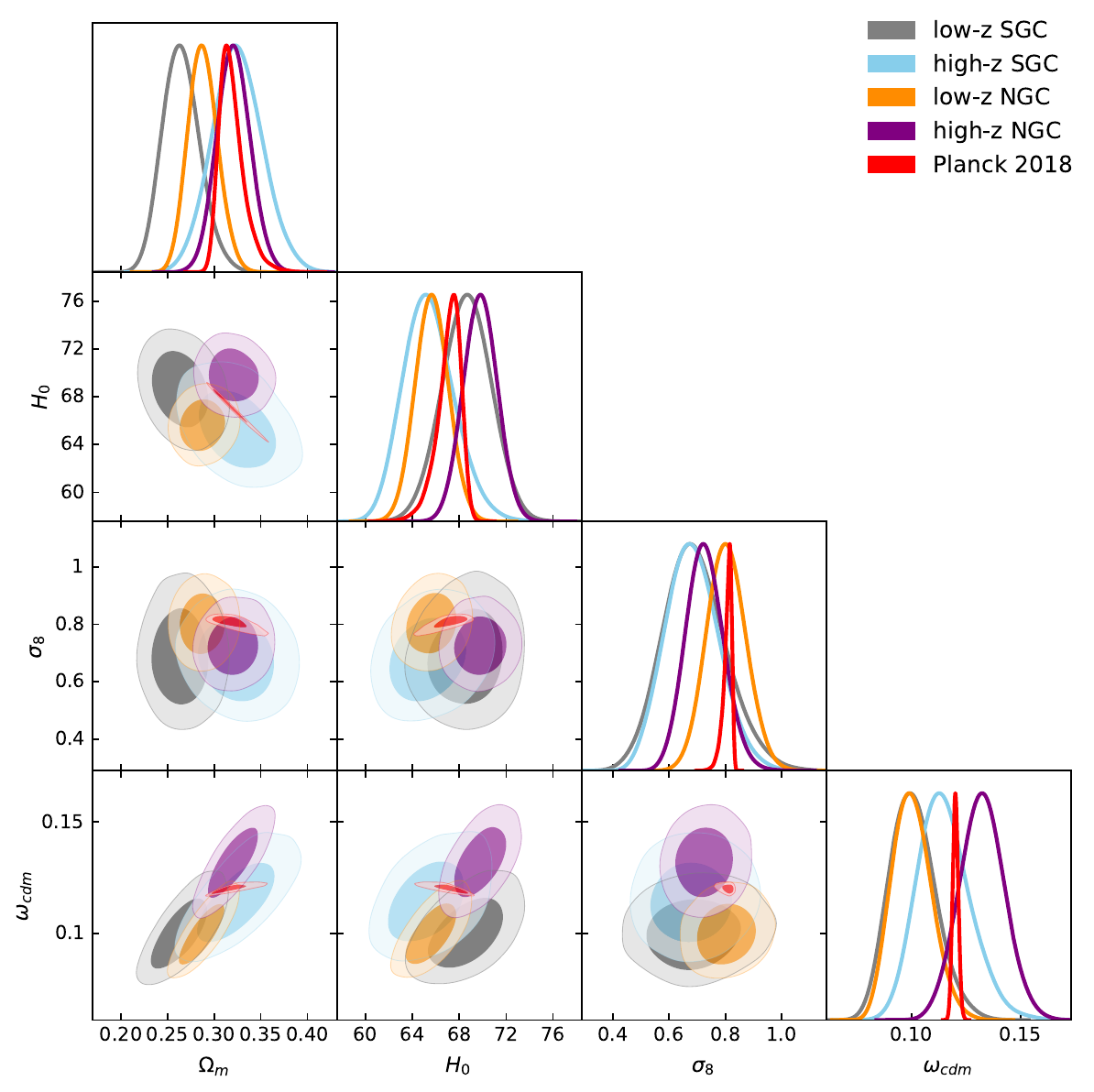}
    \caption{Cosmological constraints obtained from this work, using the CMB-independent $\nu\Lambda$CDM model, but imposing Planck priors on the spectral slope $n_s$. The FS+BAO constraints obtained from analyzing the four data chunks in combination and separately are shown in the left and right plots respectively, which have the same forms as Figs.\,\ref{fig:wns}\,\&\,\ref{fig:wns-separate}, where $n_s$ was left unconstrained.}
    \label{fig:fix-ns}
\end{figure}

The posterior distribution of cosmological parameters and their marginalized limits are presented in the right panel of Tab.~\ref{table0} and the left panel of Fig.~\ref{fig:fix-ns}, with the associated results for each chunk analyzed separately tabulated in Appendix \ref{appen: individual-chunks} and shown in the right panel of Fig.\,\ref{fig:fix-ns}. For comparison, we also show the results obtained from a similar analysis of the FS-only likelihood, without the additional BAO information provided from the AP parameters. Notably, we obtain sharper constraints on $H_0$ and $\Omega_m$ by a factor of $\sim 2$ for both FS-only and FS+BAO analyses, due to the removal of $n_s$-induced degeneracies. Furthermore, we observe that the $\sim 40\%$ improvement on $H_0$ from adding the BAO data holds even with the fixed spectral tilt, demonstrating that the BAO is still a useful source of information in this more constrained case. Remarkably, the combined FS+BAO constraint on $H_0$
in this case is stronger than that found by Planck, representing the strongest constraint on $H_0$ \resub{within $\Lambda$CDM}, albeit with
a single CMB-informed prior. 

Given the weak response of the likelihood to changes in $n_s$ (Fig.\,\ref{fig:wns}), we note that we would obtain very similar results from instead assuming that $n_s$ takes the Harrison-Zeldovich form $n_s = 1$. Such a scale-invariant power spectrum also carries significant theoretical motivation; it is a generic prediction of inflationary models that $n_s$ should be close to unity, with any strong departure completely infeasible if one assumes inflation to be correct. Our results can thus be interpreted as those arising from the imposition of a physically relevant prior on $n_s$; they are not specific to the CMB.

\section{Combining the CMB and Galaxy Surveys: Constraints on Cosmology from FS+BAO+Planck}\label{sec: results-Planck}
In this Section, we discuss cosmological implications of the combination of information provided from BOSS galaxy clustering with CMB data from Planck. From above it is clear that these are statistically compatible, and one may thus expect combination of the data-sets to yield tighter constraints on cosmology, allowing bounds to be placed on non-minimal cosmological models. In Ref.\,\citep{2019arXiv191208208I}, the Planck+FS and Planck+BAO combinations were considered (also for the BOSS DR12 data-set), and it was observed that the BAO and FS added a very similar amount of information to the CMB data-set, modulo the low clustering power observed in BOSS. It was therefore conjectured that the combination of BAO and FS likelihoods (with some proper covariance matrix) could lead to significant improvement in cosmological constraints. In this work, we have developed such a formalism, through AP parameters and theoretical error, thus it is the goal of this section to test such a claim.

Given the Planck TT,TE,EE+lowE+lowl+lensing data and the FS+BAO likelihood presented above, we may easily run a joint analysis of Planck+FS+BAO in the same manner as the Planck+FS and Planck+BAO analyses of Ref.\,\citep{2019arXiv191208208I}, to which we refer the reader for further details on the methodology and description of the Planck likelihood. In particular, we focus on two models of significant cosmological interest; the minimal $\nu\Lambda$CDM with varied neutrino masses (and unconstrained $n_s$), and the same model with additional relativistic degrees of freedom, parametrized by $N_\mathrm{eff}$ (which has previously been set to $N_\mathrm{eff}=3.046$). \resub{The motivation for the search of light relics (via $N_\mathrm{eff}$) can be found in Ref.~\cite{Green:2019glg}. See also Ref.~\cite{Baumann:2019tdh} for the attempts to measure this from the BOSS BAO
data, and Ref.~\cite{Baumann:2017gkg}
for forecasts relevant to current and future LSS and CMB experiments.}

\begin{table*}[t!]
\begin{center}
 \begin{tabular}{|c||c|c|c|c|} \hline
  & \multicolumn{4}{|c|}{$\nu\Lambda$CDM} \\  \hline
     \hline
    Parameter  &  {\small Planck } & {\small Planck~+~BAO }  &  {\small Planck~+~FS }   
     & {\small Planck + BAO + FS} \\ [0.2cm]
 \hline 
  $100~\omega_{b}$  
  &  $2.238_{-0.015}^{+0.016}$
  & 
  $2.245_{-0.014}^{+0.014}$ 
  & $2.247_{-0.013}^{+0.015}$  
  &  $2.245_{-0.013}^{+0.014}$ \\ \hline
  $\omega_{cdm}$  
  & $0.1201_{-0.0014}^{+0.0013}$ 
  & $0.11919_{-0.00099}^{+0.00099}$ 
  & $0.11893_{-0.001}^{+0.00097}$  
  & $0.11916_{-0.00089}^{+0.00089}$  \\ \hline
  $100~\theta_{s}$   
  & $1.04187_{-0.0003}^{+0.0003}$
  & $1.04195_{-0.00029}^{+0.00029}$ 
  & $1.04196^{+0.00028}_{-0.00028}$ & $1.04194_{-0.00028}^{+0.0003}$\\ \hline
$\tau$   &  
$0.0543_{-0.0079}^{+0.0074}$
& 
$0.05556_{-0.0076}^{+0.007}$ 
& $0.05539_{-0.0072}^{+0.0074}$ & $0.05453_{-0.0074}^{+0.0069}$ \\ \hline
$\ln(10^{10}A_s)$   & 
$3.045_{-0.016}^{+0.014}$
& $3.045_{-0.015}^{+0.014}$ 
& $3.044_{-0.014}^{+0.014}$ & $3.043_{-0.015}^{+0.014}$ \\ 
\hline
$n_s$  
&  $0.9646_{-0.0045}^{+0.0045}$
& $0.9669_{-0.0039}^{+0.0039}$ 
& $0.967_{-0.004}^{+0.0038}$  & $0.9664_{-0.0038}^{+0.0037}$\\ 
\hline
$M_{\text{tot}}$   
& $<0.26~$ 
& $<0.12~$
& $<0.16~$
& $<0.14~$\\  
\hline\hline
$\Omega_m$   
& $0.3188^{+0.0091}_{-0.016}$ 
& $0.3078^{+0.0060}_{-0.0071}$ 
& $0.3079^{+0.0065}_{-0.0085}$  & 
$0.3093^{+0.0060}_{-0.0072}$\\ \hline
$H_0$   
&  
$67.14_{-0.72}^{+1.3}$
& $67.97_{-0.49}^{+0.56}$ 
& $67.95_{-0.52}^{+0.66}$ & $67.87_{-0.43}^{+0.58}$ \\ \hline
$\sigma_8$   &
$0.8053_{-0.0091}^{+0.019}$ 
&$0.8135_{-0.0073}^{+0.01}$ 
& $0.8087_{-0.0072}^{+0.012}$& 
$0.8098_{-0.0069}^{+0.012}$\\ \hline
\end{tabular}
\end{center}
\begin{center}
 \begin{tabular}{|c||c|c|c|c|} \hline
   & \multicolumn{4}{|c|}{$\nu\Lambda$CDM~+~$N_{\text{eff}}$}\\  \hline
     \hline
    Parameter   &  
    Planck
     &   {\small Planck~+~BAO}
    &  {\small Planck~+~FS}
    &{\small Planck 
    + BAO + FS}
      \\ [0.2cm]
 \hline 
  $100~\omega_{b}$  
  & $2.224_{-0.023}^{+0.023}$
  & $2.240_{-0.019}^{+0.019}$
  & $2.233_{-0.019}^{+0.019}$ 
  & $2.234_{-0.019}^{+0.017}$ \\ \hline
  $\omega_{cdm}$  
  & $0.1181_{-0.0031}^{+0.003}$ 
  & $0.1182_{-0.0031}^{+0.0029}$ 
  & $0.1166_{-0.0028}^{+0.0026}$ 
  & $0.1171_{-0.0025}^{+0.0024}$ \\ \hline
  $100~\theta_{s}$   
  & $1.04220_{-0.00054}^{+0.00051}$
  & $1.04210_{-0.00052}^{+0.0005}$ 
  & $1.04234_{-0.0005}^{+0.00049}$ 
  & $1.04228_{-0.00046}^{+0.00046}$\\ \hline
$\tau$   
& $0.05341_{-0.008}^{+0.0074}$
& $0.05516_{-0.0078}^{+0.0072}$
& $0.05409_{-0.0075}^{+0.0073}$  
& $0.05372_{-0.0067}^{+0.0066}$
\\ \hline
$\ln(10^{10}A_s)$   
& $3.037_{-0.018}^{+0.018}$
& $3.042_{-0.017}^{+0.017}$ 
& $3.035_{-0.017}^{+0.016}$  
& $3.036_{-0.016}^{+0.013}$\\ 
\hline
$n_s$  
&$0.9588_{-0.0087}^{+0.0087}$ 
& $0.9647_{-0.0074}^{+0.0073}$& 
$0.9608_{-0.0072}^{+0.0074}$ 
& $0.9615_{-0.007}^{+0.0061}$\\ 
\hline
$M_{\text{tot}}$   
& $<0.27$& $<0.12$ & $<0.16$  & $<0.14$\\   \hline
$N_{\text{eff}}$   
& $2.90_{-0.19}^{+0.19}$ 
& $2.99_{-0.17}^{+0.17}$ 
& $2.88_{-0.17}^{+0.17}$ &
$2.90_{-0.16}^{+0.15}$\\  
\hline\hline
$\Omega_m$   
& $0.324^{+0.011}_{-0.019}$ 
& $0.3090^{+0.007}_{-0.0076}$ & 
$0.3127^{+0.0080}_{-0.0091}$  & 
$0.3126^{+0.0067}_{-0.0072}$\\ \hline
$H_0$    
& $66.1_{-1.6}^{+1.9}$ 
& $67.6_{-1.2}^{+1.2}$ 
& $66.8_{-1.2}^{+1.2}$  
& $67.0_{-1.0}^{+1.0}$ \\ \hline
$\sigma_8$   
& $0.798_{-0.013}^{+0.022}$
& $0.811_{-0.011}^{+0.012}$ 
& $0.8015_{-0.011}^{+0.013}$ & $0.8042_{-0.0092}^{+0.01}$ \\ \hline
\end{tabular}
\caption{Mean values and 68\% CL minimum credible
intervals for the parameters of the $\nu\Lambda$CDM (upper table) and 
$\nu\Lambda$CDM$~+~N_{\text{eff}}$ (lower table) models as extracted 
from the Planck, Planck+BAO, Planck+FS data, and 
Planck+FS+BAO data presented as ``mean$^{+1\sigma}_{-1\sigma}$'', as in Tab.\,\ref{table0}. For $M_{\text{tot}}$ we quote the 95\% CL upper limit in units of eV. $H_0$
is quoted in $\mathrm{km}\,\mathrm{s}^{-1}\mathrm{Mpc}^{-1}$ and $N_\mathrm{eff}$ is taken as $3.046$ in the first analysis.}
\label{tab:Planck}
\end{center}
\end{table*}

\begin{figure*}[ht]
\begin{center}
\includegraphics[width=1\textwidth]{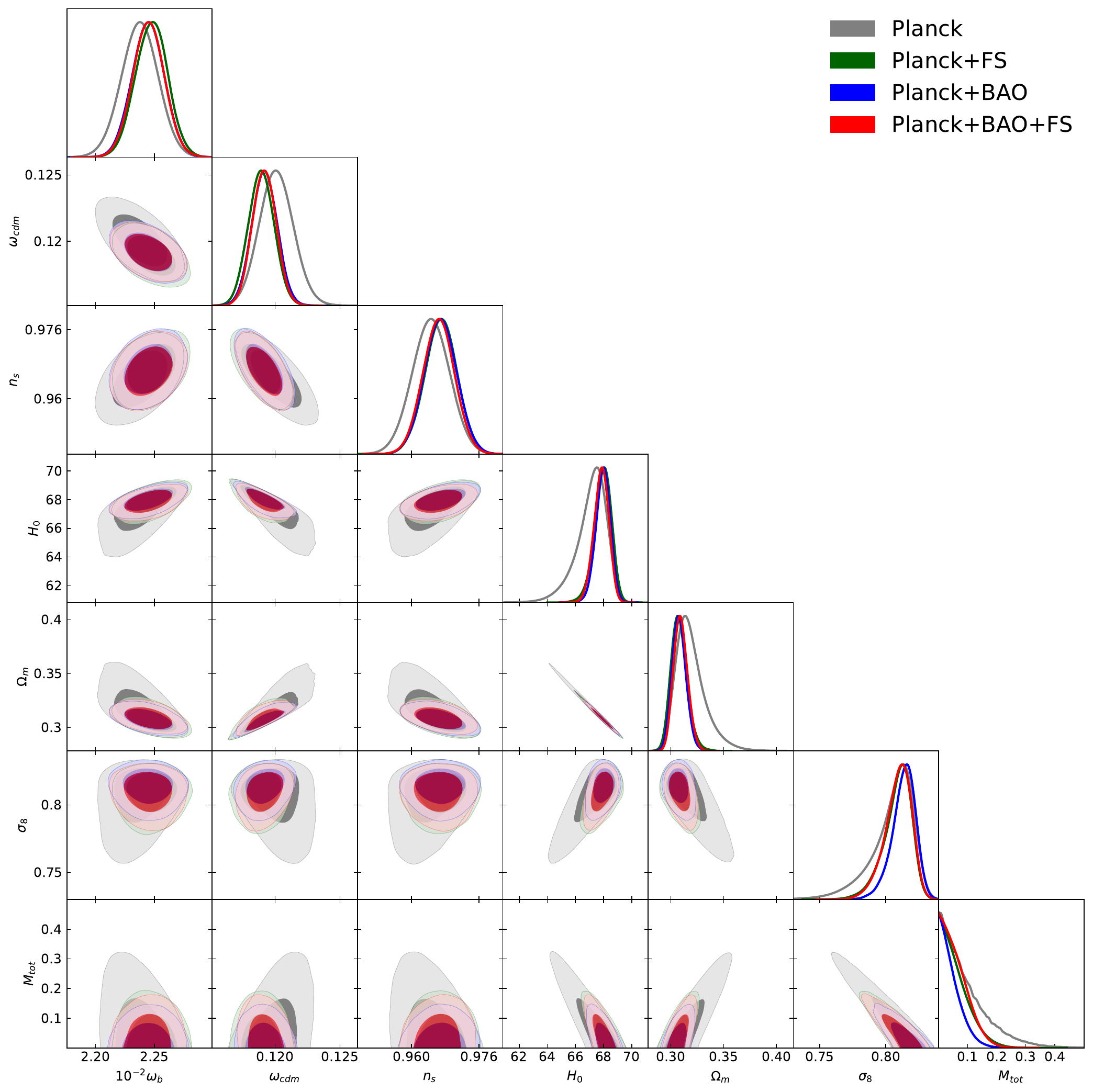}
\end{center}
\caption{Marginalized one-dimensional posterior distribution and two-dimensional probability contours (at the 68\% and 95\% CL) for the parameters of the $\nu\Lambda$CDM model, including varied neutrino masses, as obtained from analyses of the Planck likelihood \resub{separately} and in combination with BAO and FS information from BOSS. $N_{\text{eff}}$ is fixed to the standard model value $3.046$ and we quote $H_0$ in $\mathrm{km}\,\mathrm{s}^{-1}\mathrm{Mpc}^{-1}$, with $M_{\text{tot}}$ given in eV.}\label{fig:mnuPl}  
\end{figure*}

\begin{figure*}[ht]
\begin{center}
\includegraphics[width=1\textwidth]{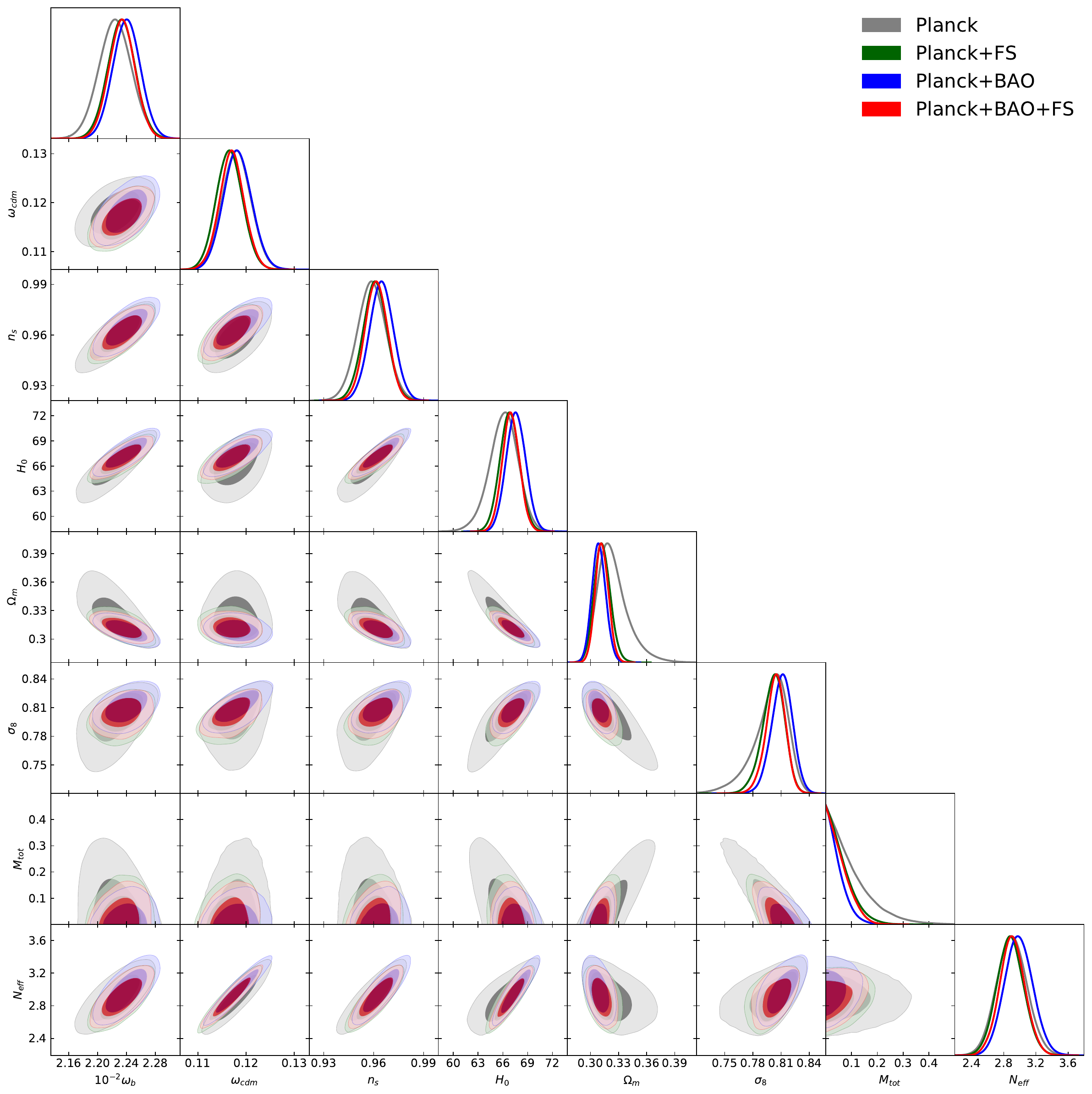}
\end{center}
\caption{As Fig.\,\ref{fig:mnuPl}, but for the cosmological parameters of the $\nu\Lambda$CDM+$N_\mathrm{eff}$ model, additionally varying the number of relativistic degrees of freedom $N_\mathrm{eff}$.}\label{fig:neffPl}
\end{figure*}

Our final results are presented in Tab.\,\ref{tab:Planck} and Figs.\,\ref{fig:mnuPl}\,\&\,\ref{fig:neffPl}, showing the one-dimensional marginalized limits along with the triangle plot for the main cosmological parameters in the $\nu\Lambda$CDM and $\nu\Lambda$CDM+$N_\mathrm{eff}$ models. In the first case, we see only marginal improvement in the joint Planck+BAO+FS likelihood over the Planck+FS (or Planck+BAO) data, indicating the the addition of BAO information does \textit{not} lead to stronger constraints on cosmology. This additionally applies to the summed neutrino mass, with joint constraints giving $M_{\rm tot}<0.14$ eV at 95$\%$ CL (though we note that the Planck+FS+BAO constraint appears to rectify a slight underestimate in the Planck+BAO likelihood). Considering the $\nu\Lambda$CDM+$N_\mathrm{eff}$ model, which has additional freedom, the improvement from adding BAO information is more sizeable, though the errorbars shrink by only $\sim 20\%$ at maximum (for $H_0$). Furthermore, we do not observe any significant change to the $N_\mathrm{eff}$ errorbar from the additional data, with the mean remaining in the same location as in the Planck+FS measurement.

Given that they are constructed from the same density fields, the BAO and FS data-sets share the same low-$k$ BAO wiggles (for $0.03\lesssim k / \hMpc\lesssim 0.15 $), but differ by additional geometric and shape information, since the BAO likelihood contains the positions of the reconstructed BAO wiggles for $k\gtrsim 0.15\hMpc$, whereas the FS data embed the broadband shape information including the power spectrum peak, but with only suppressed wiggles. Since our joint FS+BAO likelihood thus contains all the geometric information encoded in the galaxy power spectrum, one may na\"ively expect this to have a substantial improvement over the Planck+FS or Planck+BAO constraints. In practice, this is not the case, as we observe from Figs.\,\ref{fig:mnuPl}\,\&\,\ref{fig:neffPl}. This is expected to occur since the sharpening of constraints from Planck+BAO or Planck+FS is primarily a result of the breaking of geometric degeneracies that cannot be done with one data-set alone. Both the BAO and FS likelihoods are able to break this degeneracy equally efficiently, thus following its breaking, including further LSS information, produces only a minor reduction in the $H_0$ error. We thus conclude that, whilst BAO information substantially improves cosmological constraints from LSS analyses alone, it is of limited use in their combination with CMB surveys.

\section{Summary and Conclusion}\label{sec: conclusion}
In this work, we have presented new methods for the analysis of large scale structure surveys, in particular for the joint analysis of pre- and post-reconstruction galaxy power spectra in redshift space. In particular, we are able to robustly combine the Alcock-Paczynski (AP) information in the BAO peaks with the full-shape (FS) information from unreconstructed spectra by means of a joint covariance, which can be robustly estimated from mocks or simple theory.

A key outcome of this work is the development of a new technique for extracting information from reconstructed power spectra, combining simple theory with a theoretical error model \citep{2016arXiv160200674B,2019JCAP...11..034C}. This augments the usual sample covariance with an additional covariance which scales as the neglected one-loop power spectrum, and crucially has a non-zero coherence length, allowing positions of the correlated BAO peaks to be separated from the unmodeled (and poorly understood) broadband spectrum. This was shown to be highly robust and carries no free parameters besides the coherence length and amplitude, which can be fixed to physically motivated values and do not have a noticeable affect on the constraints. For this reason, the method is far simpler than conventional techniques, which involve marginalization over a number of polynomial shape parameters \citep{2020PhRvD.101d3510H,2017MNRAS.464.3409B}. We expect this to be of great use in future BAO analyses.

Applying the combined FS+BAO likelihood to the BOSS data-set, we were able to place strong constraints on $\Omega_m$ and $H_0$ which are fully independent of the CMB (using only BBN priors on $\omega_b$), achieving a $1.6\%$ constraint on the Hubble parameter in a $\Lambda$CDM model with massive neutrinos, with a $\sim 40\%$ improvement found from the addition of BAO data, due to extra geometric information being provided. In the most minimial extension to the model, we adopted a Planck prior on the spectral slope $n_s$ which is poorly constrained by BOSS; this yielded a 1\% measurement of $H_0$ with a significant improvement found from the inclusion of BAO data, as before. In this case, the constraints are consistent with, but tighter than, those obtained from the Planck analysis, though we caution that both measurements assume the same physical model and would thus be similarly affected by any esoteric new physics occurring at early times. 

In contrary to the above, the improvement found from the addition of BAO data in a joint analysis of Planck and BOSS is found to be marginal, with only $H_0$ not dominated by Planck. In combination with the CMB, the main benefit of existing galaxy surveys is to provide geometric information, and for this purpose, FS and BAO data are equally appropriate, though the constraints on non-minimal cosmological models were found to be slightly tighter using their combination. Whilst it is true that the constraints on $H_0$ are similar from Planck+FS and Planck+BAO, this does not necessarily imply that the FS is not a useful source of information; given that BAO reconstruction is a highly computationally intensive procedure for observational data-sets (due to the necessity of generating constrained field realizations and inverting matrices of size $\sim 10^9$), an FS-only analysis still requires less computation time than a BAO-only one.

To place these results in context, we consider other contemporary analyses, in particular the joint analysis of weak-lensing, BAO and BBN by the DES collaboration \citep{2018MNRAS.480.3879A}. This obtained similar constraints on $H_0$ to our free-$n_s$ analysis for a similar cosmological model, except with the fixed neutrino mass. In both analyses, $\omega_b$ is fixed by the BBN  prior (recalling that we adopt the BBN prior on $\omega_b$ as well), whilst $\Omega_m$ in constrained by weak lensing in Ref.~\citep{2018MNRAS.480.3879A} rather than the full-shape of the galaxy power spectra. 
This provides the necessary degeneracy-breaking to obtain a sharp constraint on $H_0$ from the BAO. The effects of combining BBN and BAO information are further shown in Refs.\,\citep{2019JCAP...10..029S}\,\&\citep{2019JCAP...10..044C}, demonstrating that the resulting constraints are in tension with strong lensing and supernovae Ia measurements, but consistent with the CMB. Both analyses fix $\omega_b$ from BBN, and measure the ratio of $H_0$ and $\Omega_m$ from the AP parameters, with the latter considering BAO measurements from both galaxy surveys and the Lyman-$\alpha$ forest. Crucially, it is noted that the AP effect is particularly weak in $\Lambda$CDM at low redshift (since the AP parameters are very weak functions of $\Omega_m$), and thus the BAO measurements alone are inherently poor at constraining cosmology (note also the broad BBN+BAO+Pantheon constraints in Fig.\,17 of Ref.\,\citep{2018arXiv180706209P}). At high redshift (traced by Lyman-$\alpha$ emitters), the $\Omega_m-H_0$ posterior ellipse extracted from the AP parameters takes 
a different orientation such that in combination 
with the galaxy BAO the degeneracy between these two parameters is broken and 
the constraints on them become sharp (Fig.\,1 of Ref.\,\citep{2019JCAP...10..044C}), indicating that the combination of low and high redshifts could significantly improve parameter constraints.

Given the above discussion, it is worth reflecting on the future of BAO reconstruction. Firstly, it was demonstrated that the AP parameter constraints were not improved by better modeling of the reconstructed parameter shape (leading to a more accurate estimate of the BAO damping parameter). Thus, given that the principal goal of reconstruction is to obtain extract more information from the BAO peak, which is fully encoded in the AP parameters, we should not be attempting to produce more accurate models of the reconstructed spectrum, since all relevant broadband information is contained in the pre-reconstruction spectra. If our goal is to obtain strong constraints on cosmology from galaxy surveys alone, BAO reconstruction remains an important item in the analyst's toolkit, shown by the 40\% improvements in parameter constraints in this work, \resub{though we expect this to decrease as the survey volume grows larger.} However, if our final aim is the combination with CMB data, it provides little additional information \resub{for the type of analyses presented in this paper, though we note that sharpening the power spectrum wiggles remains useful when searching for primordial features which have little signature in the broadband.}

Whilst the results in this paper represents the most complete analysis of BOSS DR12 data to date, there are a variety of ways in which they could be improved. In particular, the analysis could be extended to higher wavenumber, allowing slightly more BAO peaks to be measured with relative ease (assuming that an appropriate theoretical error kernel is chosen), though this requires more accurate modeling of the fingers-of-God effect for the full-shape model. Furthermore, the treatment of the covariance matrix is not exact, especially since it is based on mocks which are known to differ slightly from observational data. A more thorough treatment would involve recomputing the covariance matrix for the best-fit cosmology, which, whilst time-consuming, is possible via the techniques of \citep{2019JCAP...01..016L,2019arXiv191002914W}, though we do not expect this to have a significant impact on parameter constraints. The above notwithstanding, we see that robust combination of FS and BAO data allows for strong constraints to be placed on cosmology from galaxy surveys alone, though the utility of BAO is marginal when surveys are combined with CMB data.

\acknowledgments
We thank Florian Beutler for invaluable assistance \resub{and for providing public access to the BOSS data and corresponding analysis products. In addition, we thank Martin White and the anonymous referee for useful feedback.} M.I. is partially supported by the Simons Foundation's \textit{Origins of the Universe} program and by the RFBR grant 20-02-00982 A.
M.Z. is supported by NSF grants AST1409709, PHY-1521097 and PHY-1820775 the Canadian
Institute for Advanced Research (CIFAR) program on
Gravity and the Extreme Universe and the Simons Foundation Modern Inflationary Cosmology initiative.

Funding for SDSS-III has been provided by the Alfred P. Sloan Foundation, the Participating Institutions, the National Science Foundation, and the U.S. Department of Energy Office of Science. The SDSS-III web site is http://www.sdss3.org/.

SDSS-III is managed by the Astrophysical Research Consortium for the Participating Institutions of the SDSS-III Collaboration including the University of Arizona, the Brazilian Participation Group, Brookhaven National Laboratory, Carnegie Mellon University, University of Florida, the French Participation Group, the German Participation Group, Harvard University, the Instituto de Astrofisica de Canarias, the Michigan State/Notre Dame/JINA Participation Group, Johns Hopkins University, Lawrence Berkeley National Laboratory, Max Planck Institute for Astrophysics, Max Planck Institute for Extraterrestrial Physics, New Mexico State University, New York University, Ohio State University, Pennsylvania State University, University of Portsmouth, Princeton University, the Spanish Participation Group, University of Tokyo, University of Utah, Vanderbilt University, University of Virginia, University of Washington, and Yale University.

\appendix

\section{Tests on Patchy mocks}\label{appen: patchy-tests}
Here, we analyze the joint BAO+FS data from the Patchy mocks to ensure that our methods provide unbiased cosmological parameter constraints. Whilst our BAO analysis remains identical, in the FS part (Sec.\,\ref{subsec: unrecon-analysis}), our data-vector becomes the power spectrum multipoles of the unreconstructed Patchy mocks plus the best-fit AP parameters extracted from the reconstructed spectra of the same mocks. For this test, we use only a single data-chunk, low-z NGC, and use the mean power spectrum and AP parameters from 2048 and 999 mocks respectively, to reduce statistical error. It is pertinent to note that this procedure does not guarantee that the mean of the inferred parameters would exactly match the fiducial values because the Patchy mocks were generated using an approximate gravity solver
and some HOD model, which were  designed to reproduce the real data only within $1\sigma$ limits.

When fitting the joint FS+BAO datavector, we fix $\omega_b$ and $n_s$ to the fiducial values used in the simulations, for a more strict test of our analysis. This is justified since these are not the principal parameters estimated by our FS+BAO likelihood, and allowing them to vary freely leads to larger error ellipses in which it is easier to hide potential biases in the crucial parameters $\sigma_8$, $\Omega_m$ and $H_0$. Furthermore, the mocks were created without massive neutrinos, thus we additionally fix $\sum m_\nu = 0$. The remainder of our theory model and parametrization is identical to the one used in our main analysis.

\begin{figure}
    \centering
    \includegraphics[width=0.49\textwidth]{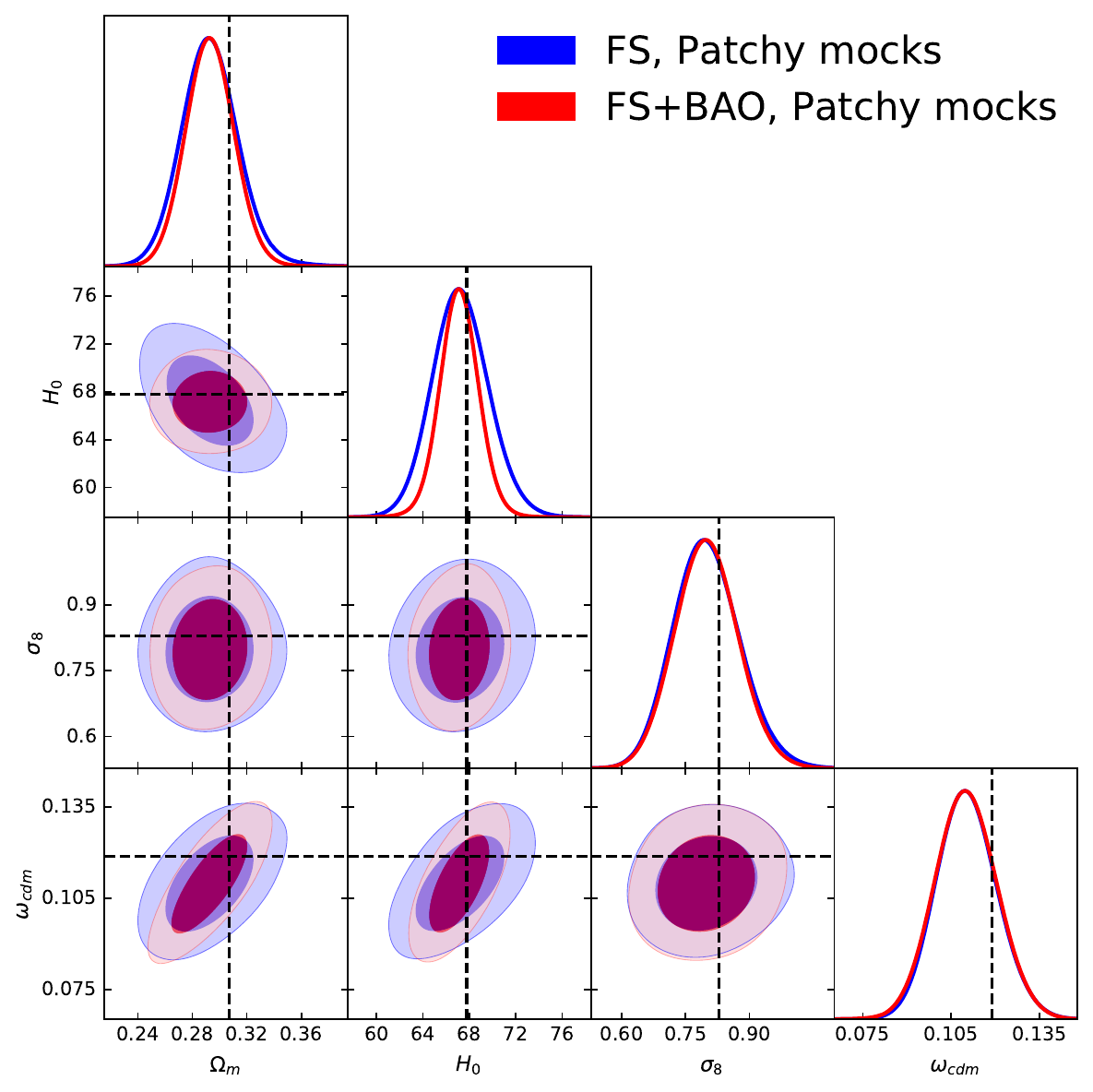}
    \caption{Parameter constraints obtained from analyzing the mean of the Patchy mocks corresponding to the low-z NGC BOSS galaxy sample, with the FS+BAO or FS-only likelihoods. Dashed lines represent the true values used in the simulations, which are $1\sigma$ consistent in all cases. This uses a slightly simplified cosmological model compared to the main analysis, as described in the text.}
    \label{fig:mocks}
\end{figure}

\begin{table*}[t!h]
    \centering
      \begin{tabular}{|c||c|c|} \hline 
      & \multicolumn{2}{|c|}{base $\Lambda$CDM} \\  \hline
         \hline
          Parameter &  FS & FS+BAO   
          \\ [0.2cm]
     \hline \hline 
         \multicolumn{3}{|c|}{Patchy mocks NGC low-z ($z_{\rm eff}=0.38$)}
         \\ \hline \hline
      $\omega_{cdm}$  & $0.1099_{-0.0098}^{+0.0086}$
      & $0.1103_{-0.0099}^{+0.0088}$
       \\ \hline
    $H_0$   & $67.28_{-2.5}^{+2}$
    & $67.18_{-1.6}^{+1.5}$
    \\ \hline
    $A^{1/2}$   & $1.019_{-0.11}^{+0.093}$
    &$1.019_{-0.1}^{+0.093}$\\
    \hline\hline
    $\Omega_m$   & $0.2929_{-0.019}^{+0.019}$ 
    & $0.2929^{+0.016}_{-0.016}$   \\ \hline
    $\sigma_8$   & $0.7990_{-0.083}^{+0.073}$
    &$0.7987_{-0.07}^{+0.071}$ \\ \hline \hline
    \end{tabular}
    \caption{Mean values and 68\% CL minimum credible
    intervals for the parameters of the base $\Lambda$CDM model as extracted 
    from the BOSS FS and BOSS FS+BAO low-z NGC data from the mean of the low-z Patchy mocks, 
    presented in the same format as before. We show only the cosmological parameters whose 
    posterior contours lie well within the priors used in the main analysis, i.e. we do not include parameters whose distributions simply reflect the priors.}
    \label{table:mocks}
\end{table*}

The results of our analysis are shown in Tab.\,\ref{table:mocks} and in the corner plot of Fig.~\ref{fig:mocks}. In full agreement with both the forecast of Ref.\,\citep{2019arXiv191208208I}, we observe that the inclusion of information from reconstructed BAO reduces the error on $H_0$ by $\sim 40\%$, while the constraints on $\sigma_8$ and $\omega_{cdm}$ remain, essentially, intact. Importantly, we report no bias from the inclusion of BAO data by our method, and observe that the posterior contours are $1\sigma$ consistent with the input values in all cases (recalling that we do not expect perfect agreement due to the approximate nature of the mock catalogs). The sharper constraints on $H_0$ also source a somewhat improved measurement of $\Omega_m$, due to the combination of the shape and geometric information. From this test, it is clear that our FS+BAO likelihood is suitable to apply to observational data.

\section{Cosmological constraints from individual BOSS data chunks}\label{appen: individual-chunks}
In Tab.\,\ref{table:allc} we display the constraints on cosmology obtained from analysis of the four individual BOSS data chunks in the FS and FS+BAO analyses, both allowing $n_s$ to be free and constraining it to lie within the Planck prior. Note that the BAO analysis is done separately for each chunk in all cases to allow a joint covariance to be constructed.

\begin{figure}
    \centering
    \includegraphics[width=0.49\textwidth]{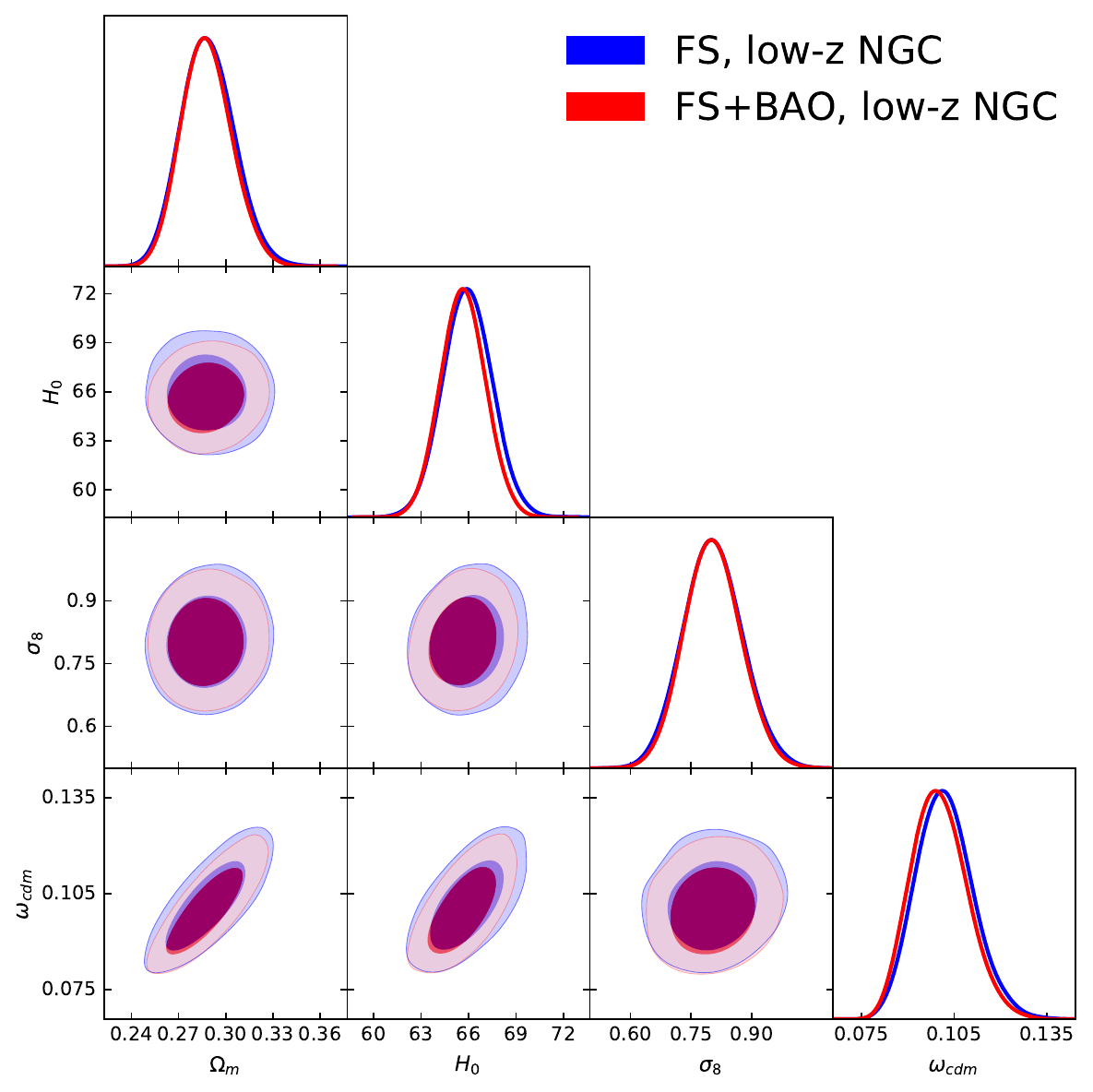}
    \includegraphics[width=0.49\textwidth]{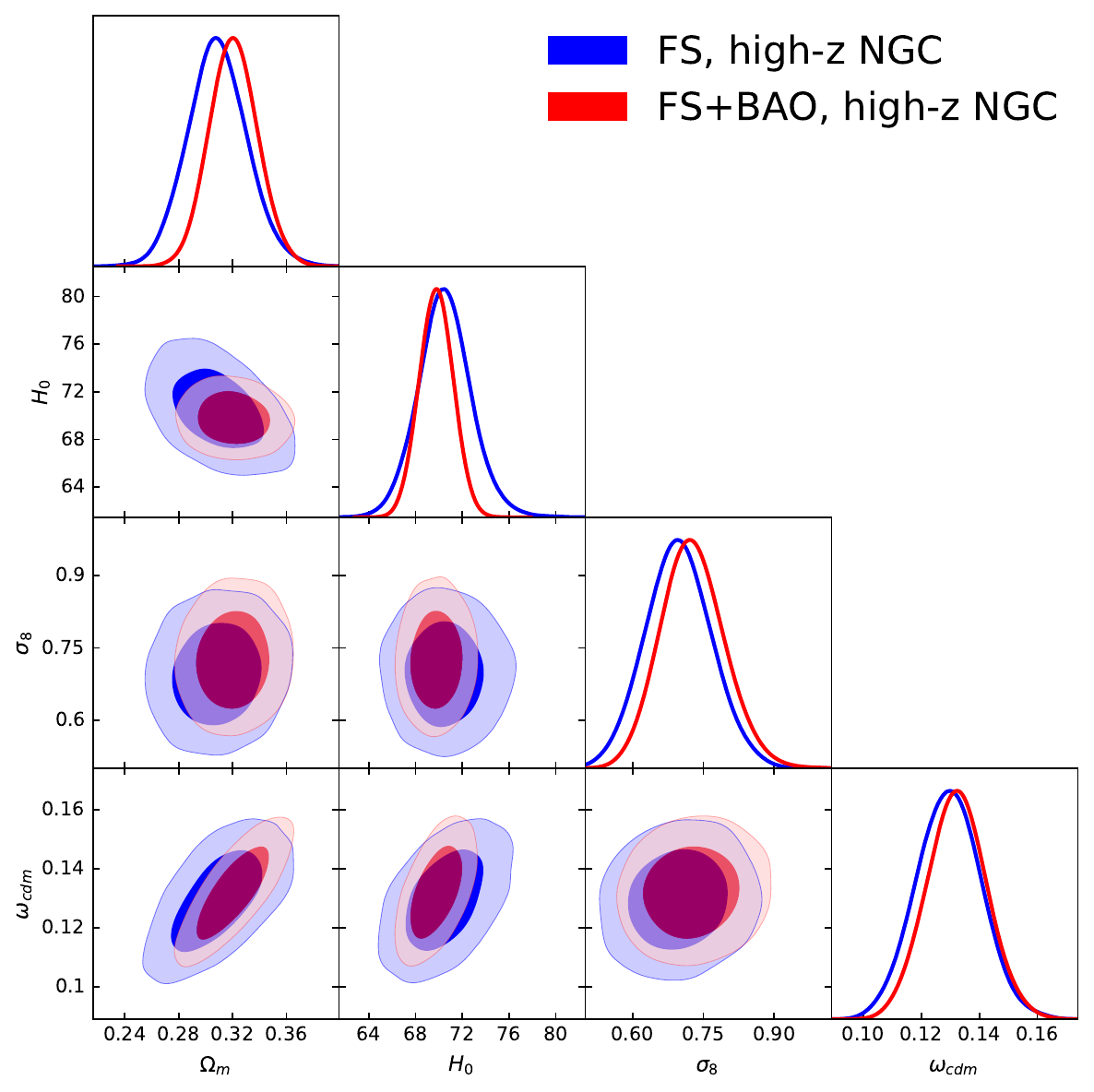}
    \caption{Parameter constraints obtained from analysis of the low-z (left panel) and high-z (right panel) NGC BOSS data using the $\nu\Lambda$CDM model assuming a fixed spectral tilt $n_s$. As discussed in the text, we observe significant improvements in $H_0$ from the inclusion of BAO data only for the high-z sample.}
    \label{fig:ngcc}
\end{figure}

As a comparison, Fig.\,\ref{fig:ngcc} shows the corner plots for the FS-only and FS+BAO analyses of the low-z and high-z NGC data in the $\nu\Lambda$CDM model with fixed $n_s$, which may be compared to the previous results on the Patchy mocks (Fig.\,\ref{fig:mocks}). Somewhat curiously, for the low-z NGC chunk, the constraints do not improve at all with the addition of BAO information from reconstructed spectra. In contrast, both the analysis of the Patchy mocks (Appendix \ref{appen: patchy-tests}) and the forecast of Ref.\,\citep{2019arXiv191208208I} found that a BAO reconstruction with 50\% efficacy would be expected to improve the constrain on $H_0$ by $\sim 40\%$. This anomaly is not observed in the high-z NGC chunk however (nor the other chunks), which yields the expected improvement, as seen in the right panel of Fig.\,\ref{fig:ngcc}.

The absence of improvement in the low-z NGC case can be explained by the known anomaly in this data chunk, as previously reported by the BOSS collaboration \citep{2017MNRAS.464.3409B}. Seemingly, the low-z NGC sample exhibits a realization of the dark matter displacement field whose amplitude is significantly reduced. This results in BAO wiggles with much less suppression than expected from theory and observed in the Patchy mocks.
Because of this reason, the constraints on $H_0$ extracted from the pre-reconstructed spectra (i.e. FS-only analyses) are observed to be $\sim 40\%$ better than expected from the analysis of Patchy mocks, cf.\,Tab.\,\ref{table:mocks} and the low-z NGC section of Tab.~\ref{table:allc}. Whilst this anomaly reduces the improvements from including BAO data it does not therefore alter the overall constraints in $H_0$.

\begin{table*}[t!h]
\centering
  \begin{tabular}{|c||c|c||c|c|} \hline 
  & \multicolumn{2}{|c|}{base $\nu\Lambda$CDM}
 & \multicolumn{2}{|c|}{base $\nu\Lambda$CDM~+ fixed $n_s$}\\  \hline
     \hline
      Parameter & FS & FS+BAO & FS & FS+BAO
      \\ [0.2cm]
 \hline \hline 
     \multicolumn{5}{|c|}{high-z NGC ($z_{\rm eff}=0.61$)}
     \\ \hline \hline
  $\omega_{cdm}$  & $0.1373_{-0.022}^{+0.016}$
  & $0.1384_{-0.02}^{+0.015}$
  & $0.1283_{-0.013}^{+0.01}$  
  & $0.1322_{-0.01}^{+0.01}$ 
   \\ \hline
$n_s$  
&  $0.9115_{-0.11}^{+0.13}$
& $0.9254_{-0.1}^{+0.11}$ 
& $-$  
& $-$   \\ 
\hline
$H_0$   & $70.79_{-2.6}^{+2.3}$
& $70.28_{-1.7}^{+1.6}$ 
& $70.17_{-2.4}^{+2.2}$
& $69.79_{-1.5}^{+1.5}$
\\ \hline
$A^{1/2}$   & $0.8107_{-0.12}^{+0.093}$
&$0.8326_{-0.12}^{+0.093}$
& $0.8346_{-0.1}^{+0.089}$
& $0.8513_{-0.095}^{+0.08}$ \\
\hline\hline
$\Omega_m$   & $0.322_{-0.038}^{+0.028}$ 
& $0.3289^{+0.025}_{-0.033}$ 
& $0.3102_{-0.021}^{+0.023}$ 
& $0.3207_{-0.018}^{+0.018}$   \\ \hline
$\sigma_8$   & $0.6879_{-0.077}^{+0.068}$
&$0.7123_{-0.08}^{+0.069}$ 
& $0.6992_{-0.072}^{+0.066}$
& $0.7274_{-0.071}^{+0.063}$ \\ \hline \hline
      \multicolumn{5}{|c|}{high-z SGC ($z_{\rm eff}=0.61$)}
     \\ \hline \hline
  $\omega_{cdm}$  & $0.1099_{-0.027}^{+0.014}$
  & $0.111_{-0.024}^{+0.014}$ 
  & $0.1119_{-0.013}^{+0.01}$  
  & $0.1151_{-0.013}^{+0.011}$
   \\ \hline
$n_s$  
&  $1.005_{-0.15}^{+0.18}$
& $1.007_{-0.13}^{+0.14}$ 
& $-$  
& $-$   \\ 
\hline
$H_0$   & $66.98_{-5.0}^{+2.7}$
& $65.46_{-2.6}^{+2.1}$
& $66.44_{-4.7}^{+2.4}$ 
& $65.45_{-2.4}^{+2}$
\\ \hline
$A^{1/2}$   & $0.9144_{-0.21}^{+0.18}$
&$0.9341_{-0.19}^{+0.16}$ 
& $0.8564_{-0.15}^{+0.14}$
& $0.8943_{-0.14}^{+0.13}$ \\
\hline\hline
$\Omega_m$   & $0.301_{-0.06}^{+0.056}$ 
& $0.3137^{+0.037}_{-0.044}$ 
& $0.3049_{-0.034}^{+0.043}$ 
& $0.3248_{-0.027}^{+0.027}$   \\ \hline
$\sigma_8$   & $0.6805_{-0.13}^{+0.11}$
&$0.700_{-0.11}^{+0.10}$ 
& $0.6463_{-0.11}^{+0.11}$
& $0.6830_{-0.10}^{+0.09}$ \\ \hline \hline
      \multicolumn{5}{|c|}{low-z NGC ($z_{\rm eff}=0.38$)}
     \\ \hline \hline
  $\omega_{cdm}$  & $0.1161_{-0.019}^{+0.012}$
  & $0.1146_{-0.018}^{+0.012}$ 
  & $0.103_{-0.011}^{+0.0089}$
  & $0.1006_{-0.01}^{+0.0084}$
   \\ \hline
$n_s$  
&  $0.751_{-0.13}^{+0.15}$
& $0.7463_{-0.15}^{+0.16}$
& $-$  
& $-$   \\ 
\hline
$H_0$   & $67.0_{-2.1}^{+1.8}$
& $66.75_{-1.9}^{+1.6}$
& $66.05_{-1.6}^{+1.6}$
& $65.7_{-1.5}^{+1.4}$
\\ \hline
$A^{1/2}$   & $1.017_{-0.15}^{+0.12}$
&$1.024_{-0.15}^{+0.11}$ 
& $1.131_{-0.12}^{+0.11}$
& $1.146_{-0.12}^{+0.11}$ \\
\hline\hline
$\Omega_m$   & $0.3115_{-0.031}^{+0.021}$ 
& $0.3115^{+0.021}_{-0.025}$ 
& $0.2905_{-0.018}^{+0.016}$ 
& $0.2879_{-0.017}^{+0.015}$   \\ \hline
$\sigma_8$   & $0.7256_{-0.09}^{+0.071}$
&$0.7212_{-0.088}^{+0.063}$ 
& $0.8075_{-0.073}^{+0.072}$
& $0.8036_{-0.069}^{+0.07}$ \\ \hline \hline
      \multicolumn{5}{|c|}{low-z SGC ($z_{\rm eff}=0.38$)}
     \\ \hline \hline
  $\omega_{cdm}$  & $0.1276_{-0.034}^{+0.02}$
  & $0.1198_{-0.031}^{+0.016}$ 
  & $0.1026_{-0.013}^{+0.01}$
  & $0.1012_{-0.012}^{+0.0097}$
   \\ \hline
$n_s$  
& $0.787_{-0.18}^{+0.17}$
& $0.8123_{-0.16}^{+0.18}$
& $-$  
& $-$   \\ 
\hline
$H_0$   & $70.73_{-5}^{+3.6}$
& $70.09_{-3}^{+2.5}$
& $69.45_{-4.6}^{+3.2}$
& $68.68_{-2.1}^{+2.2}$
\\ \hline
$A^{1/2}$   & $0.827_{-0.22}^{+0.15}$
&$0.8781_{-0.21}^{+0.16}$ 
& $0.9293_{-0.18}^{+0.17}$
& $0.9786_{-0.18}^{+0.16}$ \\
\hline\hline
$\Omega_m$   & $0.3044_{-0.056}^{+0.035}$ 
& $0.2909^{+0.03}_{-0.042}$ 
& $0.2617_{-0.026}^{+0.031}$ 
& $0.2648_{-0.022}^{+0.018}$   \\ \hline
$\sigma_8$   & $0.6241_{-0.13}^{+0.1}$
&$0.6502_{-0.13}^{+0.1}$ 
& $0.6589_{-0.13}^{+0.1}$
& $0.6588_{-0.13}^{+0.1}$ \\ \hline \hline
\end{tabular}
\caption{Mean values and 68\% CL minimum credible
intervals for the parameters of the base $\nu\Lambda$CDM (left two columns) and the base $\nu\Lambda$CDM with the fixed power spectrum tilt $n_s$ (right two columns) models as extracted from the BOSS FS and BOSS FS+BAO data, presented in the same format as before. Constraints are shown only for those parameters whose contours are significantly narrower than the priors used.}
\label{table:allc}
\end{table*}

\bibliographystyle{JHEP}
\bibliography{adslib,otherlib}

\providecommand{\href}[2]{#2}\begingroup\raggedright\begin{thebibliography}{100}

\bibitem{2018arXiv180706209P}
{Planck Collaboration}, N.~{Aghanim}, Y.~{Akrami}, M.~{Ashdown}, J.~{Aumont},
  C.~{Baccigalupi} et~al., \emph{{Planck 2018 results. VI. Cosmological
  parameters}}, {\emph{arXiv e-prints} (2018) arXiv:1807.06209}
  [\href{https://arxiv.org/abs/1807.06209}{{\ttfamily 1807.06209}}].

\bibitem{2019ApJ...876...85R}
A.~G. {Riess}, S.~{Casertano}, W.~{Yuan}, L.~M. {Macri} and D.~{Scolnic},
  \emph{{Large Magellanic Cloud Cepheid Standards Provide a 1\% Foundation for
  the Determination of the Hubble Constant and Stronger Evidence for Physics
  beyond {\ensuremath{\Lambda}}CDM}},
  \href{https://doi.org/10.3847/1538-4357/ab1422}{\emph{\apj} {\bfseries 876}
  (2019) 85} [\href{https://arxiv.org/abs/1903.07603}{{\ttfamily 1903.07603}}].

\bibitem{2019NatRP...2...10R}
A.~G. {Riess}, \emph{{The expansion of the Universe is faster than expected}},
  \href{https://doi.org/10.1038/s42254-019-0137-0}{\emph{Nature Reviews
  Physics} {\bfseries 2} (2019) 10}
  [\href{https://arxiv.org/abs/2001.03624}{{\ttfamily 2001.03624}}].

\bibitem{2019arXiv190803663K}
L.~{Knox} and M.~{Millea}, \emph{{The Hubble Hunter's Guide}}, {\emph{arXiv
  e-prints} (2019) arXiv:1908.03663}
  [\href{https://arxiv.org/abs/1908.03663}{{\ttfamily 1908.03663}}].

\bibitem{2009arXiv0912.0201L}
{LSST Science Collaboration}, P.~A. {Abell}, J.~{Allison}, S.~F. {Anderson},
  J.~R. {Andrew}, J.~R.~P. {Angel} et~al., \emph{{LSST Science Book, Version
  2.0}}, {\emph{arXiv e-prints} (2009) arXiv:0912.0201}
  [\href{https://arxiv.org/abs/0912.0201}{{\ttfamily 0912.0201}}].

\bibitem{2014arXiv1412.4872D}
O.~{Dor{\'e}}, J.~{Bock}, M.~{Ashby}, P.~{Capak}, A.~{Cooray}, R.~{de Putter}
  et~al., \emph{{Cosmology with the SPHEREX All-Sky Spectral Survey}},
  {\emph{arXiv e-prints} (2014) arXiv:1412.4872}
  [\href{https://arxiv.org/abs/1412.4872}{{\ttfamily 1412.4872}}].

\bibitem{2011arXiv1110.3193L}
R.~{Laureijs}, J.~{Amiaux}, S.~{Arduini}, J.~L. {Augu{\`e}res},
  J.~{Brinchmann}, R.~{Cole} et~al., \emph{{Euclid Definition Study Report}},
  {\emph{arXiv e-prints} (2011) arXiv:1110.3193}
  [\href{https://arxiv.org/abs/1110.3193}{{\ttfamily 1110.3193}}].

\bibitem{2011AJ....142...72E}
D.~J. {Eisenstein}, D.~H. {Weinberg}, E.~{Agol}, H.~{Aihara}, C.~{Allende
  Prieto}, S.~F. {Anderson} et~al., \emph{{SDSS-III: Massive Spectroscopic
  Surveys of the Distant Universe, the Milky Way, and Extra-Solar Planetary
  Systems}}, \href{https://doi.org/10.1088/0004-6256/142/3/72}{\emph{\aj}
  {\bfseries 142} (2011) 72} [\href{https://arxiv.org/abs/1101.1529}{{\ttfamily
  1101.1529}}].

\bibitem{2017MNRAS.466.2242B}
F.~{Beutler}, H.-J. {Seo}, S.~{Saito}, C.-H. {Chuang}, A.~J. {Cuesta}, D.~J.
  {Eisenstein} et~al., \emph{{The clustering of galaxies in the completed
  SDSS-III Baryon Oscillation Spectroscopic Survey: anisotropic galaxy
  clustering in Fourier space}},
  \href{https://doi.org/10.1093/mnras/stw3298}{\emph{\mnras} {\bfseries 466}
  (2017) 2242} [\href{https://arxiv.org/abs/1607.03150}{{\ttfamily
  1607.03150}}].

\bibitem{2017MNRAS.464.3409B}
F.~{Beutler}, H.-J. {Seo}, A.~J. {Ross}, P.~{McDonald}, S.~{Saito}, A.~S.
  {Bolton} et~al., \emph{{The clustering of galaxies in the completed SDSS-III
  Baryon Oscillation Spectroscopic Survey: baryon acoustic oscillations in the
  Fourier space}}, \href{https://doi.org/10.1093/mnras/stw2373}{\emph{\mnras}
  {\bfseries 464} (2017) 3409}
  [\href{https://arxiv.org/abs/1607.03149}{{\ttfamily 1607.03149}}].

\bibitem{2017MNRAS.469.1369S}
S.~{Satpathy}, S.~{Alam}, S.~{Ho}, M.~{White}, N.~A. {Bahcall}, F.~{Beutler}
  et~al., \emph{{The clustering of galaxies in the completed SDSS-III Baryon
  Oscillation Spectroscopic Survey: on the measurement of growth rate using
  galaxy correlation functions}},
  \href{https://doi.org/10.1093/mnras/stx883}{\emph{\mnras} {\bfseries 469}
  (2017) 1369} [\href{https://arxiv.org/abs/1607.03148}{{\ttfamily
  1607.03148}}].

\bibitem{2017MNRAS.464.1640S}
A.~G. {S{\'a}nchez}, R.~{Scoccimarro}, M.~{Crocce}, J.~N. {Grieb},
  S.~{Salazar-Albornoz}, C.~{Dalla Vecchia} et~al., \emph{{The clustering of
  galaxies in the completed SDSS-III Baryon Oscillation Spectroscopic Survey:
  Cosmological implications of the configuration-space clustering wedges}},
  \href{https://doi.org/10.1093/mnras/stw2443}{\emph{\mnras} {\bfseries 464}
  (2017) 1640} [\href{https://arxiv.org/abs/1607.03147}{{\ttfamily
  1607.03147}}].

\bibitem{2017MNRAS.464.1168R}
A.~J. {Ross}, F.~{Beutler}, C.-H. {Chuang}, M.~{Pellejero-Ibanez}, H.-J. {Seo},
  M.~{Vargas-Maga{\~n}a} et~al., \emph{{The clustering of galaxies in the
  completed SDSS-III Baryon Oscillation Spectroscopic Survey: observational
  systematics and baryon acoustic oscillations in the correlation function}},
  \href{https://doi.org/10.1093/mnras/stw2372}{\emph{\mnras} {\bfseries 464}
  (2017) 1168} [\href{https://arxiv.org/abs/1607.03145}{{\ttfamily
  1607.03145}}].

\bibitem{2017MNRAS.465.1757G}
H.~{Gil-Mar{\'\i}n}, W.~J. {Percival}, L.~{Verde}, J.~R. {Brownstein}, C.-H.
  {Chuang}, F.-S. {Kitaura} et~al., \emph{{The clustering of galaxies in the
  SDSS-III Baryon Oscillation Spectroscopic Survey: RSD measurement from the
  power spectrum and bispectrum of the DR12 BOSS galaxies}},
  \href{https://doi.org/10.1093/mnras/stw2679}{\emph{\mnras} {\bfseries 465}
  (2017) 1757} [\href{https://arxiv.org/abs/1606.00439}{{\ttfamily
  1606.00439}}].

\bibitem{2018MNRAS.478.4500P}
D.~W. {Pearson} and L.~{Samushia}, \emph{{A Detection of the Baryon Acoustic
  Oscillation features in the SDSS BOSS DR12 Galaxy Bispectrum}},
  \href{https://doi.org/10.1093/mnras/sty1266}{\emph{\mnras} {\bfseries 478}
  (2018) 4500} [\href{https://arxiv.org/abs/1712.04970}{{\ttfamily
  1712.04970}}].

\bibitem{2019JCAP...11..034C}
A.~{Chudaykin} and M.~M. {Ivanov}, \emph{{Measuring neutrino masses with
  large-scale structure: Euclid forecast with controlled theoretical error}},
  \href{https://doi.org/10.1088/1475-7516/2019/11/034}{\emph{\jcap} {\bfseries
  2019} (2019) 034} [\href{https://arxiv.org/abs/1907.06666}{{\ttfamily
  1907.06666}}].

\bibitem{1979Natur.281..358A}
C.~{Alcock} and B.~{Paczynski}, \emph{{An evolution free test for non-zero
  cosmological constant}}, \href{https://doi.org/10.1038/281358a0}{\emph{\nat}
  {\bfseries 281} (1979) 358}.

\bibitem{1987MNRAS.227....1K}
N.~{Kaiser}, \emph{{Clustering in real space and in redshift space}},
  \href{https://doi.org/10.1093/mnras/227.1.1}{\emph{\mnras} {\bfseries 227}
  (1987) 1}.

\bibitem{2008Natur.451..541G}
L.~{Guzzo}, M.~{Pierleoni}, B.~{Meneux}, E.~{Branchini}, O.~{Le F{\`e}vre},
  C.~{Marinoni} et~al., \emph{{A test of the nature of cosmic acceleration
  using galaxy redshift distortions}},
  \href{https://doi.org/10.1038/nature06555}{\emph{\nat} {\bfseries 451} (2008)
  541} [\href{https://arxiv.org/abs/0802.1944}{{\ttfamily 0802.1944}}].

\bibitem{1970ApJ...162..815P}
P.~J.~E. {Peebles} and J.~T. {Yu}, \emph{{Primeval Adiabatic Perturbation in an
  Expanding Universe}}, \href{https://doi.org/10.1086/150713}{\emph{\apj}
  {\bfseries 162} (1970) 815}.

\bibitem{1970Ap&SS...7....3S}
R.~A. {Sunyaev} and Y.~B. {Zeldovich}, \emph{{Small-Scale Fluctuations of Relic
  Radiation}}, \href{https://doi.org/10.1007/BF00653471}{\emph{\apss}
  {\bfseries 7} (1970) 3}.

\bibitem{1996ApJ...471...30H}
W.~{Hu} and M.~{White}, \emph{{Acoustic Signatures in the Cosmic Microwave
  Background}}, \href{https://doi.org/10.1086/177951}{\emph{\apj} {\bfseries
  471} (1996) 30} [\href{https://arxiv.org/abs/astro-ph/9602019}{{\ttfamily
  astro-ph/9602019}}].

\bibitem{2003ApJ...598..720S}
H.-J. {Seo} and D.~J. {Eisenstein}, \emph{{Probing Dark Energy with Baryonic
  Acoustic Oscillations from Future Large Galaxy Redshift Surveys}},
  \href{https://doi.org/10.1086/379122}{\emph{\apj} {\bfseries 598} (2003) 720}
  [\href{https://arxiv.org/abs/astro-ph/0307460}{{\ttfamily
  astro-ph/0307460}}].

\bibitem{2005ApJ...633..560E}
D.~J. {Eisenstein}, I.~{Zehavi}, D.~W. {Hogg}, R.~{Scoccimarro}, M.~R.
  {Blanton}, R.~C. {Nichol} et~al., \emph{{Detection of the Baryon Acoustic
  Peak in the Large-Scale Correlation Function of SDSS Luminous Red Galaxies}},
  \href{https://doi.org/10.1086/466512}{\emph{\apj} {\bfseries 633} (2005) 560}
  [\href{https://arxiv.org/abs/astro-ph/0501171}{{\ttfamily
  astro-ph/0501171}}].

\bibitem{1999MNRAS.304..851M}
A.~{Meiksin}, M.~{White} and J.~A. {Peacock}, \emph{{Baryonic signatures in
  large-scale structure}},
  \href{https://doi.org/10.1046/j.1365-8711.1999.02369.x}{\emph{\mnras}
  {\bfseries 304} (1999) 851}
  [\href{https://arxiv.org/abs/astro-ph/9812214}{{\ttfamily
  astro-ph/9812214}}].

\bibitem{2005ApJ...633..575S}
H.-J. {Seo} and D.~J. {Eisenstein}, \emph{{Baryonic Acoustic Oscillations in
  Simulated Galaxy Redshift Surveys}},
  \href{https://doi.org/10.1086/491599}{\emph{\apj} {\bfseries 633} (2005) 575}
  [\href{https://arxiv.org/abs/astro-ph/0507338}{{\ttfamily
  astro-ph/0507338}}].

\bibitem{2010ApJ...720.1650S}
H.-J. {Seo}, J.~{Eckel}, D.~J. {Eisenstein}, K.~{Mehta}, M.~{Metchnik},
  N.~{Padmanabhan} et~al., \emph{{High-precision Predictions for the Acoustic
  Scale in the Nonlinear Regime}},
  \href{https://doi.org/10.1088/0004-637X/720/2/1650}{\emph{\apj} {\bfseries
  720} (2010) 1650} [\href{https://arxiv.org/abs/0910.5005}{{\ttfamily
  0910.5005}}].

\bibitem{2011ApJ...734...94M}
K.~T. {Mehta}, H.-J. {Seo}, J.~{Eckel}, D.~J. {Eisenstein}, M.~{Metchnik},
  P.~{Pinto} et~al., \emph{{Galaxy Bias and Its Effects on the Baryon Acoustic
  Oscillation Measurements}},
  \href{https://doi.org/10.1088/0004-637X/734/2/94}{\emph{\apj} {\bfseries 734}
  (2011) 94} [\href{https://arxiv.org/abs/1104.1178}{{\ttfamily 1104.1178}}].

\bibitem{2007ApJ...664..675E}
D.~J. {Eisenstein}, H.-J. {Seo}, E.~{Sirko} and D.~N. {Spergel},
  \emph{{Improving Cosmological Distance Measurements by Reconstruction of the
  Baryon Acoustic Peak}}, \href{https://doi.org/10.1086/518712}{\emph{\apj}
  {\bfseries 664} (2007) 675}
  [\href{https://arxiv.org/abs/astro-ph/0604362}{{\ttfamily
  astro-ph/0604362}}].

\bibitem{2012MNRAS.427.2132P}
N.~{Padmanabhan}, X.~{Xu}, D.~J. {Eisenstein}, R.~{Scalzo}, A.~J. {Cuesta},
  K.~T. {Mehta} et~al., \emph{{A 2 per cent distance to z = 0.35 by
  reconstructing baryon acoustic oscillations - I. Methods and application to
  the Sloan Digital Sky Survey}},
  \href{https://doi.org/10.1111/j.1365-2966.2012.21888.x}{\emph{\mnras}
  {\bfseries 427} (2012) 2132}
  [\href{https://arxiv.org/abs/1202.0090}{{\ttfamily 1202.0090}}].

\bibitem{2014JCAP...02..042S}
N.~S. {Sugiyama} and D.~N. {Spergel}, \emph{{How does non-linear dynamics
  affect the baryon acoustic oscillation?}},
  \href{https://doi.org/10.1088/1475-7516/2014/02/042}{\emph{\jcap} {\bfseries
  2014} (2014) 042} [\href{https://arxiv.org/abs/1306.6660}{{\ttfamily
  1306.6660}}].

\bibitem{2017PhRvD..96b3505S}
M.~{Schmittfull}, T.~{Baldauf} and M.~{Zaldarriaga}, \emph{{Iterative initial
  condition reconstruction}},
  \href{https://doi.org/10.1103/PhysRevD.96.023505}{\emph{\prd} {\bfseries 96}
  (2017) 023505} [\href{https://arxiv.org/abs/1704.06634}{{\ttfamily
  1704.06634}}].

\bibitem{2018MNRAS.478.1866H}
R.~{Hada} and D.~J. {Eisenstein}, \emph{{An iterative reconstruction of
  cosmological initial density fields}},
  \href{https://doi.org/10.1093/mnras/sty1203}{\emph{\mnras} {\bfseries 478}
  (2018) 1866} [\href{https://arxiv.org/abs/1804.04738}{{\ttfamily
  1804.04738}}].

\bibitem{2012MNRAS.427.3435A}
L.~{Anderson}, E.~{Aubourg}, S.~{Bailey}, D.~{Bizyaev}, M.~{Blanton}, A.~S.
  {Bolton} et~al., \emph{{The clustering of galaxies in the SDSS-III Baryon
  Oscillation Spectroscopic Survey: baryon acoustic oscillations in the Data
  Release 9 spectroscopic galaxy sample}},
  \href{https://doi.org/10.1111/j.1365-2966.2012.22066.x}{\emph{\mnras}
  {\bfseries 427} (2012) 3435}
  [\href{https://arxiv.org/abs/1203.6594}{{\ttfamily 1203.6594}}].

\bibitem{2014MNRAS.441...24A}
L.~{Anderson}, {\'E}.~{Aubourg}, S.~{Bailey}, F.~{Beutler}, V.~{Bhardwaj},
  M.~{Blanton} et~al., \emph{{The clustering of galaxies in the SDSS-III Baryon
  Oscillation Spectroscopic Survey: baryon acoustic oscillations in the Data
  Releases 10 and 11 Galaxy samples}},
  \href{https://doi.org/10.1093/mnras/stu523}{\emph{\mnras} {\bfseries 441}
  (2014) 24} [\href{https://arxiv.org/abs/1312.4877}{{\ttfamily 1312.4877}}].

\bibitem{2012JHEP...09..082C}
J.~J.~M. {Carrasco}, M.~P. {Hertzberg} and L.~{Senatore}, \emph{{The effective
  field theory of cosmological large scale structures}},
  \href{https://doi.org/10.1007/JHEP09(2012)082}{\emph{Journal of High Energy
  Physics} {\bfseries 2012} (2012) 82}
  [\href{https://arxiv.org/abs/1206.2926}{{\ttfamily 1206.2926}}].

\bibitem{2012JCAP...07..051B}
D.~{Baumann}, A.~{Nicolis}, L.~{Senatore} and M.~{Zaldarriaga},
  \emph{{Cosmological non-linearities as an effective fluid}},
  \href{https://doi.org/10.1088/1475-7516/2012/07/051}{\emph{\jcap} {\bfseries
  2012} (2012) 051} [\href{https://arxiv.org/abs/1004.2488}{{\ttfamily
  1004.2488}}].

\bibitem{2015JCAP...11..007S}
L.~{Senatore}, \emph{{Bias in the effective field theory of large scale
  structures}},
  \href{https://doi.org/10.1088/1475-7516/2015/11/007}{\emph{\jcap} {\bfseries
  2015} (2015) 007} [\href{https://arxiv.org/abs/1406.7843}{{\ttfamily
  1406.7843}}].

\bibitem{2009JCAP...08..020M}
P.~{McDonald} and A.~{Roy}, \emph{{Clustering of dark matter tracers:
  generalizing bias for the coming era of precision LSS}},
  \href{https://doi.org/10.1088/1475-7516/2009/08/020}{\emph{\jcap} {\bfseries
  2009} (2009) 020} [\href{https://arxiv.org/abs/0902.0991}{{\ttfamily
  0902.0991}}].

\bibitem{2014JCAP...08..056A}
V.~{Assassi}, D.~{Baumann}, D.~{Green} and M.~{Zaldarriaga},
  \emph{{Renormalized halo bias}},
  \href{https://doi.org/10.1088/1475-7516/2014/08/056}{\emph{\jcap} {\bfseries
  2014} (2014) 056} [\href{https://arxiv.org/abs/1402.5916}{{\ttfamily
  1402.5916}}].

\bibitem{2015JCAP...05..019L}
M.~{Lewandowski}, A.~{Perko} and L.~{Senatore}, \emph{{Analytic prediction of
  baryonic effects from the EFT of large scale structures}},
  \href{https://doi.org/10.1088/1475-7516/2015/05/019}{\emph{\jcap} {\bfseries
  2015} (2015) 019} [\href{https://arxiv.org/abs/1412.5049}{{\ttfamily
  1412.5049}}].

\bibitem{2014arXiv1409.1225S}
L.~{Senatore} and M.~{Zaldarriaga}, \emph{{Redshift Space Distortions in the
  Effective Field Theory of Large Scale Structures}}, {\emph{arXiv e-prints}
  (2014) arXiv:1409.1225} [\href{https://arxiv.org/abs/1409.1225}{{\ttfamily
  1409.1225}}].

\bibitem{2016arXiv161009321P}
A.~{Perko}, L.~{Senatore}, E.~{Jennings} and R.~H. {Wechsler}, \emph{{Biased
  Tracers in Redshift Space in the EFT of Large-Scale Structure}}, {\emph{arXiv
  e-prints} (2016) arXiv:1610.09321}
  [\href{https://arxiv.org/abs/1610.09321}{{\ttfamily 1610.09321}}].

\bibitem{2014JCAP...07..057C}
J.~J.~M. {Carrasco}, S.~{Foreman}, D.~{Green} and L.~{Senatore}, \emph{{The
  Effective Field Theory of Large Scale Structures at two loops}},
  \href{https://doi.org/10.1088/1475-7516/2014/07/057}{\emph{\jcap} {\bfseries
  2014} (2014) 057} [\href{https://arxiv.org/abs/1310.0464}{{\ttfamily
  1310.0464}}].

\bibitem{2015PhRvD..92l3007B}
T.~{Baldauf}, L.~{Mercolli} and M.~{Zaldarriaga}, \emph{{Effective field theory
  of large scale structure at two loops: The apparent scale dependence of the
  speed of sound}},
  \href{https://doi.org/10.1103/PhysRevD.92.123007}{\emph{\prd} {\bfseries 92}
  (2015) 123007} [\href{https://arxiv.org/abs/1507.02256}{{\ttfamily
  1507.02256}}].

\bibitem{2015JCAP...02..013S}
L.~{Senatore} and M.~{Zaldarriaga}, \emph{{The IR-resummed Effective Field
  Theory of Large Scale Structures}},
  \href{https://doi.org/10.1088/1475-7516/2015/02/013}{\emph{\jcap} {\bfseries
  2015} (2015) 013} [\href{https://arxiv.org/abs/1404.5954}{{\ttfamily
  1404.5954}}].

\bibitem{2018JCAP...05..019S}
L.~{Senatore} and G.~{Trevisan}, \emph{{On the IR-resummation in the
  EFTofLSS}}, \href{https://doi.org/10.1088/1475-7516/2018/05/019}{\emph{\jcap}
  {\bfseries 2018} (2018) 019}
  [\href{https://arxiv.org/abs/1710.02178}{{\ttfamily 1710.02178}}].

\bibitem{2015PhRvD..92d3514B}
T.~{Baldauf}, M.~{Mirbabayi}, M.~{Simonovi{\'c}} and M.~{Zaldarriaga},
  \emph{{Equivalence principle and the baryon acoustic peak}},
  \href{https://doi.org/10.1103/PhysRevD.92.043514}{\emph{\prd} {\bfseries 92}
  (2015) 043514} [\href{https://arxiv.org/abs/1504.04366}{{\ttfamily
  1504.04366}}].

\bibitem{2016JCAP...07..028B}
D.~{Blas}, M.~{Garny}, M.~M. {Ivanov} and S.~{Sibiryakov}, \emph{{Time-sliced
  perturbation theory II: baryon acoustic oscillations and infrared
  resummation}},
  \href{https://doi.org/10.1088/1475-7516/2016/07/028}{\emph{\jcap} {\bfseries
  2016} (2016) 028} [\href{https://arxiv.org/abs/1605.02149}{{\ttfamily
  1605.02149}}].

\bibitem{2018JCAP...07..053I}
M.~M. {Ivanov} and S.~{Sibiryakov}, \emph{{Infrared resummation for biased
  tracers in redshift space}},
  \href{https://doi.org/10.1088/1475-7516/2018/07/053}{\emph{\jcap} {\bfseries
  2018} (2018) 053} [\href{https://arxiv.org/abs/1804.05080}{{\ttfamily
  1804.05080}}].

\bibitem{2016JCAP...03..057V}
Z.~{Vlah}, U.~{Seljak}, M.~{Yat Chu} and Y.~{Feng}, \emph{{Perturbation theory,
  effective field theory, and oscillations in the power spectrum}},
  \href{https://doi.org/10.1088/1475-7516/2016/03/057}{\emph{\jcap} {\bfseries
  2016} (2016) 057} [\href{https://arxiv.org/abs/1509.02120}{{\ttfamily
  1509.02120}}].

\bibitem{2019arXiv190905277I}
M.~M. {Ivanov}, M.~{Simonovi{\'c}} and M.~{Zaldarriaga}, \emph{{Cosmological
  Parameters from the BOSS Galaxy Power Spectrum}}, {\emph{arXiv e-prints}
  (2019) arXiv:1909.05277} [\href{https://arxiv.org/abs/1909.05277}{{\ttfamily
  1909.05277}}].

\bibitem{2019arXiv191208208I}
M.~M. {Ivanov}, M.~{Simonovi{\'c}} and M.~{Zaldarriaga}, \emph{{Cosmological
  Parameters and Neutrino Masses from the Final Planck and Full-Shape BOSS
  Data}}, {\emph{arXiv e-prints} (2019) arXiv:1912.08208}
  [\href{https://arxiv.org/abs/1912.08208}{{\ttfamily 1912.08208}}].

\bibitem{2019arXiv190905271D}
G.~{D'Amico}, J.~{Gleyzes}, N.~{Kokron}, D.~{Markovic}, L.~{Senatore},
  P.~{Zhang} et~al., \emph{{The Cosmological Analysis of the SDSS/BOSS data
  from the Effective Field Theory of Large-Scale Structure}}, {\emph{arXiv
  e-prints} (2019) arXiv:1909.05271}
  [\href{https://arxiv.org/abs/1909.05271}{{\ttfamily 1909.05271}}].

\bibitem{Colas:2019ret}
T.~Colas, G.~D'amico, L.~Senatore, P.~Zhang and F.~Beutler, \emph{{Efficient
  Cosmological Analysis of the SDSS/BOSS data from the Effective Field Theory
  of Large-Scale Structure}},
  \href{https://arxiv.org/abs/1909.07951}{{\ttfamily 1909.07951}}.

\bibitem{2019JCAP...02..027S}
B.~D. {Sherwin} and M.~{White}, \emph{{The impact of wrong assumptions in BAO
  reconstruction}},
  \href{https://doi.org/10.1088/1475-7516/2019/02/027}{\emph{\jcap} {\bfseries
  2019} (2019) 027} [\href{https://arxiv.org/abs/1808.04384}{{\ttfamily
  1808.04384}}].

\bibitem{2017PhRvD..96d3513H}
C.~{Hikage}, K.~{Koyama} and A.~{Heavens}, \emph{{Perturbation theory for BAO
  reconstructed fields: One-loop results in the real-space matter density
  field}}, \href{https://doi.org/10.1103/PhysRevD.96.043513}{\emph{\prd}
  {\bfseries 96} (2017) 043513}
  [\href{https://arxiv.org/abs/1703.07878}{{\ttfamily 1703.07878}}].

\bibitem{2020PhRvD.101d3510H}
C.~{Hikage}, K.~{Koyama} and R.~{Takahashi}, \emph{{Perturbation theory for the
  redshift-space matter power spectra after reconstruction}},
  \href{https://doi.org/10.1103/PhysRevD.101.043510}{\emph{\prd} {\bfseries
  101} (2020) 043510} [\href{https://arxiv.org/abs/1911.06461}{{\ttfamily
  1911.06461}}].

\bibitem{2019JCAP...09..017C}
S.-F. {Chen}, Z.~{Vlah} and M.~{White}, \emph{{The reconstructed power spectrum
  in the Zeldovich approximation}},
  \href{https://doi.org/10.1088/1475-7516/2019/09/017}{\emph{\jcap} {\bfseries
  2019} (2019) 017} [\href{https://arxiv.org/abs/1907.00043}{{\ttfamily
  1907.00043}}].

\bibitem{2017MNRAS.470.2617A}
S.~{Alam}, M.~{Ata}, S.~{Bailey}, F.~{Beutler}, D.~{Bizyaev}, J.~A. {Blazek}
  et~al., \emph{{The clustering of galaxies in the completed SDSS-III Baryon
  Oscillation Spectroscopic Survey: cosmological analysis of the DR12 galaxy
  sample}}, \href{https://doi.org/10.1093/mnras/stx721}{\emph{\mnras}
  {\bfseries 470} (2017) 2617}
  [\href{https://arxiv.org/abs/1607.03155}{{\ttfamily 1607.03155}}].

\bibitem{2018MNRAS.480.3879A}
T.~M.~C. {Abbott}, F.~B. {Abdalla}, J.~{Annis}, K.~{Bechtol}, J.~{Blazek},
  B.~A. {Benson} et~al., \emph{{Dark Energy Survey Year 1 Results: A Precise
  H$_{0}$ Estimate from DES Y1, BAO, and D/H Data}},
  \href{https://doi.org/10.1093/mnras/sty1939}{\emph{\mnras} {\bfseries 480}
  (2018) 3879} [\href{https://arxiv.org/abs/1711.00403}{{\ttfamily
  1711.00403}}].

\bibitem{2019JCAP...10..044C}
A.~{Cuceu}, J.~{Farr}, P.~{Lemos} and A.~{Font-Ribera}, \emph{{Baryon Acoustic
  Oscillations and the Hubble constant: past, present and future}},
  \href{https://doi.org/10.1088/1475-7516/2019/10/044}{\emph{\jcap} {\bfseries
  2019} (2019) 044} [\href{https://arxiv.org/abs/1906.11628}{{\ttfamily
  1906.11628}}].

\bibitem{2019MNRAS.484.3818S}
E.~{Sarpa}, C.~{Schimd}, E.~{Branchini} and S.~{Matarrese}, \emph{{BAO
  reconstruction: a swift numerical action method for massive spectroscopic
  surveys}}, \href{https://doi.org/10.1093/mnras/stz278}{\emph{\mnras}
  {\bfseries 484} (2019) 3818}
  [\href{https://arxiv.org/abs/1809.10738}{{\ttfamily 1809.10738}}].

\bibitem{2015PhRvD..92l3522S}
M.~{Schmittfull}, Y.~{Feng}, F.~{Beutler}, B.~{Sherwin} and M.~Y. {Chu},
  \emph{{Eulerian BAO reconstructions and N -point statistics}},
  \href{https://doi.org/10.1103/PhysRevD.92.123522}{\emph{\prd} {\bfseries 92}
  (2015) 123522} [\href{https://arxiv.org/abs/1508.06972}{{\ttfamily
  1508.06972}}].

\bibitem{2017MNRAS.467.2085G}
J.~N. {Grieb}, A.~G. {S{\'a}nchez}, S.~{Salazar-Albornoz}, R.~{Scoccimarro},
  M.~{Crocce}, C.~{Dalla Vecchia} et~al., \emph{{The clustering of galaxies in
  the completed SDSS-III Baryon Oscillation Spectroscopic Survey: Cosmological
  implications of the Fourier space wedges of the final sample}},
  \href{https://doi.org/10.1093/mnras/stw3384}{\emph{\mnras} {\bfseries 467}
  (2017) 2085} [\href{https://arxiv.org/abs/1607.03143}{{\ttfamily
  1607.03143}}].

\bibitem{2016MNRAS.460.2453S}
H.-J. {Seo}, F.~{Beutler}, A.~J. {Ross} and S.~{Saito}, \emph{{Modeling the
  reconstructed BAO in Fourier space}},
  \href{https://doi.org/10.1093/mnras/stw1138}{\emph{\mnras} {\bfseries 460}
  (2016) 2453} [\href{https://arxiv.org/abs/1511.00663}{{\ttfamily
  1511.00663}}].

\bibitem{2020MNRAS.tmp..347H}
S.~R. {Hinton}, C.~{Howlett} and T.~M. {Davis}, \emph{{Barry and the BAO Model
  Comparison}}, \href{https://doi.org/10.1093/mnras/staa361}{\emph{\mnras}
  (2020) } [\href{https://arxiv.org/abs/1912.01175}{{\ttfamily 1912.01175}}].

\bibitem{2016arXiv160200674B}
T.~{Baldauf}, M.~{Mirbabayi}, M.~{Simonovi{\'c}} and M.~{Zaldarriaga},
  \emph{{LSS constraints with controlled theoretical uncertainties}},
  {\emph{arXiv e-prints} (2016) arXiv:1602.00674}
  [\href{https://arxiv.org/abs/1602.00674}{{\ttfamily 1602.00674}}].

\bibitem{2011ascl.soft02026L}
A.~{Lewis} and A.~{Challinor}, \emph{{CAMB: Code for Anisotropies in the
  Microwave Background}},  Feb., 2011.

\bibitem{2011JCAP...07..034B}
D.~{Blas}, J.~{Lesgourgues} and T.~{Tram}, \emph{{The Cosmic Linear Anisotropy
  Solving System (CLASS). Part II: Approximation schemes}},
  \href{https://doi.org/10.1088/1475-7516/2011/07/034}{\emph{\jcap} {\bfseries
  2011} (2011) 034} [\href{https://arxiv.org/abs/1104.2933}{{\ttfamily
  1104.2933}}].

\bibitem{2015MNRAS.450.3822W}
M.~{White}, \emph{{Reconstruction within the Zeldovich approximation}},
  \href{https://doi.org/10.1093/mnras/stv842}{\emph{\mnras} {\bfseries 450}
  (2015) 3822} [\href{https://arxiv.org/abs/1504.03677}{{\ttfamily
  1504.03677}}].

\bibitem{1998ApJ...496..605E}
D.~J. {Eisenstein} and W.~{Hu}, \emph{{Baryonic Features in the Matter Transfer
  Function}}, \href{https://doi.org/10.1086/305424}{\emph{\apj} {\bfseries 496}
  (1998) 605} [\href{https://arxiv.org/abs/astro-ph/9709112}{{\ttfamily
  astro-ph/9709112}}].

\bibitem{2017MNRAS.464.3121W}
M.~J. {Wilson}, J.~A. {Peacock}, A.~N. {Taylor} and S.~{de la Torre},
  \emph{{Rapid modelling of the redshift-space power spectrum multipoles for a
  masked density field}},
  \href{https://doi.org/10.1093/mnras/stw2576}{\emph{\mnras} {\bfseries 464}
  (2017) 3121} [\href{https://arxiv.org/abs/1511.07799}{{\ttfamily
  1511.07799}}].

\bibitem{2000MNRAS.312..257H}
A.~J.~S. {Hamilton}, \emph{{Uncorrelated modes of the non-linear power
  spectrum}},
  \href{https://doi.org/10.1046/j.1365-8711.2000.03071.x}{\emph{\mnras}
  {\bfseries 312} (2000) 257}
  [\href{https://arxiv.org/abs/astro-ph/9905191}{{\ttfamily
  astro-ph/9905191}}].

\bibitem{2013JCAP...02..001A}
B.~{Audren}, J.~{Lesgourgues}, K.~{Benabed} and S.~{Prunet},
  \emph{{Conservative constraints on early cosmology with MONTE PYTHON}},
  \href{https://doi.org/10.1088/1475-7516/2013/02/001}{\emph{\jcap} {\bfseries
  2013} (2013) 001} [\href{https://arxiv.org/abs/1210.7183}{{\ttfamily
  1210.7183}}].

\bibitem{2018arXiv180407261B}
T.~{Brinckmann} and J.~{Lesgourgues}, \emph{{MontePython 3: boosted MCMC
  sampler and other features}}, {\emph{arXiv e-prints} (2018) arXiv:1804.07261}
  [\href{https://arxiv.org/abs/1804.07261}{{\ttfamily 1804.07261}}].

\bibitem{doi:10.1080/10618600.1998.10474787}
S.~P. Brooks and A.~Gelman, \emph{General methods for monitoring convergence of
  iterative simulations},
  \href{https://doi.org/10.1080/10618600.1998.10474787}{\emph{Journal of
  Computational and Graphical Statistics} {\bfseries 7} (1998) 434}.

\bibitem{gelman1992}
A.~Gelman and D.~B. Rubin, \emph{Inference from iterative simulation using
  multiple sequences},
  \href{https://doi.org/10.1214/ss/1177011136}{\emph{Statist. Sci.} {\bfseries
  7} (1992) 457}.

\bibitem{2018arXiv180512394F}
L.~{Fonseca de la Bella}, D.~{Regan}, D.~{Seery} and D.~{Parkinson},
  \emph{{Impact of bias and redshift-space modelling for the halo power
  spectrum: Testing the effective field theory of large-scale structure}},
  {\emph{arXiv e-prints} (2018) arXiv:1805.12394}
  [\href{https://arxiv.org/abs/1805.12394}{{\ttfamily 1805.12394}}].

\bibitem{2018JCAP...04..030S}
M.~{Simonovi{\'c}}, T.~{Baldauf}, M.~{Zaldarriaga}, J.~J. {Carrasco} and J.~A.
  {Kollmeier}, \emph{{Cosmological perturbation theory using the FFTLog:
  formalism and connection to QFT loop integrals}},
  \href{https://doi.org/10.1088/1475-7516/2018/04/030}{\emph{\jcap} {\bfseries
  2018} (2018) 030} [\href{https://arxiv.org/abs/1708.08130}{{\ttfamily
  1708.08130}}].

\bibitem{2017MNRAS.464.1493S}
A.~G. {S{\'a}nchez}, J.~N. {Grieb}, S.~{Salazar-Albornoz}, S.~{Alam},
  F.~{Beutler}, A.~J. {Ross} et~al., \emph{{The clustering of galaxies in the
  completed SDSS-III Baryon Oscillation Spectroscopic Survey: combining
  correlated Gaussian posterior distributions}},
  \href{https://doi.org/10.1093/mnras/stw2495}{\emph{\mnras} {\bfseries 464}
  (2017) 1493} [\href{https://arxiv.org/abs/1607.03146}{{\ttfamily
  1607.03146}}].

\bibitem{2016MNRAS.457..993P}
D.~W. {Pearson} and L.~{Samushia}, \emph{{Estimating the power spectrum
  covariance matrix with fewer mock samples}},
  \href{https://doi.org/10.1093/mnras/stw062}{\emph{\mnras} {\bfseries 457}
  (2016) 993} [\href{https://arxiv.org/abs/1509.00064}{{\ttfamily
  1509.00064}}].

\bibitem{2020MNRAS.492.1214P}
O.~H.~E. {Philcox} and D.~J. {Eisenstein}, \emph{{Computing the small-scale
  galaxy power spectrum and bispectrum in configuration space}},
  \href{https://doi.org/10.1093/mnras/stz3335}{\emph{\mnras} {\bfseries 492}
  (2020) 1214} [\href{https://arxiv.org/abs/1912.01010}{{\ttfamily
  1912.01010}}].

\bibitem{2019MNRAS.490.5931P}
O.~H.~E. {Philcox} and D.~J. {Eisenstein}, \emph{{Estimating covariance
  matrices for two- and three-point correlation function moments in Arbitrary
  Survey Geometries}},
  \href{https://doi.org/10.1093/mnras/stz2896}{\emph{\mnras} {\bfseries 490}
  (2019) 5931} [\href{https://arxiv.org/abs/1910.04764}{{\ttfamily
  1910.04764}}].

\bibitem{2013AJ....145...10D}
K.~S. {Dawson}, D.~J. {Schlegel}, C.~P. {Ahn}, S.~F. {Anderson},
  {\'E}.~{Aubourg}, S.~{Bailey} et~al., \emph{{The Baryon Oscillation
  Spectroscopic Survey of SDSS-III}},
  \href{https://doi.org/10.1088/0004-6256/145/1/10}{\emph{\aj} {\bfseries 145}
  (2013) 10} [\href{https://arxiv.org/abs/1208.0022}{{\ttfamily 1208.0022}}].

\bibitem{2015PhRvD..92h3532S}
R.~{Scoccimarro}, \emph{{Fast estimators for redshift-space clustering}},
  \href{https://doi.org/10.1103/PhysRevD.92.083532}{\emph{\prd} {\bfseries 92}
  (2015) 083532} [\href{https://arxiv.org/abs/1506.02729}{{\ttfamily
  1506.02729}}].

\bibitem{2015MNRAS.453L..11B}
D.~{Bianchi}, H.~{Gil-Mar{\'\i}n}, R.~{Ruggeri} and W.~J. {Percival},
  \emph{{Measuring line-of-sight-dependent Fourier-space clustering using
  FFTs}}, \href{https://doi.org/10.1093/mnrasl/slv090}{\emph{\mnras} {\bfseries
  453} (2015) L11} [\href{https://arxiv.org/abs/1505.05341}{{\ttfamily
  1505.05341}}].

\bibitem{2015arXiv150906384V}
M.~{Vargas-Maga{\~n}a}, S.~{Ho}, S.~{Fromenteau} and A.~J. {Cuesta}, \emph{{The
  clustering of galaxies in the SDSS-III Baryon Oscillation Spectroscopic
  Survey: Effect of smoothing of density field on reconstruction and
  anisotropic BAO analysis}}, {\emph{arXiv e-prints} (2015) arXiv:1509.06384}
  [\href{https://arxiv.org/abs/1509.06384}{{\ttfamily 1509.06384}}].

\bibitem{2014MNRAS.439L..21K}
F.~S. {Kitaura}, G.~{Yepes} and F.~{Prada}, \emph{{Modelling baryon acoustic
  oscillations with perturbation theory and stochastic halo biasing.}},
  \href{https://doi.org/10.1093/mnrasl/slt172}{\emph{\mnras} {\bfseries 439}
  (2014) L21} [\href{https://arxiv.org/abs/1307.3285}{{\ttfamily 1307.3285}}].

\bibitem{2016MNRAS.456.4156K}
F.-S. {Kitaura}, S.~{Rodr{\'\i}guez-Torres}, C.-H. {Chuang}, C.~{Zhao},
  F.~{Prada}, H.~{Gil-Mar{\'\i}n} et~al., \emph{{The clustering of galaxies in
  the SDSS-III Baryon Oscillation Spectroscopic Survey: mock galaxy catalogues
  for the BOSS Final Data Release}},
  \href{https://doi.org/10.1093/mnras/stv2826}{\emph{\mnras} {\bfseries 456}
  (2016) 4156} [\href{https://arxiv.org/abs/1509.06400}{{\ttfamily
  1509.06400}}].

\bibitem{2016MNRAS.457.4340K}
A.~{Klypin}, G.~{Yepes}, S.~{Gottl{\"o}ber}, F.~{Prada} and S.~{He{\ss}},
  \emph{{MultiDark simulations: the story of dark matter halo concentrations
  and density profiles}},
  \href{https://doi.org/10.1093/mnras/stw248}{\emph{\mnras} {\bfseries 457}
  (2016) 4340} [\href{https://arxiv.org/abs/1411.4001}{{\ttfamily 1411.4001}}].

\bibitem{2016MNRAS.460.1173R}
S.~A. {Rodr{\'\i}guez-Torres}, C.-H. {Chuang}, F.~{Prada}, H.~{Guo},
  A.~{Klypin}, P.~{Behroozi} et~al., \emph{{The clustering of galaxies in the
  SDSS-III Baryon Oscillation Spectroscopic Survey: modelling the clustering
  and halo occupation distribution of BOSS CMASS galaxies in the Final Data
  Release}}, \href{https://doi.org/10.1093/mnras/stw1014}{\emph{\mnras}
  {\bfseries 460} (2016) 1173}
  [\href{https://arxiv.org/abs/1509.06404}{{\ttfamily 1509.06404}}].

\bibitem{2007A&A...464..399H}
J.~{Hartlap}, P.~{Simon} and P.~{Schneider}, \emph{{Why your model parameter
  confidences might be too optimistic. Unbiased estimation of the inverse
  covariance matrix}},
  \href{https://doi.org/10.1051/0004-6361:20066170}{\emph{\aap} {\bfseries 464}
  (2007) 399} [\href{https://arxiv.org/abs/astro-ph/0608064}{{\ttfamily
  astro-ph/0608064}}].

\bibitem{2014MNRAS.439.2531P}
W.~J. {Percival}, A.~J. {Ross}, A.~G. {S{\'a}nchez}, L.~{Samushia},
  A.~{Burden}, R.~{Crittenden} et~al., \emph{{The clustering of Galaxies in the
  SDSS-III Baryon Oscillation Spectroscopic Survey: including covariance matrix
  errors}}, \href{https://doi.org/10.1093/mnras/stu112}{\emph{\mnras}
  {\bfseries 439} (2014) 2531}
  [\href{https://arxiv.org/abs/1312.4841}{{\ttfamily 1312.4841}}].

\bibitem{2019JCAP...01..016L}
Y.~{Li}, S.~{Singh}, B.~{Yu}, Y.~{Feng} and U.~{Seljak}, \emph{{Disconnected
  covariance of 2-point functions in large-scale structure}},
  \href{https://doi.org/10.1088/1475-7516/2019/01/016}{\emph{\jcap} {\bfseries
  2019} (2019) 016} [\href{https://arxiv.org/abs/1811.05714}{{\ttfamily
  1811.05714}}].

\bibitem{2019arXiv191002914W}
D.~{Wadekar} and R.~{Scoccimarro}, \emph{{The Galaxy Power Spectrum Multipoles
  Covariance in Perturbation Theory}}, {\emph{arXiv e-prints} (2019)
  arXiv:1910.02914} [\href{https://arxiv.org/abs/1910.02914}{{\ttfamily
  1910.02914}}].

\bibitem{2015JCAP...07..011A}
E.~{Aver}, K.~A. {Olive} and E.~D. {Skillman}, \emph{{The effects of He I
  {\ensuremath{\lambda}}10830 on helium abundance determinations}},
  \href{https://doi.org/10.1088/1475-7516/2015/07/011}{\emph{\jcap} {\bfseries
  2015} (2015) 011} [\href{https://arxiv.org/abs/1503.08146}{{\ttfamily
  1503.08146}}].

\bibitem{2018ApJ...855..102C}
R.~J. {Cooke}, M.~{Pettini} and C.~C. {Steidel}, \emph{{One Percent
  Determination of the Primordial Deuterium Abundance}},
  \href{https://doi.org/10.3847/1538-4357/aaab53}{\emph{\apj} {\bfseries 855}
  (2018) 102} [\href{https://arxiv.org/abs/1710.11129}{{\ttfamily
  1710.11129}}].

\bibitem{2019JCAP...10..029S}
N.~{Sch{\"o}neberg}, J.~{Lesgourgues} and D.~C. {Hooper}, \emph{{The BAO+BBN
  take on the Hubble tension}},
  \href{https://doi.org/10.1088/1475-7516/2019/10/029}{\emph{\jcap} {\bfseries
  2019} (2019) 029} [\href{https://arxiv.org/abs/1907.11594}{{\ttfamily
  1907.11594}}].

\bibitem{2016JCAP...02..018L}
T.~{Lazeyras}, C.~{Wagner}, T.~{Baldauf} and F.~{Schmidt}, \emph{{Precision
  measurement of the local bias of dark matter halos}},
  \href{https://doi.org/10.1088/1475-7516/2016/02/018}{\emph{\jcap} {\bfseries
  2016} (2016) 018} [\href{https://arxiv.org/abs/1511.01096}{{\ttfamily
  1511.01096}}].

\bibitem{2019arXiv191013970L}
A.~{Lewis}, \emph{{GetDist: a Python package for analysing Monte Carlo
  samples}}, {\emph{arXiv e-prints} (2019) arXiv:1910.13970}
  [\href{https://arxiv.org/abs/1910.13970}{{\ttfamily 1910.13970}}].

\bibitem{2002PhRvD..66j3511L}
A.~{Lewis} and S.~{Bridle}, \emph{{Cosmological parameters from CMB and other
  data: A Monte Carlo approach}},
  \href{https://doi.org/10.1103/PhysRevD.66.103511}{\emph{\prd} {\bfseries 66}
  (2002) 103511} [\href{https://arxiv.org/abs/astro-ph/0205436}{{\ttfamily
  astro-ph/0205436}}].

\bibitem{2013PhRvD..87j3529L}
A.~{Lewis}, \emph{{Efficient sampling of fast and slow cosmological
  parameters}}, \href{https://doi.org/10.1103/PhysRevD.87.103529}{\emph{\prd}
  {\bfseries 87} (2013) 103529}
  [\href{https://arxiv.org/abs/1304.4473}{{\ttfamily 1304.4473}}].

\bibitem{2016MNRAS.460.4188G}
H.~{Gil-Mar{\'\i}n}, W.~J. {Percival}, J.~R. {Brownstein}, C.-H. {Chuang},
  J.~N. {Grieb}, S.~{Ho} et~al., \emph{{The clustering of galaxies in the
  SDSS-III Baryon Oscillation Spectroscopic Survey: RSD measurement from the
  LOS-dependent power spectrum of DR12 BOSS galaxies}},
  \href{https://doi.org/10.1093/mnras/stw1096}{\emph{\mnras} {\bfseries 460}
  (2016) 4188} [\href{https://arxiv.org/abs/1509.06386}{{\ttfamily
  1509.06386}}].

\bibitem{Green:2019glg}
D.~Green et~al., \emph{{Messengers from the Early Universe: Cosmic Neutrinos
  and Other Light Relics}}, {\emph{Bull. Am. Astron. Soc.} {\bfseries 51}
  (2019) 159} [\href{https://arxiv.org/abs/1903.04763}{{\ttfamily
  1903.04763}}].

\bibitem{Baumann:2019tdh}
D.~D. Baumann, F.~Beutler, R.~Flauger, D.~R. Green, A.~Slosar,
  M.~Vargas-Magaña et~al., \emph{{First constraint on the neutrino-induced
  phase shift in the spectrum of baryon acoustic oscillations}},
  \href{https://doi.org/10.1038/s41567-019-0435-6}{\emph{Nature Phys.}
  {\bfseries 15} (2019) 465}
  [\href{https://arxiv.org/abs/1803.10741}{{\ttfamily 1803.10741}}].

\bibitem{Baumann:2017gkg}
D.~Baumann, D.~Green and B.~Wallisch, \emph{{Searching for light relics with
  large-scale structure}},
  \href{https://doi.org/10.1088/1475-7516/2018/08/029}{\emph{JCAP} {\bfseries
  1808} (2018) 029} [\href{https://arxiv.org/abs/1712.08067}{{\ttfamily
  1712.08067}}].

\end{thebibliography}\endgroup

\end{document}